\documentclass[aps,prx,superscriptaddress,nofootinbib,notitlepage,showpacs,floatfix,twocolumn]{revtex4-1}

\usepackage{graphicx,graphics,epsfig,subfigure,times,bm,bbm,amssymb,amsmath,amsfonts,mathrsfs}
\usepackage[scr=boondoxo,scrscaled=1.05]{mathalfa}
\usepackage[matrix,frame,arrow]{xypic}
\usepackage[pdfstartview=FitH]{hyperref}
\usepackage[pdftex]{color}

\newcommand{\beq}{\begin{equation}}
\newcommand{\eneq}{\end{equation}}
\newcommand{\beqnn}{\begin{equation*}}
\newcommand{\eneqnn}{\end{equation*}}
\newcommand{\beqy}{\begin{eqnarray}}
\newcommand{\eneqy}{\end{eqnarray}}
\newcommand{\beqynn}{\begin{eqnarray*}}
\newcommand{\eneqynn}{\end{eqnarray*}}

\newcommand{\ket}[1]{ | #1 \rangle  }
\newcommand{\bra}[1]{ \langle  #1 | }

\newcommand{\expect}[1]{\langle #1 \rangle}

\newcommand{\erf}[1]{Eq.~(\ref{#1})}

\def\XXint#1#2#3{{\setbox0=\hbox{$#1{#2#3}{\int}$}
     \vcenter{\hbox{$#2#3$}}\kern-.5\wd0}}

\newcommand{\bes} {\begin{subequations}}
\newcommand{\ees} {\end{subequations}}
	\newcommand{\bea} {\begin{eqnarray}}
	\newcommand{\eea} {\end{eqnarray}}

\newcommand{\Tr}{\mathrm{Tr}}

\newenvironment{proof}[1][Proof]{\noindent\textbf{#1.} }{\ \rule{0.5em}{0.5em}}

\newcommand{\ignore}[1]{}
\begin{document}

\title{Multiqubit Spectroscopy of Gaussian Quantum Noise}

\author{Gerardo~A. Paz-Silva}
\affiliation{{ Centre for Quantum Dynamics \& 
Centre for Quantum Computation and Communication Technology,  
Griffith University, Brisbane, Queensland 4111, Australia}} 

\author{Leigh M. Norris}
\affiliation{ \mbox{Department of Physics and Astronomy, Dartmouth College,
6127 Wilder Laboratory, Hanover, New Hampshire 03755, USA}}

\author{Lorenza Viola}
\affiliation{ \mbox{Department of Physics and Astronomy, Dartmouth College,
6127 Wilder Laboratory, Hanover, New Hampshire 03755, USA}}

\begin{abstract}

We introduce multi-pulse quantum noise spectroscopy protocols for spectral estimation of the noise affecting 
multiple qubits coupled to Gaussian dephasing environments including both classical and quantum sources. 
Our protocols are capable of reconstructing all the noise auto- and cross-correlation spectra 
entering the multiqubit dynamics.  We argue that this capability is crucial not only for metrological purposes, 
as it provides access to the asymmetric spectra associated with non-classical environments, but ultimately 
for achieving quantum fault-tolerance,  as it enables the characterization of bath correlation functions. 
Our result relies on (i) an exact analytic solution for the reduced multiqubit dynamics that holds in the presence of 
an arbitrary Gaussian environment and dephasing-preserving control; (ii) the use of specific timing symmetries 
in the control, which allow for a frequency comb to be engineered for all filter functions of interest, and for the 
spectra to be related to experimentally accessible qubit observables. 
We show that quantum spectra have distinctive dynamical signatures, which we explore in two paradigmatic 
open-system models describing spin and charge qubits coupled to bosonic environments. 
Complete multiqubit noise spectroscopy is demonstrated numerically in a realistic setting consisting of 
two-exciton qubits coupled to a phonon bath. The estimated spectra allow us to accurately predict the 
exciton dynamics as well as extract the temperature and spectral density of the quantum environment.
\end{abstract}

\date{\today}
\maketitle
\tableofcontents

\section{Introduction}
\vspace*{-2mm}
\subsection{Context and motivation}

Quantum systems are naturally susceptible to interactions with external, classical or quantum, degrees of freedom. 
To the extent that such ``environment'' (or ``bath'') degrees of freedom are typically largely unknown and not directly accessible, these unwanted interactions pose a major challenge for the implementation of coherence-enabled quantum technologies and scalable quantum information processing. 
A number of techniques have been developed to address this challenge, ranging from physical-layer 
dynamical error suppression strategies to full-fledged fault-tolerant quantum error correction \cite{Lidar:book}. 
While general-purpose error control protocols may be constructed without 
making reference to a complete specification of the underlying noise sources, 
this high degree of robustness against ``model uncertainty'' tends to come at the cost of inefficient 
scaling with the dimension of the system one wants to protect~\cite{CDD,DCG}. Likewise, 
no optimal performance can be guaranteed, in terms of achievable fidelities and required overheads, 
in a specific setting of interest.  In fact, precise knowledge 
of the open-system model describing the interaction of the target system with its environment is a 
prerequisite for optimal control methods to be viable~\cite{OptCtrl,Calarco}. 
Ideally, one would want that the relevant noisy environment be fully characterized, so that 
control design can be optimally tailored and error suppression achieved as efficiently as possible.

On the plus side, the exquisite sensitivity of qubits to their surrounding environment is a boon that can be exploited 
for sensing purposes, and is helping to unlock unprecedented opportunities in single- and multi-parameter quantum estimation and metrology, see e.g.~\cite{QMetro,QMagneto,Misha,DattaReview} for representative contributions.
It has in fact long been appreciated that qubits can in principle be used as ``spectrometers of quantum noise''~\cite{QSpectroOld,Lara2004}. Loosely speaking, measuring a qubit's response to the interaction with an  environment of interest, information about the noise properties may be inferred, 
much in the same way that light-matter interactions are used in traditional spectroscopy.
The idea of using a qubit as a probe for noise has been recently formalized into 
open-loop {\it quantum noise spectroscopy} (QNS) protocols~\cite{OlderRefs, Alvarez2011, Norris2016},
by leveraging the fact that the controlled dynamics of an open quantum system can be 
characterized in the frequency domain through convolution integrals that involve purely control-dependent 
{\it filter functions} (FFs)~\cite{Kurizki,MikeFF, Paz2014} and the {\it noise power spectra} -- 
as determined by the Fourier transform of the relevant bath correlation functions~\cite{Breuer:book}. 

Single-qubit QNS protocols designed to characterize a classical Gaussian noise source 
in the dephasing regime have been successfully demonstrated in experimental platforms including  
solid-state nuclear magnetic resonance \cite{Alvarez2011}, superconducting and spin qubits 
\cite{Bylander2011, Muhonen2014, Oliver, Yan2013, SpinQubits} to as well as nitrogen vacancy centers 
in diamond~\cite{NVs}.  Characterization of discrete non-Gaussian phase noise has, likewise, been 
implemented in trapped ions \cite{Kotler}, whereas general QNS protocols for reconstructing high-order 
spectra of non-Gaussian classical and quantum dephasing environments have been proposed in 
\cite{Norris2016}. Central to QNS protocols is the idea that, by suitably tailoring the external 
control, and hence the FFs describing the ensuing modulation in the frequency domain, one may engineer 
a frequency comb which makes it possible to  ``deconvolve'' the effect of the noise and sample 
a desired spectrum at a set of control-dependent harmonic frequencies. Notably, from a system-identification 
standpoint, the resulting spectral estimates are {\em non-parametric} in the sense that 
no specific functional form is assumed \cite{Percival}.

Despite the above advances, QNS protocols that use a single qubit as a probe face intrinsic limitations 
-- even in the simplest yet important scenario where noise may be taken to be stationary and 
obey Gaussian statistics. 
First, as remarked in \cite{Norris2016}, although a general spectrum $S(\omega)$ is {\em asymmetric}
about $\omega=0$, only the even contribution $S_{1,1}^+(\omega)\equiv S(\omega) +S(-\omega) $
enters the dynamics (and can thus be reconstructed) in generic single-qubit dephasing scenarios.
Second, noise may exhibit non-trivial {\em spatial correlations}, which may only become manifest in the 
coherence dynamics of multiple qubit probes at different locations.  
Let $ S_{\ell,\ell'}(\omega) \equiv S_{\ell,\ell'}^+(\omega) + S_{\ell,\ell'}^-(\omega) $
denote the spectrum of the noise affecting a pair of qubits $\ell, \ell'$, where the ``classical'' ($+$)
and the ``quantum'' ($-$) components depend on Fourier-transformed
commutators and anti-commutators of bath operators, respectively. 
Since $S_{\ell,\ell}^-(\omega)  = S_{\ell,\ell'}(\omega) - S_{\ell',\ell}(-\omega)$,
any quantum contribution to the ``self-spectrum'' 
is undetected, as noted, by single-qubit QNS in a generic dephasing setting. 
While a protocol capable of accessing the {\em classical}  ``cross-correlation 
spectrum'', $S_{\ell,\ell'}^+(\omega), \ell \ne \ell'$, has been recently put forward~\cite{Cywinski2}, 
the assumption of classical noise cannot be expected to be {\em a priori} or universally valid. 
This motivates the search for QNS protocols able to characterize arbitrary, {\em quantum and classical}, 
noise sources simultaneously influencing {\em multiple qubits}. 

Such a complete spectral estimation is crucial for a variety of reasons. Most obviously, 
in the context of developing improved techniques for characterizing  
quantum information processing systems of increasing scale and complexity, 
it would enable, as noted, application of optimal control methods to multiqubit operations, as well as 
validation of engineered noise environments --  in particular in the context of (analog) 
open-system quantum simulators \cite{Mostame, Gross}.  A number of other implications across quantum 
science may be envisioned, however; in particular:

$\bullet$ {\it The environment as a resource.} Several schemes have been proposed for using quantum 
environments as a resource, notably, for entangling two or more qubits via their interaction with a common 
bath~\cite{Braun2002,Benatti2003,Oh2006,An2007,McCutcheon2009,Bhaktavatsala2011,multiDD,Krzywda:2016}.
While no detailed knowledge of the bath is necessary to generate entanglement, full knowledge of the power 
spectra is instrumental to generate {\em controlled} entanglement, i.e., to retrieve it on-demand or 
perform a precise entangling gate. 
Along similar lines, it has recently been shown that suitable noise can make a set of commuting 
Hamiltonians universal for quantum computation~\cite{Arenz2016}. 
While knowledge of the noise process is assumed in this proposal, use of a suitable QNS protocol 
would enable such information to be directly extracted from measurable quantities. 

$\bullet$ {\it Correlations and quantum fault-tolerance.} 
While initial versions of the accuracy threshold theorem were derived 
under restrictive noise assumptions~\cite{FaultTolerance},  
the types of noise under which the theorem holds have expanded over the years~\cite{PreskillSufficient}. 
In particular, it has been established that a threshold still exists in the presence of quantum Gaussian 
{\em correlated} noise~\cite{NgPreskill}, 
provided that the two-point correlations of the relevant bath operators,  
say, $\langle B_\ell(t) B_{\ell'}(t')) \rangle$, decay sufficiently fast 
as a function of the qubit spatial separation, $|\vec{r}_\ell-\vec{r}_{\ell'}|$.
The slower correlations decay with distance, the lower the threshold value and the higher 
the necessary gate fidelity required to guarantee fault-tolerance~\cite{Novais, Hutter2014}. 
Consequently, even if a physical system supports high single-qubit gate fidelities, 
it will most likely not be a good candidate for a scalable quantum computer if the bath correlations are
insufficiently well-behved.
Similar arguments can be made for the effect of bath correlations in quantum metrology protocols that use multiple probes to achieve the Heisenberg limit~\cite{Jeske2014}. Thus, using multiqubit QNS to quantitatively characterize the spatial dependence of bath correlations should be one of the first tests to determine the suitability of a platform for a given quantum technology. 

$\bullet$ {\it Quantum metrology and thermometry.} Metrology is a task of fundamental significance for 
science and technology. The simplest scenario is one where the response of a probe is used to infer 
information about a physical degree of freedom. In a typical magnetometry setting, 
for instance, the Hamiltonian ruling the evolution of the probe qubit under the influence of a magnetic field 
of unknown strength $\mu$ is given by $  H(t) =  \mu  Z.$ By suitably preparing the probe, and 
tracking the expectation value of a particular observable over time, it is possible to extract the parameter $\mu$ 
\cite{QMagneto}.  In essence, this is a limiting case of a single-qubit QNS protocol, for a classical 
(deterministic) signal 
$\beta(t)$ whose mean is $\langle \beta(t) \rangle = \mu$ and higher-order cumulants vanish. 
More generally, QNS may be regarded as a form of multi-parameter estimation, where the 
noise spectra (rather than just the mean) grant access to information stored in the correlations of the 
bath operators. 

When the information about the noise spectra is augmented with prior knowledge 
about the noise origin, it is possible to further infer physical parameters of interest. 
A relevant example is extracting the temperature, $T_B$, of a bosonic 
bath~\cite{Ivette,BosonicThermo}. 
Taking the ratio of the above-mentioned symmetric and antisymmetric spectral components, 
$S^{+}_{\ell,\ell'} (\omega) /S^{-}_{\ell,\ell'} (\omega)$, for $\omega >0$,  gives access to 
$\coth(\omega \beta/2)$, from which one may extract $\beta=\hbar/k_B T_B$ and thereby 
$T_B$. In turn, estimating $S^{-}_{\ell,\ell'} (\omega)$ gives one access to the spectral density 
function $J_{\ell,\ell'} (\omega)$ describing the coupling of qubits $\ell,\ell'$ to the oscillator bath.
It is interesting to contrast a single- vs. multiple-qubit QNS setting in the dephasing regime.
As remarked, only a multiqubit QNS protocol can estimate {\it both} $S^{+}_{\ell,\ell'} (\omega)$ 
and $S^{-}_{\ell,\ell'} (\omega)$ for a generic system-bath coupling operator.
In contrast, with access to $S^{+}_{\ell,\ell'} (\omega)$ alone, 
an estimate of $T_B$ may be given only by assuming a functional form for the 
spectral density. Using two probes removes the need for these extra assumptions, 
thus showcasing the power of multiqubit QNS.

\subsection{Summary of main results} 

In this work, we introduce QNS protocols capable of fully characterizing {\em classical and quantum 
Gaussian noise} 
on a set of $N$ qubits in the dephasing regime. We first lay out, in Sec.~\ref{sec:frame}, the 
necessary open-quantum system and quantum control background.  In particular, we introduce 
the two classes of dephasing models we focus attention on (\ref{sec::background}), 
distinguished by the different nature (generic, full-rank vs rank-one) of the system-bath coupling
-- as well as the relevant control resources (\ref{sec::ControlAssumpt}). 
Two settings of increasing complexity are examined, depending on whether control operations are 
restricted to purely local (single-qubit) $\pi$ pulses, or, additionally, non-local (swap) gates  
are allowable. Sec.~\ref{sec::RD} contains our first result, namely, an {\em exact} analytical 
expression for the time-dependent expectations of arbitrary $N$-qubit Pauli observables, and
hence the controlled reduced dynamics 
-- valid under the sole assumptions that the noise 
has {\em Gaussian} statistics and the {\em dephasing} nature of the system-bath interaction is 
preserved by the applied control. To the best of our knowledge, this generalizes exact results
derived under the explicit assumption of bosonic environments \cite{Doll2008,Paz2014,multiDD}.

Before delving into the construction of the QNS protocols, we devote Sec.~\ref{sec::physSig} to 
elucidate the physical significance of both self- and cross- quantum spectra, $S^-_{\ell, \ell}(\omega)$ 
and $S^-_{\ell, \ell'}(\omega)$ with $\ell'\ne \ell$, respectively. We find that, for otherwise identical 
environments, the details of how the system couples to the environment play an 
important role in determining different spectral signatures in the reduced dynamics. Notably, 
even in a single-qubit setting, $S^{-}_{1,1} (\omega)$ may result in {\em observable phase evolution} if 
the coupling operator has rank-one.  Regardless, we show that the quantum self-spectra are crucial 
in determining the {\em steady-state} behavior in relaxation dynamics, and argue that  two-qubit QNS 
provides a minimal setting for reconstructing these spectra in an exactly solvable dephasing regime.
We further show how quantum spectra are ultimately responsible for the ability of the environment to 
mediate entangling interactions between uncoupled qubits and, more generally, generate quantum 
correlations.  In particular, at variance with existing approaches where a common bath is assumed,
we illustrate how entanglement generation may be possible also for qubits coupled to {\em independent} 
baths -- as long as suitable swap-based (non-entangling) control is applied.

Sec.~\ref{sec::protocols} is the core section of the paper, presenting in detail both the design principles 
and implementation steps of the proposed multiqubit QNS protocols. In particular, special emphasis is
given to introducing and analyzing the key enabling {\em symmetry requirements} (\ref{CtrlKit}), and to detail the  
execution of the protocol in the simplest yet practically relevant two-qubit setting (\ref{sec::2qsetting}).  
In the process, we show how it is possible to construct dynamical decoupling (DD) sequences which 
combine local and non-local (swap) gates and achieve {\em arbitrarily high cancellation order} 
through concatenation, in principle -- a result that may be of independent interest. 
We stress that even in their most general form, our QNS protocols do {\em not} assume entangling 
unitary gates nor initially entangled qubit states. Remarkably, by employing only {\em local} pulses, 
all spectra except $S^-_{\ell, \ell}(\omega)$ can be reconstructed -- the latter, however, becoming also accessible 
if prior knowledge about the nature (e.g. bosonic) of the environment is available. 

The proposed protocols are numerically implemented in a realistic setting of two-exciton qubits 
coupled to an equilibrium phonon bath in Sec.~\ref{sec:Exciton}, by assuming access to 
{\em local qubit-selective control alone}. The numerical reconstructions of the spectra are used, 
in particular, to implement quantum thermometry of the phonon bath, as outlined 
above.  To further test the accuracy of the results, we also use the obtained spectral 
estimates to predict the qubit dynamics under both free evolution and representative 
DD control, by specifically tracking the influence of quantum vs. classical spectral signatures.
Our results demonstrate the need to properly account for the quantum spectra in order to 
accurately predict dynamical behavior in general.

\section{Quantum noise spectroscopy framework}
\label{sec:frame}

\subsection{Open-system model}
\label{sec::background}
 
We consider an open quantum system $S$, consisting of $N$ qubits, coupled to an uncontrollable environment (bath) $B$. The joint system is described by the Hamiltonian $H=H_S+H_B+H_{SB}$, where $H_S$ and $H_B$ are the internal Hamiltonians of $S$ and $B$, respectively, and $H_{SB}$ is the interaction between the two. We restrict ourselves to dephasing noise models, i.e., $[H_S,H_{SB}]=0$. While our analysis may  be extended to more general dephasing interactions, we assume for concreteness that $H$ contains at most two-body coupling terms between the qubits. In the interaction picture associated with $H_S$ and $H_B$, we may write the relevant Hamiltonian in the form 
\begin{align}\label{NqH} 
H_I(t)  &= \sum_{\ell=0}^N Z_\ell \otimes B_\ell(t) + \sum_{\substack{\ell,\ell'=1\\\ell\neq\ell'}}^N Z_{\ell\ell'} \otimes B_{\ell\ell'}(t), 
\end{align}
where $Z_0=\mathbf{1}$, $Z_{\ell}$ ($\ell\neq 0$) is the Pauli $Z$ operator acting on qubit $\ell$, 
$Z_{\ell\ell'} \equiv  Z_\ell\otimes Z_{\ell'}$ ($\ell, \ell'\neq 0$),  
and 
\begin{align}
\label{eq::Bnq}
B_\ell(t) = \zeta_{\ell} (t) \mathbf{1} + \tilde{B}_{\ell}(t), \quad
B_{\ell\ell'}(t)= \zeta_{\ell\ell'}(t) \mathbf{1} .
\end{align}
Here, $B_\ell(t)$ and $ \tilde{B}_{\ell}(t)$ are time-dependent bath operators coupled to qubit $\ell$, and $\zeta_{\ell} (t)$, 
$\zeta_{\ell\ell'}(t)$ are classical stochastic processes  coupled to qubit $\ell$ and qubit pair $\ell\ell'$, respectively.  
In this way, we allow for single-qubit combined noise sources of 
both classical and quantum nature, along with classically fluctuating inter-qubit couplings. For simplicity, we assume that 
$ \tilde{B}_{\ell}(t)$ is statistically independent of both $\zeta_{\ell} (t) $ and $\zeta_{\ell\ell'}(t)$.  

Two special cases of $H_I(t)$ frequently arise in physical systems. Most commonly, each qubit corresponds to a (pseudo)spin-$1/2$ 
degree of freedom, which couples to the bath by full-rank Pauli operators, such as $Z_{\ell}$ and $Z_{\ell\ell'}$.  Alternatively, for qubits
described in terms of the presence/absence of a (quasi)particle in one of two states, coupling to the bath occurs via rank-1 projectors, say, 
$\ket{0}\bra{0}_\ell$ or $\ket{00}\bra{00}_{\ell\ell'}$.  We formally account for these two scenarios by allowing for a ``pure-bath'' term 
proportional to $Z_0$ in  \erf{NqH} and letting  

(i) $B_0(t) = 0$ when coupling operators are full-rank (``{\em M1 models}'' henceforth); 

(ii) $B_0(t)=\sum_{\ell=1}^NB_\ell(t)$ when coupling operators have rank-1 (``{\em M2 models}'' henceforth).  

A paradigmatic M1 model is the well-known purely dephasing linear spin-boson model \cite{Leggett}
in which case, relative to the interaction picture associated with the free oscillator-bath Hamiltonian 
$H_B= \sum_k \Omega_k a_k^\dagger a_k$, $\Omega \geq 0$, 
the relevant time-dependent bath operators are \cite{multiDD}
\begin{align}
\label{eq::SBBt}
B_{\ell}(t) = \sum_k ( e^{i \Omega_k t }g_k^{\ell} a_k^\dagger  +e^{-i \Omega_k t }g_k^{\ell*} a_k),  
\end{align}
with  $g_k^\ell \in {\mathbb C}$ quantifying the strength of the coupling between qubit 
$\ell$ and the $k$th bosonic mode. 
Likewise, the recent work on cross-correlation QNS in \cite{Cywinski2} corresponds to a M1 model 
where noise is purely classical and single-qubit: specifically,  
$\zeta_{\ell\ell'}(t)\equiv 0$ and $\zeta_{\ell}(t)$ models a Gaussian random telegraph noise process, as 
relevant to superconducting systems.  

M2 models are characteristic, in particular, of excitonic qubit systems 
\cite{Hodgson,Cotlet:2014} in which case, for the same $H_B$ given above and by associating the 
computational-basis state $\ket{0}\bra{0}_\ell = (Z_\ell +  \mathbf{1}_\ell)/2$ to the presence of an exciton, 
the relevant interaction Hamiltonian may be written as 
\begin{align} 
\label{eq::ExBt}
H_I(t) &\!= \!\sum_{\ell =1}^N 
|0\rangle \langle 0|_\ell \otimes B_\ell(t) +\!\!\! \sum_{\ell, \ell'\ne \ell=1}^N \!\!\!\!\!|00\rangle \langle 00|_{\ell \ell'} \, B_{\ell \ell'}(t) , 
\end{align}
with $B_\ell(t)$ having the same form given in Eq.~(\ref{eq::SBBt}).

In order to treat single- and two-qubit terms on similar footing, we will often write \erf{NqH}  in the more compact form 
\beq
\label{Dephase}
H_I(t) = \sum_{a\in \mathcal{I}_N} Z_{a} \otimes {B}_a(t), 
\eneq
where $\mathcal{I}_N \equiv \big\{ 0, \ell,\ell{\ell'}|\;\ell,{\ell'}\in\{1,\ldots ,N\},\;\ell\neq {\ell'}\,\big\}$, and each $Z_a$ 
has its associated bath operator ${B}_a(t)$, as per Eq.~\eqref{NqH}.  Occasionally, we will use the notation 
$\bar{\ell}$ to automatically imply $\bar{\ell}\ne \ell$ and, if necessary to distinguish between the indices 
$\ell$ and $\ell \bar{\ell}$, it will be understood that $\ell,\bar{\ell}  \in \{1,\cdots, N\}$.

\subsection{Control resources}
\label{sec::ControlAssumpt}

Beside interacting with the bath, the $N$ qubits are subject to external control generated by a Hamiltonian 
$H_\text{ctrl}(t)$.  We restrict ourselves to control that preserves the dephasing character of the noise in the interaction picture 
associated with $H_\text{ctrl}(t)$ (aka the ``toggling frame").  Upon introducing the control propagator 
$U_{\text{ctrl}}(t) \equiv \mathcal{T}_+[\text{exp}(-i\int_0^tdsH_\text{ctrl}(s))]$, 
\erf{Dephase} can be written in the toggling frame as  
\beq
\label{eq::togglingH} 
\tilde{H}(t) \!=\! U_{\text{ctrl}}^\dag(t)H_I(t) U_{\text{ctrl}}(t)\!=\!\! \sum_{a,a'\in\mathcal{I}_N} \!\!y_{a,a'} (t) Z_{a} \otimes {B}_{a'}(t),
\eneq
where the assumed dephasing property implies that all system operators in $\tilde{H}(t)$ still commute, as in \erf{NqH}. 
The $y_{a,a'} (t)$ are ``switching functions" induced by the control, the exact form of which depend on 
$H_{\textrm{ctrl}}(t)$, as we specify next.

\subsubsection{Local and non-local control sequences}

While we work in the idealized limit where control operations are perfect, we 
consider two types of dephasing-preserving control of increasing complexity.
The first is sequences of instantaneous $\pi$-pulses, which are built as products of operators $X_\ell$ and $Y_\ell$ 
and act \emph{locally} on the qubits. This family of control includes single-qubit 
(``bang-bang'') DD sequences.  Each $\pi$-pulse, denoted $\Pi_{\mathcal{A}}$, has the action 
$\Pi_{\mathcal{A}}^\dag Z_a\Pi_{\mathcal{A}}=-Z_a$ for $a\in\mathcal{A}\subseteq\mathcal{I}_N-\{0\}$. For $N=3$, 
for example, $X_1X_2=\Pi_{\{1,2,13,23\}}$ since
 \begin{align*}
&(X_1X_2)Z_\ell (X_1X_2)^\dag=-Z_\ell, \\
 &(X_1X_2)Z_\ell Z_3 (X_1X_2)^\dag=-Z_\ell Z_3, 
 \end{align*}
for $\ell=1,2$. A control sequence of total duration $T$, composed of $n$ instantaneous $\pi$-pulses, takes the form 
 $$U_{\text{ctrl}}(T)=U_{f}(t_{n+1},t_n)\prod_{i=1}^n\Pi_{\mathcal{A}_i}U_{f}(t_i,t_{i-1}),$$
\noindent
where $t_0=0$, $t_{n+1}=T$, and $U_{f}(t_i,t_j)$ denotes free evolution under ${H_I}(t)$ from time $t_j$ to $t_i$. Transforming $H_I(t)$ into the toggling frame implies that the switching functions $y_{a,a'} (t)$ in \erf{eq::togglingH} are nonzero only when $a=a'$, as the  $\pi$-pulses act locally on the qubits. We thus refer to control schemes involving only instantaneous $\pi$-pulses as \emph{diagonal} control. The $y_{a,a}(t)$, assume values of $\pm 1$, changing sign with the application of a pulse $\Pi_\mathcal{A}$ such that $a\in\mathcal{A}$. Clearly, since $Z_0=\mathbf{1}$, $y_{0,0}(t)=1$ for all $t$, with no sign changes.

The second form of dephasing-preserving control we consider are instantaneous swap gates between any pair of qubits.  The gate SWAP$_{\ell,\ell'}$ acts \emph{non-locally} on qubits $\ell$ and $\ell'$, effecting the transformation 
$\text{SWAP}_{\ell,\ell'}^\dag\, Z_\ell\, \text{SWAP}_{\ell,\ell'}=Z_{\ell'}$. A sequence consisting of both instantaneous 
$\pi$-pulses and swap gates has a control propagator of the form 
$$U_{\text{ctrl}}(T)=U_{f}(t_{n+1},t_n)\prod_{i=1}^nP_iU_{f}(t_i,t_{i-1}),$$
where $P_i$ is either $\Pi_{\mathcal{A}_i}$ or $\text{SWAP}_{\ell_i,\ell_i'}$. The inclusion of swap gates in the control sequence makes the switching functions in \erf{eq::togglingH} \emph{non-diagonal}, that is, there exist $a,a'\in\mathcal{I}_N$ with $a\neq a'$  such that $y_{a,a'}(t)\neq 0$ for some $t$. Additionally, the switching functions take values of $\pm 1$ and $0$, rather than just $\pm 1$.  For illustration, consider $N=2$, 
and suppose we apply the control sequence described by the propagator 
$U_{\text{ctrl}}(T)=U_{f}(T,T/2)\Pi_{\{1,12\}}$ SWAP$_{1,2}U_{f}(T/2,0)$. 
The toggling-frame Hamiltonian becomes
$$ \tilde{H}(t) \!=\!\! \begin{cases} 
\! B_{0}(t)\!+\!Z_1\!B_{1}\!(t) \!+\! Z_2B_{2}(t)\!+\! Z_{12}B_{\!12}(t),\;t\!\in\!\!\left[0,\!\frac{T}{2}\right) ,\\
\! B_{0}(t)\!-\!Z_1\!B_{2}(t) \!+\! Z_2B_{1}\!(t)\!-\!Z_{12}B_{\!12}(t),\;t\!\in\!\!\left[\!\frac{T}{2},T\right) .\\
\end{cases}\!\!\!\!\!
$$  
When this Hamiltonian is written in the form of \erf{eq::togglingH}, 
it is straightforward to see that 
\begin{align*}
y_{\ell,\ell}(t) &= \begin{cases} 
+1 & t\in\left[0,T/2\right),\\
 0 & t\in\left[T/2,T\right),
\end{cases} \\
y_{1,2}(t) &= \begin{cases} 
0 & t\in\left[0,T/2\right),\\
 -1 & t\in\left[T/2,T\right),
\end{cases}\\
y_{2,1}(t) &= 
\begin{cases} 0 & t\in\left[0,T/2\right),\\
 +1 & t\in\left[T/2,T\right),
\end{cases}\\
y_{12,12}(t) &= 
\begin{cases} +1 & t\in\left[0,T/2\right),\\
 -1 & t\in\left[T/2,T\right). 
\end{cases}
\end{align*}
A compact way to represent a non-diagonal control sequence is 
via a corresponding ``switching matrix" with elements $[y(t)]_{a,a'}=y_{a,a'}(t)$. For example, the switching matrix corresponding to the above two-qubit sequence is 
\begin{align}
[y(t)] & = \begin{cases}
\left(\begin{array}{cccc}
+1& 0& 0& 0\\
0& +1 &0 &0\\
0& 0& +1& 0\\
0& 0& 0& +1
\end{array}\right)\vspace*{2mm},  \;\;t\in\left[0,T/2\right),\\ 
\left(\begin{array}{cccc}
+1& 0& 0& 0\\
0& 0& -1& 0\\
0& +1 & 0 & 0\\
0& 0& 0& -1
\end{array}\right), \;\; t\in\left[T/2,T\right),\\
\end{cases}
\end{align}
where the rows and columns are ordered by 0, 1, 2, and 12.

From an experimental standpoint, non-local control via swap gates is clearly more taxing than purely local control via 
$\pi$-pulses. As we will show in Sec.~\ref{sec::protocols}, complete spectral characterization of dephasing models requires
both $\pi$-pulses and swap gates in general.  Under prior knowledge that the bath is bosonic and thermal, however, protocols employing $\pi$-pulses alone suffice to reconstruct all classical spectra as well as the quantum cross-spectra, making it possible to also infer the quantum self-spectra in a way to be made more precise later. As expected, without prior knowledge of the bath or noise model, there is a trade-off between the complexity of the available control and the spectral quantities of the bath that can be directly accessed and reconstructed.

\subsubsection{Filter functions}

Since we are concerned with the spectral properties of the bath, we will work primarily in the frequency domain.
In the frequency domain, the effects of the applied control are described by transfer FFs, which are 
related to the Fourier transforms of the switching functions.
Using the general formalism developed in ~\cite{Paz2014,multiDD}, all relevant FFs can be written in terms of a set of 
easily computable {\em fundamental FFs}. 
The fundamental FFs for the controlled dephasing setting of interest are the first- and second- order, given by
\begin{align}
\label{eq::F1}&
F_{a,a'}^{(1)} (\omega,t) =  \int_{0}^t ds\, y_{a,a'} (s) e^{i \omega s }\;\;\;\text{and}\\
\label{eq::F2}&
F_{a,a';b,b'}^{(2)} (\omega,t) = \!\! \int_{0}^t\!\! ds \int_0^{s} \!\!ds' \, y_{a,a'} (s) y_{b,b'} (s') e^{i \omega (s-s') }.
\end{align}
Note that the first-order fundamental FF, $F_{a,a'}^{(1)} (\omega,t)$, is simply the finite Fourier transform of the 
switching function $y_{a,a'}(t)$.

\subsection{Noise assumptions and spectra}

Statistical features of the noise are compactly described by the cumulants of the bath operators \cite{Kubo,Kardar:book,multiDD}. For the zero-mean Gaussian noise we consider\footnote{We stress that the notion of {\em statistical Gaussianity} used here is not to be confused with a Gaussian functional form of the power spectra, i.e., the power spectra of a (statistically) Gaussian noise process can have an arbitrary (non-Gaussian) functional form.}, the only non-vanishing cumulants are the second-order cumulants, equivalent to two-point connected correlation functions. For a bath operator $B_a(t)=\tilde{B}_a(t)+\zeta_a(t)$, containing statistically independent quantum and classical noise sources as in \erf{eq::Bnq}, the second cumulant reduces to
\begin{align*}
C^{(2)}&(B_a(t_1)B_b(t_2)) \notag \\
&=\langle \tilde{B}_a(t_1)\tilde{B}_b(t_2)\rangle_q+\langle \zeta_a(t_1)\zeta_b(t_2)\rangle_c
\\&=C^{(2)}(\tilde{B}_a(t_1)\tilde{B}_b(t_2))+C^{(2)}(\zeta_a(t_1)\zeta_b(t_2)).
\end{align*}
Here, $\langle \cdot \rangle_q \equiv  \Tr_B[ \cdot \rho_B]$ indicates a quantum expectation value with respect to the initial bath 
state $\rho_B$, while  $\langle \cdot \rangle_c$ indicates a classical ensemble average. Stationarity of the bath 
implies time-translational invariance, in that a second-order cumulant at times $t_1$ and $t_2$ is fully specified by the lag time 
$\tau=t_1-t_2$, 
hence $\langle B_a(t_1)B_b(t_2)\rangle_{c,q} = \langle B_a(\tau)B_b(0)\rangle_{c,q}$.

The aim of QNS is characterizing the spectral properties of noise affecting a quantum system. Our QNS protocols estimate the power spectra, defined as the Fourier transforms of the second-order cumulants with respect to the lag time $\tau$, 
\begin{align}
S_{a,b}(\omega)= \int_{-\infty}^\infty d \tau e^{- i \omega \tau} C^{(2)} (B_{a}(\tau),B_{b}(0)).
\end{align}
Distinctions between classical and quantum noise emerge when we consider the ``quantum spectra"
\begin{align}
S_{a,b}^-(\omega)&\equiv \int_{-\infty}^\infty d \tau e^{- i \omega \tau} \langle[B_{a}(\tau),B_{b}(0)]\rangle_{q}
\\&=S_{a,b}(\omega)-S_{b,a}(-\omega).
\label{eq::qspectradef}
\end{align}
Because the commutator above vanishes for classical noise, 
$S_{a,b}^-(\omega)$ is non-zero {\em only} when the bath is quantum. In contrast, the ``classical spectra"
\begin{align}
S_{a,b}^+(\omega)&\equiv  \int_{-\infty}^\infty d \tau e^{- i \omega \tau} \langle \{B_{a}(\tau),B_{b}(0)\} \rangle_{c,q}
\\&=S_{a,b}(\omega)+S_{b,a}(-\omega). 
\label{eq::cspectradef}
\end{align}
can be non-zero for both classical and quantum baths. 
We refer to the $S_{a,b}^\pm(\omega)$ as ``self-spectra" when $a=b$ 
and as ``cross-spectra'' when $a\neq b$. Physically, these spectra describe, in the 
frequency domain, the ``auto-correlation'' of a noise operator with itself -- or, respectively, 
its ``cross-correlation'' with another one -- at two different points in time.
Mathematically, the self-spectra are real, whereas the cross-spectra are in general complex. 
All spectra satisfy the following symmetry properties:
\begin{align}
\label{eq::spsymmetry}
(S^{\pm}_{a,b} (\omega))^* = \pm \,(S^{\pm}_{a,b} (-\omega)) =  S^{\pm}_{b,a} (\omega).
\end{align}
 
For bosonic baths, considered as an example of M1-M2 models in Eqs. (\ref{eq::SBBt})-(\ref{eq::ExBt}),  
Gaussianity conditions are satisfied when the the bath is initially at thermal equilibrium. 
In the continuum limit, a thermal bath at temperature $T_B$ has spectra  
\begin{align}
\label{eq::SBllSpectra} 
S_{\ell,\ell'}(\omega)=\pi J_{\ell,\ell'}(\omega) \left\{ \begin{array}{ll}
         \text{coth}(\beta\omega/2)+1, &\omega\geq 0\\
        \text{coth}(-\beta\omega/2)-1, &\omega< 0 \end{array} \right.,
\end{align}
where as usual $\beta\equiv\hbar/k_BT_B$ denotes inverse temperature and 
\begin{equation}
J_{\ell,\ell'}(\omega) \!=\! \sum_{k} [\delta(\omega+\Omega_k) g_{k}^{\ell} g_{k}^{\ell'*} + \delta(\omega-\Omega_k) g_{k}^{\ell'} g_{k}^{\ell*}]
\label{eq::sd}
\end{equation}
is the spectral density function for qubits $\ell,\ell'$. From $S_{\ell,\ell'}(\omega)$, the quantum and classical spectra can be determined from Eqs. (\ref{eq::qspectradef}) and (\ref{eq::cspectradef}), yielding
\begin{align}
\label{eq::cqspectra}
S^{+}_{\ell,\ell'} (\omega) &=  2\pi J_{\ell,\ell'} (\omega) \coth( \beta |\omega| /2), \\\label{eq::cqspectra2}
S^{-}_{\ell,\ell'} (\omega) &=  2\pi J_{\ell,\ell'} (\omega) \textrm{sign} (\omega).
\end{align}

\subsection{Reduced qubit dynamics}
\label{sec::RD}

Our QNS protocols obtain information about the bath spectra by using the $N$ qubits as probes of their environment. Extracting this information requires knowledge of how the bath spectra enter the reduced 
qubit dynamics.   For Gaussian dephasing, this hinges on an {\em exact} analytic expression which 
relates expectation values of a relevant class of observables to bath cumulants, 
and which may be  of independent interest.

Assume a factorizable joint state at the initial time $t=0$, say, $\rho_{SB}(0) = \rho_S (0) \otimes \rho_B 
\equiv \rho_0 \otimes \rho_B$.  Then the expectation value of any invertible operator $O$ resulting 
from evolution under the time-dependent Hamiltonian $\tilde{H}(t)$ may be formally expressed as follows:
\begin{align}
\label{eq::EOt}
E_{\rho_0} ( O(t) ) &\equiv \langle \Tr_S [  \rho_{S}(t) O] \rangle_{c}  
 = \langle \Tr_{SB} [  \rho_{SB}(t) O] \rangle_{c}\notag \\
&= \langle \Tr_S[ \Tr_B ( O^{-1} \tilde{U}(t)^\dagger O \tilde{U}(t) \rho_B) \rho_{0} O ] \rangle_c\notag \\
&\equiv \Tr_S [ \langle \mathcal{T}_+ e^{- i \int_{-t}^{t} \tilde{H}_O(s) ds}\rangle_{c,q} \,\rho_0 O] , 
\end{align}
where  $\tilde{U}(t)=\mathcal{T}_+\text{exp}[-i\int_0^t\tilde{H}(s)ds]$ and in the last line we have introduced an 
operator-dependent (not necessarily Hermitian) effective Hamiltonian given by
\beq
\label{eq::HO}
\tilde{H}_O(s)\equiv \begin{cases}  -O^{-1}  \tilde{H}(t- s) O  & \textrm{   for  }  0 < s \leq t,  \\ 
\,\,\,\,\,\,\,\,\,\,\tilde{H}(t+s) &\textrm{   for  }  -t \leq s < 0 .
\end{cases}
\eneq
The calculation in Eq.~(\ref{eq::EOt}) can be carried out exactly if $O$ is, additionally, dephasing-preserving
in the sense that 
$$O^{-1}Z_a O = \sum_b V_{ab}Z_b, \quad \forall a,b \in {\mathcal I}_N,  {V_{ab} \in {\mathbb C}.}$$  
\noindent 
As proved in Appendix \ref{App_Gauss_Exact}, the following result holds:

\vspace*{1mm}

{\bf Theorem.} 
{\em The time-dependent expectation value of a dephasing-preserving 
invertible operator on $N$-qubits evolving under 
controlled Gaussian dephasing dynamics is } 
\beq
\label{Gauss-sol}
 E_{\rho_0} (O(t)) = \Tr \Big[ e^{-i \mathcal{C}^{(1)}_O(t) - \frac{\mathcal{C}^{(2)}_O(t)}{2!}} \rho_{0} O \Big],
\eneq
{\em where the time-dependent cumulants are}
\begin{align}
\mathcal{C}^{(1)}_O(t)  &= \int_{-t}^t ds \langle \tilde{H}_O(s)\rangle_{c,q}, 
\label{eq::CO1} \\
\mathcal{C}^{(2)}_O(t)  &
=  2 \int_{-t}^t ds_1 \int_{-t}^{s_1} ds_2 \langle \tilde{H}_O(s_1)\tilde{H}_O(s_2)\rangle_{c,q}
\label{eq::CO2}\\
&\,\,\,\,\, -  \int_{-t}^t ds_1 \langle \tilde{H}_O(s_1)\rangle_{c,q} \int_{-t}^t ds_2 \langle \tilde{H}_O(s_2)\rangle_{c,q}\notag. 
\end{align} 

Remarkably, the above result relies solely on the dephasing character of the effective time-dependent Hamiltonian 
[Eq.~(\ref{eq::HO})] and the Gaussianity of the noise, regardless of the specific nature (e.g., bosonic or not) of the bath.  In fact, the theorem applies more generally (see Appendix~\ref{App_Gauss_Exact})
to controlled quantum systems of arbitrary dimension coupled to Gaussian baths, 
as long as the dephasing (commuting) requirement is preserved.  
In this sense, it generalizes existing results for free (uncontrolled) Gaussian 
models \cite{Kardar:book,Doll2008}, as well as DD-controlled one- and two-qubit Gaussian models, see e.g. \cite{Paz2014} (supplement), \cite{multiDD}, and references therein. 

In the $N$-qubit setting under consideration for QNS, 
the zero-mean assumption, $\expect{\tilde{H}_O(t)}_{c,q}=0$, 
implies that $\mathcal{C}^{(1)}_O(t) \equiv 0$. 
The reduced qubit dynamics are, thus, governed by 
$\mathcal{C}^{(2)}_O(t)$ which, using Eq.~(\ref{eq::togglingH}),
takes the following form: 
\begin{widetext}
\begin{align}
\frac{\mathcal{C}^{(2)}_O(t)}{2!}   
= \sum_{a,b,a',b'\in\mathcal{I}_N}\!\!\!\Big[& Z_aZ_b
\!\int_0^t\!\!ds_1\!\!\int_0^{s_1}\!\!\!\!\!ds_2\,y_{a,a'}(s_1)y_{b,b'}(s_2)
\expect{B_{a'}(s_1)B_{b'}(s_2)}_{c,q} \notag \\ 
& + O^{-1}Z_aZ_b O \!\int_0^t\!\!ds_1\!\!\int_0^{s_1}\!\!\!\!\!ds_2\,y_{a,a'}(s_2)y_{b,b'}(s_1)
\expect{B_{a'}(s_2)B_{b'}(s_1)}_{c,q}  \notag \\
&-O^{-1}Z_aOZ_b\!\int_0^t\!\!ds_1\!\!\int_0^{t}\!\!\!\!\!
ds_2\,y_{a,a'}(s_1)y_{b,b'}(s_2)\expect{B_{a'}(s_1)B_{b'}(s_2)}_{c,q}\Big]. 
\label{eq::CO22}  
\end{align} 
\end{widetext}
Since the dephasing-preserving property is automatically obeyed by 
observables that are in the $N$-qubit Pauli group (that is, up to an 
irrelevant phase, are a product of Pauli operators on the qubits), it also 
follows that 
$O^{-1} Z_aZ_b O =  \pm\, Z_{a}Z_b$ for all $a,b\in\mathcal{I}_N$.  
Let $\text{sign}(O,a,b)$ be a function defined as
\begin{align}
\label{eq::sgn}
\text{sign}(O,a,b)\equiv \left\{ \begin{array}{ccc}
        + & \;\text{if} \;&O^{-1} Z_aZ_b O = +Z_{a}Z_b\\
        - & \;\text{if} \;&O^{-1} Z_aZ_b O = -Z_{a}Z_b \end{array}\right. .
\end{align}
In order to make contact with the bath spectra of interest, we transform \erf{eq::CO22}  
into the frequency domain which, after straightforward algebraic manipulations, yields
\begin{align} 
\notag
&\frac{\mathcal{C}^{(2)}_O(t)}{2!} = 
-\!\!\sum_{a,b,a',b'\in\mathcal{I}_N}\!\!\frac{Z_a Z_b}{2}  \int_{-\infty}^\infty \frac{d\omega}{2 \pi} S^{\text{sign}(O,a,b)}_{a',b'} (\omega)\\
\label{C2}
 &\times\big[ \text{sign}(O,a,0) G_{a,a';b,b'}^{+} (\omega,t) - G_{a,a';b,b'}^{\text{sign}(O,a,b)} (\omega,t) \big]. 
\end{align}
Here, $G_{a,a';b,b'}^{\pm }(\omega,t)$ are FFs that capture effects of the external control in the frequency domain. In terms of the 
fundamental FFs given in Eqs. (\ref{eq::F1}) and (\ref{eq::F2}), they read 
\begin{align*}
G_{a,a';b,b'}^{+} (\omega,t)&= F_{a,a';b,b'}^{(2)} (\omega,t) +  F_{b,b';a,a'}^{(2)} (-\omega,t)\\
&= F_{a,a'}^{(1)} (\omega,t) F_{b,b'}^{(1)} (-\omega,t),\\
G_{a,a';b,b'}^{-} (\omega,t)&= F_{a,a';b,b'}^{(2)} (\omega,t) -  F_{b,b';a,a'}^{(2)} (-\omega,t).
\end{align*}
\erf{C2} makes it explicit that the reduced qubit dynamics are completely ruled by convolutions between the power 
spectra of the bath and the FFs generated by external control.

\section{Physical significance of quantum noise spectra}
\label{sec::physSig}

A central feature of our QNS protocols is the ability to reconstruct spectra associated with quantum baths in a dephasing setting. Section \ref{sec::protocols} will explain how using multiple qubits as probes makes this possible. Before delving into the technical details of the protocols, however, we further motivate interest in the quantum spectra by examining their unique dynamical signatures. 

\label{sec::physSigA}

As discussed in Sec.~\ref{sec::background}, the quantum spectra are nonzero only when the bath operators $B_a(t)$ are non-classical, 
i.e., they do not commute at all times. The quantum spectra have physical significance beyond their absence in the classical case, however. Useful insight may be gained by considering a model Hamiltonian that interpolates between models M1 and M2, that is, in the toggling frame, we let 
\begin{align}
\label{eq::dephasingH}
\tilde{H}(t)=\sum_{\ell,\ell'=1}^N
[y_{\ell,\ell'}(t)Z_\ell+c \, I_\ell]B_{\ell'}(t).
\end{align}
where $c\in[0,1]$ and, in order to offer a clearer picture of the quantum dynamics, 
we assume no classical noise contribution, $\zeta_a(t)\equiv 0$. 
Recall that in the M1 model ($c=0$), the pure-bath term $B_0(t)=0$, whereas 
$B_0(t)=\sum_{\ell=1}^N B_\ell(t)\neq 0$ for the M2 model ($c=1$). Despite 
commonalities and the seemingly minor distinction between M1 and M2, their dynamics under 
$\tilde{H}(t)$ in Eq.~(\ref{eq::dephasingH}) display striking differences.

\subsection{Dynamical influence of the quantum self-spectra}
\label{sec::singleQdephase}

Even at the level of single-qubit dephasing, signatures of the quantum bath as well as differences between the M1 and M2 models are evident. For $N=1$, the dynamical effects of dephasing are encapsulated in the qubit's coherence element 
\begin{align*}
\bra{1}\rho_S(t)\ket{0}=\bra{1}\rho_S(0) \ket{0} \,e^{-\chi(t)+i\phi(t)},
\end{align*}
where the decay rate depends on the classical self-spectrum,
\begin{align}
\chi(t)=\frac{1}{2\pi}\int_{-\infty}^{\infty}\!\! d\omega \, G_{1,1;1,1}^{+}(\omega,t)S_{1,1}^{+}(\omega),
\end{align}
and the phase angle depends on the quantum self-spectrum,
\begin{align}
\label{eq::phase1qubit}
i\phi(t)=\frac{c}{2\pi}\int_{-\infty}^{\infty}\!\!d\omega \, (G^{-}_{1,1;01}(\omega,t)+G^{+}_{1,1;0,1}(\omega,t))S_{1,1}^{-}(\omega).
\end{align}
Because $\phi(t)$ 
is proportional to $c$, the quantum self-spectrum is entirely absent from the coherence element of the qubit in the M1 model. 
The quantumness of the bath, thus, has no dynamical implications.
For M2, on the contrary, quantum noise exerts a substantial influence, in that the presence of the quantum self-spectrum in Eq.~(\ref{eq::phase1qubit}) causes observable rotation of the qubit about $Z_1$. Interestingly, similar phase evolution is observed in a classical {\em non-Gaussian} 
or {\em non-stationary} dephasing on a single qubit \cite{Norris2016}. In a regime where bath statistics are stationary and 
well-approximated as Gaussian as we assume here, however, the presence of non-trivial phase evolution is a signature of quantum noise. 

The absence of the quantum self-spectrum from the reduced single-qubit dephasing dynamics of M1 may lead one to conclude that a quantum bath has no observable effect on this model. This is far from the case, however. ``Tilting" the quantization axis by adding a driving term in a direction orthogonal to $z$ reveals signatures of quantum noise in \emph{both} the M1 and M2 models.  Consider a single-qubit Hamiltonian of the form in Eq.~(\ref{eq::dephasingH}) with the addition of a continuous driving term with amplitude $g$ 
and no other external control, 
\begin{align}
\label{eq::relaxationH}
H_g(t)= 
(Z_1+cI_1)B_{1}(t)+\frac{g}{2} X_1, 
\end{align}
in suitable units. The presence of the continuous drive effectively sets a new quantization axis for the qubit along $x$. While the bath induces pure dephasing when the qubit is quantized along $z$ absent the drive, the bath causes both dephasing and relaxation with respect to the qubit's new quantization axis \cite{Yan2013}. Let $\tilde{\rho}_S(t)$ and $\tilde{H}_g(t)$ denote the state of the system and the Hamiltonian in the interaction picture with respect to the drive term.  In the limit of {\em weak coupling} $|\!|B_1(t)|\!|t\ll 1$, the relaxation dynamics are evident in the qubit 
master equation obtained through the second order time-convolutionless projection method, 
\begin{align*}
\frac{d\tilde{\rho}_S (t)}{dt}=-\!\!\int_0^t\!\!dt'\text{Tr}_B\left([\tilde{H}_g(t),[\tilde{H}_g(t'),\tilde{\rho}_S(t)\otimes{\rho}_B ]]\right).
\end{align*}
Letting $\ket{\pm}$ denote the eigenstates of $X_1$ and returning to the frame of \erf{eq::relaxationH}, where 
$\rho_{ij} \equiv \bra{i}\rho_S(t)\ket{j}$, $i,j\in\{+,-\}$ denotes the state of the qubit, we have
\begin{align}
\dot{\rho}_{++}=&\frac{1}{2\pi}\int_{-\infty}^{\infty}\Big\{-\frac{2\text{sin}[t(\omega+g)]}{\omega +g}S_{1,1}(\omega)\rho_{++}
\label{eq::rhodotplus}\\\notag&
+\frac{2\text{sin}[t(\omega-g)]}{\omega-g}S_{1,1}(\omega)\rho_{--}
\label{eq::rhodotminus}\\\notag&
+ic \,\frac{e^{it\omega}-1}{\omega}S_{1,1}^{-}(\omega)(\rho_{-+}-\rho_{+-})\Big\},\\
\dot{\rho}_{--}=&-\dot{\rho}_{++}, 
\end{align}
\begin{align}
\dot{\rho}_{+-}=&\frac{i}{2\pi}\int_{-\infty}^{\infty}\Big\{-\frac{1-e^{it(\omega+g)}}{\omega+g}S_{1,1}^+(\omega)\rho_{+-}\\\notag&
+\frac{1-e^{it(\omega-g)}}{\omega-g}S_{1,1}^+(\omega)\rho_{-+}\\\notag&
+c\,\frac{e^{it\omega}-1}{\omega}S_{1,1}^-(\omega)(\rho_{--}-\rho_{++})\Big\}-ig\rho_{+-}, \\ 
\dot{\rho}_{-+}=&-\dot{\rho}_{+-}+ig(\rho_{-+}-\rho_{+-}).
\label{eq::rhodotminusplus}
\end{align}
From these equations, we see that the evolution of populations and coherences are coupled by terms proportional to $c\,S_{1,1}^-(\omega)$. As these terms vanish for a classical bath or for a M1 model, the dynamical signatures of the quantum bath are more prominent in the M2 model, similar to the case of dephasing-preserving dynamics discussed above. 

The M1 model is not immune to the effects of the quantum bath, however. For both M1 and M2, the quantum nature of the 
bath enters in determining the {\em steady-state populations}. In the steady-state limit of Eqs. (\ref{eq::rhodotplus})-(\ref{eq::rhodotminusplus}), 
letting $\text{sin}(t\Omega)/\Omega\approx \pi\delta(\Omega)$ for large $t$, we obtain $\rho_{+-}^{\text{ss}}/\rho_{-+}^{\text{ss}}=1$ and 
$\rho_{++}^{\text{ss}}/\rho_{--}^{\text{ss}}=S_{1,1}(g)/S_{1,1}(-g)$ for both M1 and M2. While the steady-state populations are always equal for a classical, spectrally symmetric bath, 
this is generally {\em not} true when the bath is quantum, as dictated by the requirement of detailed balance at equilibrium \cite{Kubo}. 
In particular, for the thermal bosonic spectra given in 
\erf{eq::SBllSpectra},
$\rho_{++}^{\text{ss}}/\rho_{--}^{\text{ss}}=S_{1,1}(g)/S_{1,1}(-g)=e^{\beta g}$, which exceeds $1$, for every finite temperature. 

Given the extent to which the quantum self-spectrum influences the qubit dynamics during driven evolution, it is not surprising that off-axis driving can be used to perform spectroscopy on quantum noise sources. In a variety of other platforms, including NMR and 
superconducting qubits, interaction-frame Hamiltonians of the form in Eq.~(\ref{eq::relaxationH}) arise in ``spin-locking" or ``$T_{1\rho}$" experiments \cite{SpinLocking}.  Approaches based on spin-locking and $T_{1\rho}$ have 
in fact been employed to characterize classical noise sources \cite{Yan2013,Bylander2011}. These strategies can also be extended to quantum noise sources, as we describe in Appendix \ref{sec::spinlocking}.   We emphasize, however, that the non-commuting nature 
of Eq.~(\ref{eq::relaxationH}) prohibits exact solutions for the reduced qubit dynamics. Performing QNS in this setting, 
therefore, inevitably entails approximations (such as weak coupling and, in practice, weak driving \cite{Yan2013}), which need not be well-controlled or whose range of validity may be unclear a priori. In contrast, at the cost of an additional qubit, working in the dephasing setting affords 
an exact analytic description of the reduced dynamics, allowing for QNS to be carried out 
beyond the regime of validity of the weak-coupling or similar assumptions.

\subsection{Quantum spectra and bath-induced entanglement}

The dynamics of multiple qubits coupled to a quantum bath are considerably more rich than those of a single qubit. Notably, interaction with a quantum bath can generate quantum correlations and entangle the qubits, even in the absence of direct coupling between them, see e.g. \cite{Braun2002,Benatti2003,Oh2006,An2007,McCutcheon2009,Bhaktavatsala2011,multiDD,Krzywda:2016}. The quantum spectra relate directly to the ability of a quantum bath to mediate such entangling interaction. Consider the Magnus expansion of the qubit-bath propagator, 
$U_I(t)=\mathcal{T}_+\{\text{exp}[-i\int_0^t ds H_I(s)]\}\equiv \text{exp}[\sum_{\alpha=1}^\infty\Omega_\alpha(t)]$, 
where the second term is given by
\begin{align}
\label{eq::Magnus}
\Omega_2(t)=&-\frac{1}{2}\!\sum_{\ell,m,\ell',m' =1}^N
\int_{0}^t\!\!\!ds\!\!\int_{0}^s\!\!\!ds'(y_{\ell,\ell'}(s)Z_\ell+cI_\ell)
\\&\times (y_{m,m'}(s')Z_{m}+cI_{m})[B_{\ell'}(s),B_{m'}(s')].\notag
\end{align}
The nonlinear term in $\Omega_2(t)$ that is proportional to $Z_\ell \otimes Z_{m}$ serves to couple the pair of qubits $\ell$ and $m$. The quantity $\langle\Omega_2(t)\rangle_q$ arises in the reduced dynamics of the qubits given in Eq.~(\ref{C2}), producing bilinear (Ising) coupling terms which, in the frequency domain, have the structure 
\begin{align}
\label{eq::CouplingTerm}
H_2^{\text{eff}} \sim Z_\ell \otimes Z_{m} \!\!\int_{-\infty}^{\infty}\!\!d\omega\, G_{\ell,\ell';m,m'}^{-}(\omega,t)S_{\ell',m'}^-(\omega).
\end{align}
Being explicitly proportional to the quantum self- and cross-spectra, these coupling terms clearly vanish in the case of a classical bath, 
in accordance with the expectation that coupling to a classical bath cannot induce quantum correlations. 

An interesting example of bath-induced entanglement arises when each qubit is coupled to its own {\em independent} bath, i.e., for $\ell',m'\in\{1,\ldots,N\}$, $[B_{\ell'}(t),B_{m'}(t')]=0$, $\forall \ell'\neq m'\;\forall t,\,t'$. Bath-induced entanglement has been typically examined in the context of a common bath, where there exists $\ell'\neq m'$ such that $[B_{\ell'}(t),B_{m'}(t')]\neq0$ for at least one pair of $t$ and $t'$.  Yet, a common bath is {\em not} required to mediate entanglement between the qubits.  This can be seen from Eq.~(\ref{eq::CouplingTerm}), where all quantum cross-spectra vanish in the case of independent baths, leaving only terms containing quantum self-spectra, of the form 
\begin{align}
\label{eq::CouplingTerm2}
H_2^{\text{eff}} \sim Z_\ell \otimes Z_{m} \!\!\int_{-\infty}^{\infty}\!\!d\omega\, G_{\ell,m';m,m'}^{-}(\omega,t)S_{m',m'}^-(\omega).
\end{align}
With access to only diagonal control, the FF $G_{\ell,m';m,m'}^{-}(\omega,t)$ is zero whenever $\ell\neq m$.  Consequently, no entanglement is generated between qubits $\ell$ and $m$. Remarkably, non-diagonal control via (non-entangling) swap gates can produce a non-zero FF when 
$\ell\neq m$, which does allow for bath-induced entanglement. Qualitatively, through a swap gate, both qubits $\ell$ and $m$ can couple to the same bath degrees of freedom $m'$ at different times. As long as the time correlations of the bath 
decay sufficiently slowly, information about qubit $\ell$ or $m$ ``imprinted'' in $m'$ persists even after qubits $\ell$ and $m$ are ``swapped". 
This mechanism, similar in spirit to generation of collective (permutationally-symmetric) decoherence via repeated swaps \cite{ViolaSymm}, 
explains how independent baths can mediate entanglement in principle. 

To further illustrate entanglement generation via independent baths, consider $N=2$ qubits but imagine that 
only qubit 1 is interacting with the bosonic bath, so that in Eq.~(\ref{eq::dephasingH}) 
$B_1(t)$ has the standard form [Eq.~(\ref{eq::SBBt})], while 
$B_2(t)\equiv 0$. Clearly, these baths are trivially independent. 
Suppose we allow the qubits to freely evolve for a time $T/2$,  
then apply a SWAP to qubits 1 and 2, and let them freely evolve again for another $T/2$ duration.  The 
overall evolution is fully specified by the two switching functions
\begin{align*}
y_{1,1}(t)=\left\{ \begin{array}{ll}
        1, &t\in[0,T/2]\\
        0, &t\in[T/2,T] \end{array} \right., \\
y_{2,1}(t)=\left\{ \begin{array}{ll}
        0, &t\in[0,T/2]\\
        1, &t\in[T/2,T] \end{array} \right., 
\end{align*}
and, correspondingly, 
the coupling term in Eq.~(\ref{eq::CouplingTerm}) becomes
\begin{align*}
H_2^{\text{eff}} \sim 4 Z_1\otimes Z_2\int_{-\infty}^{\infty}\!\!\!d\omega\, \frac{\text{cos}(\omega T/2)\,\text{sin}^2(\omega T/4)}{\omega^2}\,S_{1,1}^-(\omega).
\end{align*}
This term allows qubits 1 and 2 to interact with a strength that depends on the overlap between the free-evolution FF 
and the quantum self-spectrum of qubit 1. Although qubits 1 and 2 never interact with a common bath at the same time, 
they can nonetheless become non-trivially entangled.

\section{Multiqubit noise spectroscopy protocols} 
\label{sec::protocols}

In general terms, spectroscopy protocols infer information about an external parameter by measuring the response of a probe system under different controllable experimental conditions. In our QNS protocols, the noise spectra are the external parameter, the  $N$ qubits are the probe system, and control sequences applied to the qubits generate different experimental conditions. The essential steps of the procedure will be (1) prepare the qubits in a known state; (2) let the qubits evolve under both bath-induced noise and external control; (3) measure a set of observables on the qubits that quantifies their response to the bath and control; (4) extract information about the bath spectra from the measured value of the observables. 

From a theoretical standpoint, the most challenging step is the last one. Section \ref{sec::RD} demonstrated that expectation values of qubit observables depend on the spectra of the bath.  This dependence, however, takes the form of a convolution between the bath spectra and FFs, as seen in \erf{C2}. Obtaining an estimate of the spectra requires that we invert or ``deconvolve" this convolution, which is non-trivial in general. In our spectroscopy protocols, the FFs are instrumental in accomplishing this. Specific timing symmetries of the control sequences enable us to engineer frequency combs in {\em all} the relevant FFs. As we shall show, this reduces the problem of obtaining the spectra to solving a system of linear equations.   

\subsection{The deconvolution problem}
\label{sec::deconvolution}

It is instructive to first revisit the case of a single qubit. Let $\ket{\pm}$ denote the eigenstates of $X$.  Suppose we prepare a qubit in the initial state $\ket{\psi_1}=\ket{+}_1$ and allow it to evolve under bath-induced noise and control for a time $T$. We repeat this process, each time measuring either $X_1$ or $Y_1$ at time $T$. After collecting a large number of measurements, we compute $\overline{X}_1^2+\overline{Y}_1^2$, where $\overline{O}$ denotes the average measured value of observable $O$. The expected value of this quantity is
\begin{align}
E[X_1(T)]^2+E[Y_1(T)]^2=e^{-2\chi(T)}.
\end{align}
As discussed in Sec.~\ref{sec::physSigA}, the decay constant $\chi(T)$ is given by
\begin{align}
\label{eq::convolution}
\chi_{\color{red}}(T)=\frac{1}{2\pi}\int_{-\infty}^\infty d\omega\, G_{1,1;1,1}^{+} (\omega,T)S^{+}_{1,1} (\omega).
\end{align}
Through repeated measurements of $X_1$ and $Y_1$, we can obtain $\chi_{\color{red}}(T)$, which depends on the classical self-spectrum $S^{+}_{1,1} (\omega)$. However, this spectral dependence is buried in a convolution between the FF $G_{1,1;1,1}^{+} (\omega,T)$ and $S^{+}_{1,1} (\omega)$. Extracting $S^{+}_{1,1} (\omega)$
 requires that we deconvolve the integral in \erf{eq::convolution}.

In Ref. \cite{Alvarez2011}, Alvarez and Suter devised a solution to this problem for a single qubit subject to classical dephasing,  
based on repetition of fixed control sequences.
Consider a ``base" control sequence of duration $T$. Repeating this sequence a total of $M\gg 1$ times creates a frequency comb in the associated FF, i.e.,
\begin{align*}
G_{1,1;1,1}^{+} (\omega,MT)&=\frac{\sin^2(\frac{M\omega T}{2})}{\sin^2(\frac{\omega T}{2})}G_{1,1;1,1}^{+} (\omega,T)
\\&\simeq \frac{2 \pi M}{T}\!\!\sum_{k=-\infty}^\infty\!\!\!\delta(\omega\!-\!k\omega_0)G_{1,1;1,1}^{+} (\omega,T), 
\end{align*}
where the ``teeth" of the frequency comb are centered at the harmonic frequencies, integer multiples of $\omega_0=2\pi/T$. The frequency comb effectively discretizes the integral in \erf{eq::convolution}, producing a linear equation 
\begin{align*}
\chi(MT)&\simeq\frac{M}{T}\sum_{k=-\infty}^\infty G_{1,1;1,1}^{+} (k\omega_0,T)S^{+}_{1,1} (k\omega_0)\\
&\approx \frac{M}{T}\sum_{k\in\mathcal{K}} G_{1,1;1,1}^{+} (k\omega_0,T)S^{+}_{1,1} (k\omega_0), 
\end{align*}
where in the second line the summation has been restricted to a finite set of harmonics, $\{k\omega_0|k\in\mathcal{K}\}$. This truncation is justified by the decay of the spectrum and FFs at high frequencies. 
In this expression, both $\chi(MT)$ and $G_{1,1;1,1}^{+} (k\omega_0,T)$ are known, the former from measurement and the latter from the control sequence. Repeating this procedure for $N_c\geq |\mathcal{K}|$ distinct control sequences generates a set of linear equations, which we can invert to obtain $\{S^{+}_{1,1} (k\omega_0)|k\in\mathcal{K}\}$. The frequency comb technique transforms the deconvolution of the integral in \erf{eq::convolution} to an inverse problem.  

Generalizing this method to multiple qubits and to quantum noise sources entails a number of complications. First, the number of dynamically relevant spectra grows considerably as the number of qubits increases. The cross-spectra, furthermore, can have both real and imaginary components. To reconstruct the expanded number of spectral quantitites, we must measure an expanded number of observables, whose expectation values are sums of convolutions involving all of the different filters and spectra. Terms containing the FF $G_{a,a';b,b'}^{+} (\omega,t)$ for arbitrary $a,\,a',b,\,b'\in\mathcal{I}_N$ can be deconvolved via frequency comb in a mannner similar to $\chi(T)$. With the exception of certain non-generic cases, which are examined in the following section, $M$ repetitions of a base control sequence produces
\begin{align*}
G_{a,a';b,b'}^{+} (\omega,MT)&\!\simeq\! \frac{2 \pi M}{T}\!\!\sum_{k=-\infty}^\infty\!\!\!\delta(\omega\!-\!k\omega_0)G_{a,a';b,b'}^{+} (\omega,T).\notag
\end{align*}
In addition to  $G_{a,a';b,b'}^{+} (\omega,t)$, the multiqubit dynamics depend on the second-order FFs $G_{a,a';b,b'}^{-} (\omega,t)$. Because $G_{a,a';b,b'}^{-} (\omega,t)$ involves nested time integrals, repetition of an arbitrary control sequence is {\em not} sufficient to generate a comb. We next identify timing symmetries in the control sequences that enable us to generate combs in all filters, as needed.

\subsection{Control timing symmetries} 
\label{CtrlKit}

For multiple qubits, timing symmetries in the applied control sequences will not only be essential 
to overcome the deconvolution problem, 
but they will also enable us to 
alter the real and imaginary character of the FFs, and thereby selectively extract the real and imaginary components of 
the cross-spectra. Building on our work in \cite{multiDD}, we introduce below 
the concepts of \emph{displacement symmetry} and \emph{mirror symmetry}, 
which are essential to our spectroscopy procedure.

\begin{figure}[t]
\centering
\includegraphics[width = 0.95 \linewidth]{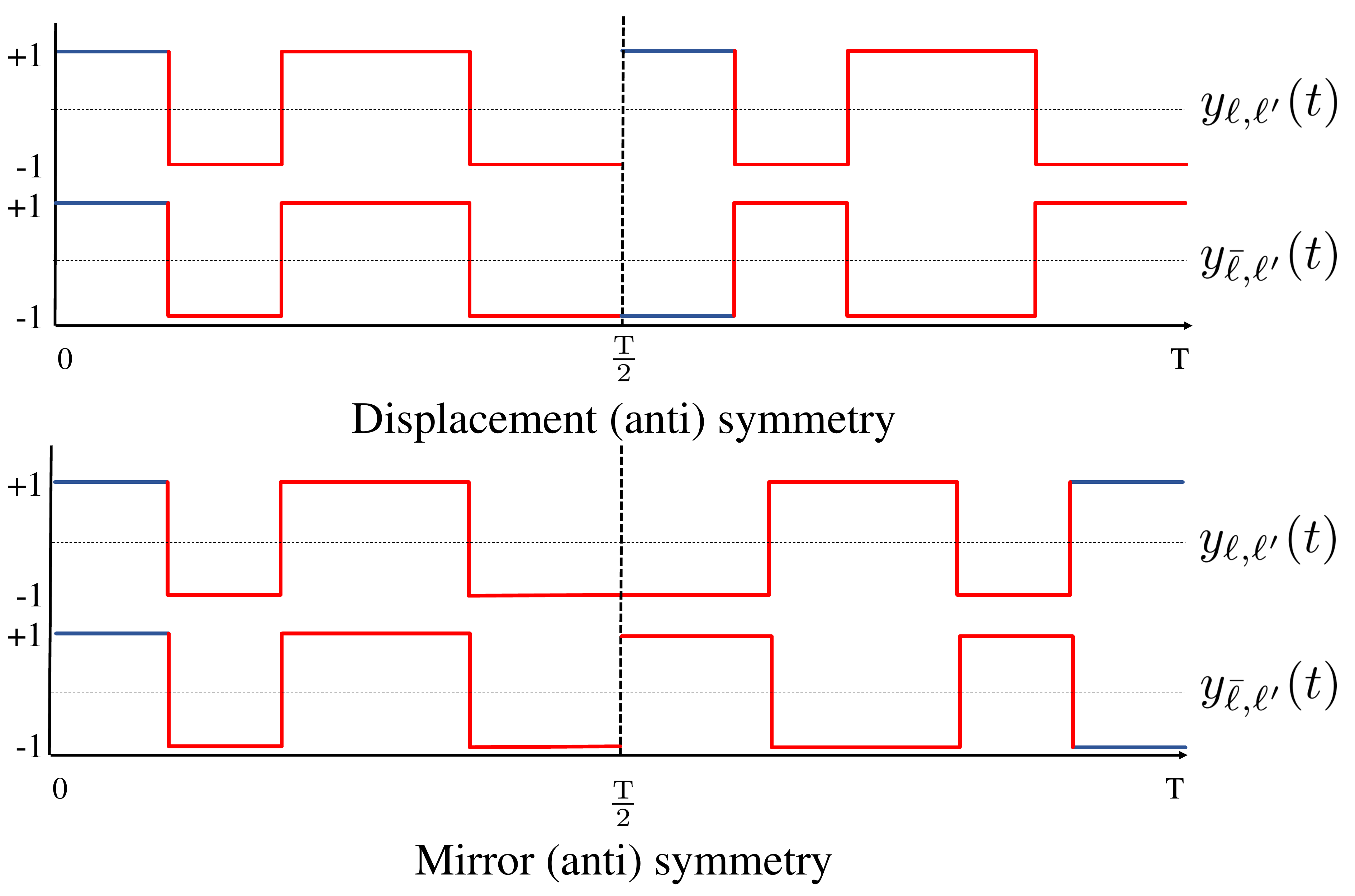}
\vspace*{-1mm}
\caption{(Color online) Sample pulse patterns of pairs of switching functions $y_{\ell,\ell'}(t)$ and $y_{\bar{\ell},\ell'} (t)$, obeying mirror and displacement (anti)symmetry in an interval $[0,T]$. For each symmetry (displacement or mirror), $y_{\ell,\ell'}(t)$ has been chosen to be symmetric and $y_{\bar{\ell},\ell'}(t)$ antisymmetric, so that that their product is antisymmetric.}

\label{fig:Symmetry}
\end{figure}

\subsubsection{Displacement (anti)symmetry}

Consider a control sequence with cycle time $T$ and an associated switching function $y_{a,a'}(t)$. We will say that the switching function is {\em displacement symmetric (antisymmetric)}, corresponding to the $+\,-$ cases below, if   
\begin{align} 
y_{a,a'}(t) =\pm y_{a,a'}(T/2+ t)\;\;\text{for}\; t\in[0,T/2].
\end{align}
Switching functions with displacement symmetry or antisymmetry are easily manufactured from arbitrary control sequences. Consider a sequence $Q(T/2)$ of duration $T/2$. Repeating this sequence twice forms $U_\text{ctrl}(T)=Q(T/2)Q(T/2)$ whose switching functions naturally satisfy $y_{aa'}(t) = y_{aa'}(T/2+t)$ for all $a,a'\in\mathcal{I}_N$. Suppose we conjugate the second repetition by a $\pi$-pulse, $\Pi_{\mathcal{A}}$, creating $U_\text{ctrl}(T)=\Pi_{\mathcal{A}}^\dag Q(T/2)\Pi_{\mathcal{A}}Q(T/2)$. The switching function associated with this sequence takes the form 
\begin{align*}
y_{aa'}(t) = \left\{\! \begin{array}{ll} - y_{aa'}(T/2+t)  & \textrm{  if  } \;a\in\mathcal{A} \\
 + y_{aa'}(T/2+t)  & \textrm{  if  } \; a\not\in\mathcal{A} \end{array} \right..
\end{align*}
Because displacement symmetry or antisymmetry of $y_{aa'}(t)$  depends on $\mathcal{A}$, i.e. the local operations that comprise the $\pi$-pulse, we can selectively control the symmetry characteristics of multiple switching functions associated with the same control sequence. Consider e.g. two switching functions $y_{aa'}(t)$ and $y_{bb'}(t)$ with $a,a',b,b'\in\mathcal{I}_N$ and $a\neq b$, which are associated with the sequence $U_\text{ctrl}(T)=\Pi_{\mathcal{A}}^\dag Q(T/2)\Pi_{\mathcal{A}}Q(T/2)$. By choosing $\Pi_{\mathcal{A}}$ such that $ a\in\mathcal{A} $ and $  b\not\in\mathcal{A} $, $y_{aa'}(t)$ is displacement antisymmetric and $y_{bb'}(t)$ is displacement symmetric. Because the switching functions associated with a single control sequence can possess different symmetries, the timing symmetries are best viewed as properties of the switching functions and not the control sequence.

The notions of displacement symmetry and antisymmetry can be extended to products of switching functions. The switching functions $y_{a,a'}(t)$ and $y_{b,b'}(t)$ are {\em product-displacement symmetric (antisymmetric)} on the interval $[0,T]$, corresponding to the $+\, -$ cases below,  if 
\begin{equation}
\label{proddisp} y_{a,a'}(t) y_{b,b'}(t') =\pm y_{a,a'}(T/2+ t) y_{b,b'}(T/2+ t'),
\end{equation}
for $t,t'\in[0,T/2]$. The joint symmetry of a pair of switching functions depends on the symmetries of the individual switching functions. For example, $y_{a,a'}(t)$ and $y_{b,b'}(t)$ are product displacement symmetric when both switching functions are individually displacement symmetric or displacement antisymmetric, i.e. when
\begin{align} \label{eq::pmSym}
&y_{a,a'}(t) = \pm y_{a,a'}(T/2+ t)\;\;\;\text{and}\\
&y_{b,b'}(t) = \pm y_{b,b'}(T/2+ t)\notag
\end{align} 
hold simultaneously. We refer to the $\pm$ cases above as product-displacement $\pm$-symmetry. Similarly, $y_{a,a'}(t)$ and $y_{b,b'}(t)$ are product-displacement antisymmetric when $y_{a,a'}(t)$ is individually displacement symmetric and $y_{b,b'}(t)$ is individually displacement antisymmetric or vice versa, i.e.
\begin{align} \label{eq::pmAntisym} 
&y_{a,a'}(t) = \pm y_{a,a'}(T/2+ t)\;\;\;\text{and}\\
&y_{b,b'}(t) = \mp y_{b,b'}(T/2+ t).\notag
\end{align} 
We refer to the $\pm$ cases above as {\em product-displacement $\pm$-antisymmetry}. 
We see that controlling whether individual switching functions are displacement symmetric or antisymmetric can create pairs of switching functions that are jointly displacement-product symmetric or antisymmetric.

\subsubsection{Mirror (anti)symmetry} 

A switching function in $[0,T]$ is {\em mirror symmetric (antisymmetric)}, corresponding to the $+-$ cases below, if
$$y_{a,a'}(T/2-t) = \pm y_{a,a'}(T/2+t).$$ 
Like displacement (anti) symmetry, switching functions with mirror symmetry or antisymmetry are easy to engineer from arbitrary subsequences.
Let $Q(T/2)$ be a sequence of $\pi$-pulses and/or swap gates. The sequence $U_{\text{ctrl}}(T)=Q(T/2)^\dag Q(T/2)$ has 
mirror symmetric switching functions satisfying $y_{a,a'}(T/2-t) = y_{a,a'}(T/2+t)$ for all $a,a'\in\mathcal{I}_N$. Mirror antisymmetric switching functions can be generated by conjugating the second half of the evolution with a $\pi$-pulse, forming $U_{\text{ctrl}}(T)= \Pi_\mathcal{A}^\dag Q(T/2)^\dag\Pi_\mathcal{A}Q(T/2)$. The switching function associated with this sequence is 
\begin{align*}
y_{aa'}(T/2-t) = \left\{\! \begin{array}{ll} - y_{aa'}(T/2+t)  & \textrm{  if  }\;  a\in\mathcal{A} \\
 + y_{aa'}(T/2+t)  & \textrm{  if  } \;  a\not\in\mathcal{A}  \end{array} \right..
\end{align*}
Similar to the case of displacement symmetry and antisymmetry, conjugation by $\pi$-pulses enables us to control whether individual switching functions associated with the same control sequence are mirror symmetric or antisymmetric.

\subsubsection{Symmetry-enhanced control design}  

As mentioned, repetition of an arbitrary control sequence does {\em not} generate a frequency comb in $G^{-}_{a,a';b,b'}(\omega,t)$. Repetition of a control sequence with FFs that are product-displacement antisymmetric, however, \emph{does} generate a comb\footnote{Note that $+$-antisymmetry in $[0,T]$ also generates a comb, but it does not lend itself to be combined with other types of symmetries in $[0,T/2]$ or $[0,T/4]$, that we need to execute our protocol.}. To see this, suppose a control sequence has associated switching functions, $y_{a,a'}(t)$ and $y_{b,b'}(t)$, which are product-displacement antisymmetric in $[0,T]$. Applying $M$ repetitions of the sequence produces the FF
\begin{align*}
&\nonumber G^{-}_{a,a';b,b'}(\omega,M T)  \\
& =\pm  \frac{\sin (M \omega T)}{\sin (\omega T/2)} F^{(1)}_{a,a'}\!\left(\!\omega,\frac{T}{2}\right) F^{(1)}_{b,b'}\!\left(\!-\omega,\frac{T}{2}\right) \\
& \simeq  \pm \frac{2 \pi}{T} \!\sum_{k=-\infty}^\infty\!\! \!(\!-1)^k \delta(\omega \!-\! k \omega_0) G^+_{a,a';b,b'}\left(\!\omega,\frac{T}{2}\right),
\end{align*}
where the sign $\pm$ depends on whether  $y_{a,a'}(t)$ and $y_{b,b'}(t)$ are product-displacement $\pm$-antisymmetric. This FF contains the alternating frequency comb,
\begin{align*}
\frac{\sin (M \omega T)}{\sin (\omega T/2)} \simeq \frac{2 \pi}{T} \!\sum_{k=-\infty}^\infty\!\! \!(-1)^k \delta(\omega \!-\! k \omega_0),
\quad M \gg 1.
\end{align*}
Thus, through product-displacement antisymmetry, we can deconvolve integrals containing $G^{-}_{a,a';b,b'}(\omega,M T) $. 

A word of caution is in order, however. While product-displacement antisymmetry generates a comb in $G^{-}_{a,a';b,b'}(\omega,M T) $, this is not so for $G^{+}_{a,a';b,b'}(\omega,M T)$.  For $G_{a,a';b,b'}^{+} (\omega,T)$, the comb fails to emerge under sequence repetition for certain non-generic cases in which $G_{a,a';b,b'}^{+} (\omega,T) \sim \mathcal{O} (( \omega- 2\pi k/T)^{p>0})$, for $k\in\mathbb{Z}$. If switching functions $y_{a,a'}(t)$ and  $y_{b,b'}(t)$ are product-displacement $\pm$-antisymmetric on $[0,T]$, then 
$$G^{+}_{a,a';b,b'}(\omega,T) =\pm 2 i \sin\bigg(\frac{\omega T}{2}\bigg) G^{+}_{a,a';b,b'}(\omega,T/2),$$ 
\noindent
which is necessarily $\mathcal{O} (\omega -2 k\pi/T)$.  As a consequence, it is impossible to generate combs in $G^{-}_{a,a';b,b'}(\omega,M T)$ and  $G^{+}_{a,a';b,b'}(\omega,M T)$ {\em simultaneously}. It should also be highlighted that the comb approximation for both $G^{\pm}_{a,a';b,b'}(\omega,M T) $ holds only if the spectrum appearing in the convolution does not diverge at any point, implying that only sufficiently smooth power spectra can be reconstructed \cite{Norris2016}.

Timing symmetries can also control the real or imaginary character of the FFs, a technique that allows us to efficiently extract the real and imaginary components of the spectra. Suppose the switching functions $y_{a,a'}(t)$ and $y_{b,b'}(t)$ have-product displacement $\pm$-symmetry on the intervals $[0,T]$ and $[0,T/2]$. These switching functions generate the FF
\begin{align}
&\nonumber G^{+}_{a,a';b,b'}(\omega,T) = 4 \left[ 1 \pm \cos\left(\frac{\omega T}{2}\right) \right] \times \\
&\label{eq::prodsym1} 
\left[ 1 \pm \cos\left( \frac{\omega T}{4}\right) \right]  F^{(1)}_{a,a'}\!\!\left(\omega,\frac{T}{4}\right) F^{(1)}_{b,b'}\!\!\left(-\omega,\frac{T}{4}\right).   
\end{align}
This FF is real provided $F^{(1)}_{a,a'}(\omega,T/4)F^{(1)}_{b,b'}(-\omega,T/4)$ is real, a condition easily satisfied with the appropriate choice of subsequence. For example, consider a  subsequence $Q(T/4)$ of duration $T/4$ with switching functions $y_{a,a'}(t)$ and $y_{b,b'}(t)$ satisfying $y_{a,a'}(t)=y_{b,b'}(t)$ on the interval $[0,T/4]$. Note that $F^{(1)}_{a,a'}(\omega,T/4)F^{(1)}_{b,b'}(-\omega,T/4)$ is real for such a sequence. Consequently, the sequence $U_{\text{ctrl}}(T)=Q(T/4)^4,$
which is product-displacement $+$-symmetric on $[0,T]$ and $[0,T/2]$, produces a real FF $G^{+}_{a,a';b,b'}(\omega,T) $. Likewise, the sequence 
$$U_{\text{ctrl}}(T)\!=\!Q(T/4)\Pi_{\{a,b\}}^\dag Q(T/4)Q(T/4)\Pi_{\{a,b\}}Q(T/4),$$  
is product-displacement $-$-symmetric on $[0,T]$ and $[0,T/2]$, also generating a real filter. 

Suppose instead that the switching functions $y_{a,a'}(t)$ and $y_{b,b'}(t)$ have product-displacement $-$-symmetry on $[0,T]$ and 
$\pm$-antisymmetry on  $[0,T/2]$. This produces the filter
\begin{align}
&\nonumber  G^{+}_{a,a';b,b'}(\omega,T) = 4 i \left[1 - \cos\left(\frac{\omega T}{2}\right) \right] \times \\
&\label{eq::prodsym2}\left[ \pm  \sin \left(\frac{\omega T}{4}\right) \right]  F^{(1)}_{a,a'}\!\!\left(\omega,\frac{T}{4}\right) F^{(1)}_{b,b'}\!\!\left(-\omega,\frac{T}{4}\right),
\end{align}
which is imaginary provided $F^{(1)}_{a,a'}(\omega,T/4)F^{(1)}_{b,b'}(-\omega,T/4)$ is real.  A sequence satisfying these conditions can be constructed from the subsequence $Q(T/4)$ in a manner similar to the case above.  

Real and imaginary filters can also be generated from switching functions that are individually mirror symmetric or antisymmetric. If $y_{a,a'}(t)$ is mirror symmetric in an interval $[0,T]$, the associated first-order fundamental FF is 
\begin{align}
F_{a,a'}^{(1)}(\omega,T)&=  2 e^{i\frac{ \omega T}{2}}\textrm{Re} \left[e^{-i\frac{ \omega T}{2}} F_{a,a'}^{(1)}\left(\omega,\frac{T}{2}\right)\right]. 
\end{align}
On the other hand, if $y_{a,a'}(t)$ is mirror antisymmetric in $[0,T]$, the first-order fundamental FF is
\begin{align}
F_{a,a'}^{(1)}(\omega,T)&=  2 i e^{i\frac{\omega T}{2}}\textrm{Im} \left[e^{-i\frac{\omega T}{2}} F_{a,a'}^{(1)}\left(\omega,\frac{T}{2}\right)\right]. 
\end{align}
Thus, $G^{+}_{a,a';b,b'}(\omega,T) =F_{a,a'}^{(1)}(\omega,T)F_{b,b'}^{(1)}(-\omega,T)$  is real if both $y_{a,a'}(t)$ and $y_{b,b'}(t)$ are mirror symmetric or antisymmetric on $[0, T]$. If $y_{a,a'}(t)$ is mirror symmetric and $y_{b,b'}(t)$ is 
mirror antisymmetric on $[0,T]$ or vice versa, $G^{+}_{a,a';b,b'}(\omega,T)$ is imaginary. Designing mirror symmetry or antisymmetry into switching functions on $[0,T/4]$ can also be used to set the real or imaginary character of $F^{(1)}_{a,a'}(\omega,T/4) F^{(1)}_{b,b'}(-\omega,T/4)=G^+_{a,a';b,b'}(\omega, T/4)$ in Eqs. (\ref{eq::prodsym1}) and (\ref{eq::prodsym2}).

\subsection{Two-qubit noise spectroscopy setting}
\label{sec::2qsetting}

Having introduced the control symmetries essential for multiqubit QNS, we now explore the protocol in detail. 

\subsubsection{Qubit initialization and observables} 

In the single-qubit example in Sec.~\ref{sec::deconvolution}, the decay constant $\chi(t)$ depends on a convolution between the spectrum $S^+_{1,1}(\omega)$ and the FF $G^{+}_{1,1;1,1}(\omega,t)$. Measuring $\chi(MT)$ after repetition of multiple control sequences allows the spectrum to be reconstructed through linear inversion.  In the multiqubit case, the integral terms of \erf{C2} take the place of $\chi(t)$. Here, we demonstrate how the integral terms can be determined in the two-qubit case through state preparation of the qubits and measurement of particular observables.

To more clearly differentiate the integral terms, it is useful to formally expand \erf{C2} linearly in the 
operators $Z_a$,
\begin{align}
\label{eq::coefficients}
&\frac{\mathcal{C}^{(2)}_O(t)}{2!} =\sum_{a\in\mathcal{I}_N}\mathcal{C}^{}_{O,a}(t)\,Z_a\,,
\end{align}
where the expansion coefficient for observable $O$ along $a$ is
\begin{align}
&\mathcal{C}^{}_{O,a}(t)=\frac{1}{2^N}\Tr\Big[\,\frac{\mathcal{C}^{(2)}_O(t)}{2!}\,Z_a\Big]\label{eq::trC}.
\end{align}
Note from \erf{C2} that $\mathcal{C}^{(2)}_O(t)=\mathcal{C}^{(2)}_{O'}(t)$ for any two observables $O$ and $O'$ satisfying  $\text{sign}(O,a,b)=\text{sign}(O',a,b)$ for all $a,b\in\mathcal{I}_N$. As a consequence, $\mathcal{C}^{(2)}_O(t)$ is identical when $O\in\{X_\ell,Y_\ell\}$ or when $O\in\{X_\ell X_{\bar{\ell}},\,X_\ell Y_{\bar{\ell}},\,Y_\ell X_{\bar{\ell}},\,Y_\ell Y_{\bar{\ell}}\}$, where recall that the bar signifies  $\bar{\ell}\neq\ell$. The expansion coefficients in \erf{eq::trC} are likewise identical with
\begin{align*}
&\mathcal{C}^{}_{{}\ell,a}(t)\equiv\mathcal{C}^{}_{X_\ell,a}(t) = \, \mathcal{C}^{}_{Y_\ell,a}(t)\;\;\text{and}
\\&\mathcal{C}^{}_{{}{\ell\bar{\ell}},a} \!(t)\!\equiv\!\mathcal{C}^{}_{X_\ell X_{\bar{\ell}},a}\!(t)\!=\! \mathcal{C}^{}_{Y_\ell Y_{\bar{\ell}},a} \!(t)
\!= \!\mathcal{C}^{}_{X_\ell Y_{\bar{\ell}},a}\!(t)
\!=\!\mathcal{C}^{}_{Y_\ell X_{\bar{\ell}},a}\!(t).
\end{align*} 
We refer to the expansion coefficients by this shorthand notation for the remainder of the text.

\begin{table}
\begin{tabular}{| c | c | }
 \hline
{\bf Initial two-qubit state} & {\bf Two-qubit observables}  \\[3pt]
\hline 
$\;\ket{\psi_1^{\pm}} =\ket{+}_1 \otimes \ket{\pm z}_2\;$ & $\;X_1,\, Y_1\;$\\[3pt]
\hline 
$\;\ket{\psi_2^{\pm}} = \ket{\pm z}_1 \otimes \ket{+}_2\;$\; & $X_2,\, Y_2\;$\\[3pt]
\hline 
$\;\ket{\psi_{12}} = \ket{+}_1 \otimes \ket{+}_2\;$ & $\;X_1 X_{2},\,Y_1 Y_{2},\,Y_1 X_{2},\,X_1 Y_{2}\;$ \\[3pt]
   \hline
\end{tabular}
\caption{State preparations and observables for noise spectroscopy on two qubits. Here, $\ket{\pm}$ denote the eigenstates of $X$, while $\ket{\!+\!z}\equiv\ket{0}$ and
$\ket{\!-\!z}\equiv\ket{1}$  denote eigenstates of $Z$.}
\label{tab::2qubit}
\end{table}

In the two-qubit case, the expansion coefficients $ \mathcal{C}^{}_{{}\ell,a}(t)$ and $\mathcal{C}^{}_{{}{\ell\bar{\ell}},a}(t)$ for $\ell,\,\bar{\ell}\in\{1,2\}$ and $a\in\mathcal{I}_2$ serve as the analogues to $\chi(t)$ for a single qubit. These coefficients can be obtained experimentally by preparing the qubits in the initial states and measuring the corresponding observables given in Table \ref{tab::2qubit}. This choice of states and observables is not unique and can be refined with prior knowledge of the system. For two qubits initially prepared in state $\ket{\psi}$,  the expectation value of observable $O$ at time $t$ is given by $E_{\psi}[O(t)]= {\rm Tr} [ \tilde{U}(t) \ket{\psi}\bra{\psi} \otimes \rho_B \tilde{U}^\dagger(t) O]$. Let us then introduce the quantities 
\begin{align*}
&A_\ell^{\pm}(t)\equiv -\frac{1}{4}\text{log}\{E_{\psi_\ell^\pm}[X_\ell(t)]^2+E_{\psi_\ell^\pm}[Y_\ell(t)]^2\}, \;\ell=1,2, \\
&B_\ell^{\pm}(t)\equiv\frac{1}{2}\text{tanh}^{-1}\bigg(i\frac{E_{\psi_\ell^\pm}[Y_\ell(t)]}{E_{\psi_\ell^\pm}[X_\ell(t)]}\bigg), \;\ell=1,2,\\
&D^\pm(t)\equiv-\frac{1}{4}\text{log}\big(\{E_{\psi_{12}}[X_1Y_2(t)]\pm E_{\psi_{12}}[Y_1X_2(t)]\}^2+\\
&\;\;\;\;\;\;\;\;\;\;\;\;\;\;\;\;\;\;\;\;\;\;\;\;\{E_{\psi_{12}}[X_1X_2(t)]\mp E_{\psi_{12}}[Y_1Y_2(t)]\}^2\big).
\end{align*} 
From Eq.~\eqref{C2}, we find
\begin{align}
\label{eq::c1}\mathcal{C}_{\ell,\bar{\ell}}(t)&=A_\ell^+(t)\!+\!A_\ell^-(t),\\
\label{eq::c2}\mathcal{C}_{\ell,0}(t)&=A_\ell^+(t)\!-\!A_\ell^-(t),\\
\label{eq::c3}\mathcal{C}_{\ell,\ell\bar{\ell}}(t)&=B_\ell^+(t)\!+\!B_\ell^-(t),\\
\label{eq::c4}\mathcal{C}_{\ell,\ell}(t)&=B_\ell^+(t)\!-\!B_\ell^-(t),\\
\label{eq::c5}\mathcal{C}_{12,0}(t)&=D^+(t)\!+\!D^-(t),\\
\label{eq::c6}\mathcal{C}_{12,12}(t)&=D^+(t)\!-\!D^-(t),
\end{align} 
where $\ell,\bar{\ell}\in\{1,2\}$ with $\ell\neq\bar{\ell}$. The expectation values of the observables in Table \ref{tab::2qubit} are, thus, sufficient to determine {\em all} expansion coefficients.

\subsubsection{Base sequences for repetition}

The introduction of appropriate control timing symmetries  generates frequency combs in all the relevant filters, transforming the expansion coefficients into linear equations, rather than sums of convolutions. To proceed, we need a large number of sequences with the required symmetries in order to create a system of linear equations that can be inverted to obtain the spectra. Here, we provide criteria for selecting these sequences under both local and non-local control.

Generating the frequency combs requires repetition of base control sequences, $U_{\text{ctrl}}(T)$, which can be built as compositions of shorter subsequences, $Q(t_{i+1},t_i)$. An important factor to consider when building the base sequences is the presence of experimentally motivated constraints. 
Following our work in \cite{Norris2016}, we consider two: the minimum switching time $\tau_0$ and the time resolution $\delta $.  The former captures the fact that there is a finite pulse bandwidth, resulting in an unavoidable minimal waiting time between the application of two pulses. The latter is a constraint on our ability to apply a pulse at an arbitrary time. These constraints establish that the time separation between any two pulses, $\tau$, must satisfy
$$ k \delta = \tau \geq \tau_0 >0, \quad k \in {\mathbb N}.$$
As discussed in detail in Ref.~\cite{Norris2016}, the above condition implies the existence of a natural upper bound to the frequencies we can sample via the comb, namely $\omega \leq \pi/ 	\delta$. 

Another important factor is the spectral profile of the FFs associated with the base control sequences. Consider a spectrum $s(\omega)\in\{S_{a,b}^{\pm}(\omega)|a,b\in\mathcal{I}_N\}$. Each base sequence generates a different FF that couples to $s(\omega)$ in its convolution term. In order to reconstruct $s(\omega)$ at a set of harmonics, $\{k\omega_0|k\in\mathcal{K}\}$, at least one member of this FF set must nonzero at each harmonic $k\omega_0$. This ensures that $s(k\omega_0)$ contributes to the dynamics, allowing it to be sampled. While the set of FFs must have spectral weight at each of the harmonics, we would like to minimize the spectral overlap between the individual FFs. In other words, if $G_{a,a';b,b'}^{(1)}(\omega,t)$ and  $G_{c,c';d,d'}^{(2)}(\omega,t)$ are two filters associated with base sequences (1) and (2), respectively, the quantity $\int_{-\infty}^\infty d\omega G_{a,a';b,b'}^{(1)}(\omega,t)G_{c,c';d,d'}^{(2)}(\omega,t)$ should be as small as possible. This ensures a well-conditioned linear inversion.  The approach taken in Ref. \cite{Alvarez2011} to reduce spectral overlap was employing base sequences of different durations, $\{T/n,\ldots,T/2,T\}$ for $n\in\mathbb{Z}^+$. A base sequence of duration $T/n$ produces a comb that is non-zero at every $n$th harmonic.  Another approach is using base sequences whose FFs have different values of their {\em filtering order} (FO) and {\em cancellation order} (CO) \cite{Paz2014}. Generally, filter FFs with larger values of FO and CO have more support at higher frequencies, while those with smaller values of FO and CO have support at lower frequencies. A key ingredient to characterize the spectra is a FF with vanishing FO, which is nonzero at $\omega=0$, allowing the DC component of the spectra to be reconstructed. This may be achieved by using subsequences of free evolution \cite{Norris2016}.  

For local control by instantaneous $\pi$-pulses, we can choose base sequences from an extensive library of single- and multiqubit DD sequences with well-known filtering and cancellation properties~\cite{CDD,multiDD,NUDD}. 
Although low-order non-local control sequences have been considered, for instance in the context of open-loop pointer-state
engineering \cite{Pointer}, their filtering and cancellation properties have not been studied, to our knowledge.
Here, we demonstrate that it is possible to create FFs with arbitrarily high CO through concatenation \cite{CDD, multiDD}. 
Consider the following  two-qubit non-local DD sequence:
\begin{align*}
U_{1}(T)\! &=\! \Pi_{\!\{2,12\}} \textrm{SWAP}_{\!1,2} U_{f}\,\!\!\bigg(\frac{T}{4}\bigg)\! \textrm{SWAP}_{\!1,2}  \Pi_{\!\{1,12\}}  \!U_{f}\,\!\!\bigg(\frac{T}{4}\bigg)\\
&\;\;\times\! \Pi_{\!\{2,12\}}  \textrm{SWAP}_{\!1,2} U_{f}\,\!\!\left(\frac{T}{4}\right)\! \textrm{SWAP}_{\!1,2}  \Pi_{\!\{1,12\}}  \!U_{f}\,\!\!\left(\frac{T}{4}\right)\!\!,
\end{align*}
where $U_{f}(T/4)$ denotes free evolution for duration $T/4$. Using the formalism developed in Ref. \cite{Paz2014}, the FO and CO are completely determined by the fundamental FFs that compose a given FF. Direct calculation shows that 
\begin{align*}
F^{(1)}_{a,a'}(\omega,T) \sim \mathcal{O} (\omega^{1} T^{2}),\;
F^{(k\geq 2)}_{a,a';b,b'}(\omega,T) \sim \mathcal{O} (\omega^{0} T^{k}).	
\end{align*}
That is, the proposed non-local DD sequence has CO$ \,=1$ for the error basis $\{Z_1,Z_2,Z_1 Z_2\}$ relevant to our problem. Concatenating this sequence a total of $k$ times via the recursion
\begin{align*}
U_k(T) &= X_2\, \textrm{SWAP}_{12}\, U_{k-1}\!\!\left(\frac{T}{4}\right) \textrm{SWAP}_{12}\, X_1 \,U_{k-1}\!\!\left(\frac{T}{4}\right)\\
& \times  X_2\, \textrm{SWAP}_{12}\, U_{k-1}\!\!\left(\frac{T}{4}\right) \textrm{SWAP}_{12} \,X_1\, U_{k-1}\!\!\left(\frac{T}{4}\right)
\end{align*}
achieves CO$\,=k$, as desired.  

\subsubsection{Spectroscopy protocols}  

With the ability to measure the expansion coefficients and generate frequency combs in all relevant filters, the necessary tools are 
in place. We next present a detailed QNS protocol to reconstruct all quantum, classical, cross- and self-spectra. The accessible spectra depend on the level of control complexity, i.e. diagonal (purely local) vs. non-diagonal (non-local) control.  
Before delving into specifics, we outline the essential procedure. 
All expansion coefficients take the general form
\begin{align}
\label{eq::genEC}
\mathcal{C}_{a,b}(t)=\int_{-\infty}^\infty \!\!\!\!d\omega\!\!\!\sum_{\substack{g_i(\omega,t)\in\mathcal{G}_i\\s(\omega)\in\mathcal{S}_2}}\!\!\!\!k_{a,b}(g_i,s)g_i(\omega,t)s(\omega),
\end{align}
where $\mathcal{G}_i$ is the set of all FFs for a control sequence $i$, $\mathcal{S}_2$ is the set of all spectra for $N=2$, and the constants $k_{a,b}(g_i,s)$ are specific to $\mathcal{C}_{a,b}(t)$. Most of the expansion coefficients contain either $G_{a,a';b,b'}^+(\omega,t)$ or $G_{a,a';b,b'}^-(\omega,t)$, but not both. In the case where $\mathcal{C}_{a,b}(t)$ depends only on $G_{a,a';b,b'}^+(\omega,t)$ FFs, a comb can be generated through repetition provided that the non-generic cases are avoided for the $g_i(\omega,T)$ involved,
\begin{align}
\label{eq::GpInvert}
\overline{\mathcal{C}}_{a,b}(MT)_i\!\simeq\!\frac{2\pi M}{T}\!\!\!\!\!\!\!\!\sum_{\substack{k\in\mathcal{K}\\g_i(k\omega_0,T)\in\mathcal{G}_i\\s(k\omega_0)\in\mathcal{S}_2}}\!\!\!\!\!\!\!\!\!\!k_{a,b}(g_i,s)g_i(k\omega_0,T)s(k\omega_0).
\end{align}
Here, we have replaced $\mathcal{C}_{a,b}(t)$ with $\overline{\mathcal{C}}_{a,b}(MT)_i$, its measured value after $M$ repetitions of $i$, and we have truncated the summation to a finite set of harmonics.
In Eq.~(\ref{eq::GpInvert}), every quantity is known except for the $s(k\omega_0)$: $\overline{\mathcal{C}}_{a,b}(MT)_i$ through measurement, $k_{a,b}(g_i,s)$ through the explicit form of the expansion coefficient, and $g_i(k\omega_0,T)$ through the control sequence $i$. Obtaining $\overline{\mathcal{C}}_{a,b}(MT)_i$ for a set of sequences $i\in\{1,\ldots,N_c\}$, with $N_c$ greater or equal to  the number of unique $s(k\omega_0)$ in \erf{eq::GpInvert}, thus creates a system of linear equations that can be solved for the $s(k\omega_0)$. 

For expansion coefficients containing $G_{a,a';b,b'}^-(\omega,t)$ filters, product-displacement antisymmetry generates a comb through repetition, forming 
\begin{align}
\overline{\mathcal{C}}_{a,b}(MT)_i\!\simeq\!\frac{2\pi}{T}\!\!\!\!\!\!\!\!\sum_{\substack{k\in\mathcal{K}\\g_i(k\omega_0,T)\in\mathcal{G}_i\\s(k\omega_0)\in\mathcal{S}_2}}\!\!\!\!\!\!\!\!\!\!(-1)^kk_{a,b}(g_i,s)g_i(k\omega_0,T)s(k\omega_0).
\end{align}
In this case, we can solve for for the spectra just as above. Determining $\overline{\mathcal{C}}_{a,b}(MT)_i$ for a sufficient number of control sequences creates a system of linear equations, inverting which returns the $s(k\omega_0)$. Finally, some expansion coefficients contain both $G_{a,a';b,b'}^\pm(\omega,t)$. Because it is impossible to generate combs in these FFs simultaneously, we take linear combinations of the expansion coefficients to isolate the terms containing either $G_{a,a';b,b'}^+(\omega,t)$ or $G_{a,a';b,b'}^-(\omega,t)$. This procedure will be described in detail below, in treating diagonal control.

\vspace*{1mm}

{\bf{Diagonal (local) control.}} The simplest scenario involves diagonal or, equivalently, purely local control.  Recall that diagonal control consists of $\pi$-pulses, which are products of the Pauli operators $\{ X_\ell, Y_\ell\}$ on the individual qubits $\ell\in\{1,2\}$. The expansion coefficients then take the explicit form
\begin{align}
\nonumber \!&  \mathcal{C}_{\ell,12}(t)\! = i \!\int \limits_{-\infty}^\infty \!\frac{d \omega}{2\pi} \textrm{Im} \{S_{1,2}^-(\omega)[G_{1,1;2,2}^{-}(\omega,t)\\
\label{ddc1} \! & \quad \quad \quad\quad\quad\quad\quad\quad\quad\quad\quad- {(-1)^\ell} G_{1,1;2,2}^{+}(\omega,t)]\},  \\
\nonumber \!&  \mathcal{C}_{\ell,\ell}(t)\! = i \int \limits_{-\infty}^\infty \frac{d \omega}{2\pi} \textrm{Im}\{S_{\ell,0}^{-} (\omega) [ G^{-}_{\ell,\ell;0,0} (\omega,t) \\
\label{ddc2} & \quad \quad\quad\quad\quad\quad\quad\quad\quad\quad\quad + G^{+}_{\ell,\ell;0,0} (\omega,T)]\},\\
\label{ddc3}\!&{ \mathcal{C}_{\ell,0}(t)\! =\int \limits_{-\infty}^\infty \frac{d\omega}{2\pi}\sum_{a \in \{\ell,12\}}  \textrm{Re}[ S_{a,a}^{+}(\omega) G_{a,a;a,a}^{+} (\omega,T) ],} \\
\label{ddc4a}\!&{{ \mathcal{C}_{1,2}(t)\! = 2 \int \limits_{-\infty}^\infty \frac{d\omega}{2\pi} \textrm{Re}[\ignore{\sum_{a,b \in \{(\ell,12)\}{a > b\neq \ell{'},0}}} S_{1,12}^{+}(\omega) G_{1,1;12,12}^{+} (\omega,T) ],}}\\
\label{ddc4b}\!&{{ \mathcal{C}_{2,1}(t)\! = 2 \int \limits_{-\infty}^\infty \frac{d\omega}{2\pi} \textrm{Re}[\ignore{\sum_{a,b \in \{(\ell,12)\}{a > b\neq \ell{'},0}}} S_{2,12}^{+}(\omega) G^{+}_{2,2;12,12} (\omega,T) ],}} \\
\label{ddc5}\!&{{ \mathcal{C}_{12,0}(t)\! =  \int \limits_{-\infty}^\infty \frac{d\omega}{2\pi} \sum_{\ell=1}^2\textrm{Re}[ S_{\ell,\ell}^{+}(\omega) G_{\ell,\ell;\ell,\ell}^{+} (\omega,T)],}} \\
\label{ddc6}\!&{ \mathcal{C}_{12,12}(t)\! = 2 \int \limits_{-\infty}^\infty \frac{d\omega}{2\pi} \textrm{Re}[ S_{1,2}^{+}(\omega) G_{1,1;2,2}^{+} (\omega,T)].}
\end{align}
Note that another distinction between the M1 and M2 noise models is evident in these expansion coefficients. Because $B_0(t)=0$ in the M1 model, $S^-_{1,0}(\omega)=0=S^-_{2,0}(\omega)$, implying the expansions coefficients $\mathcal{C}_{1,1}(t)$ and 
$\mathcal{C}_{2,2}(t)$ vanish. For the M2 model, where $B_0(t)=B_1(t)+B_{2}(t)$, 
\begin{align*}
S^-_{\ell,0}(\omega)&=\int_{-\infty}^\infty d \tau e^{- i \omega \tau} \langle[B_{\ell}(\tau),B_{0}(0)]\rangle_{c,q}\\
&=S^-_{\ell,1}(\omega) + S^-_{\ell,2}(\omega).
\end{align*}
The quantum self-spectra $S^-_{1,1}(\omega)$ and $S^-_{2,2}(\omega)$, thus, enter the qubit dynamics through the expansion coefficients $\mathcal{C}_{1,1}(t)$ and $\mathcal{C}_{2,2}(t)$ in the M2 model. In contrast, the quantum self-spectra have {\em no} dynamical influence in the M1 model, as anticipated in the Introduction. This is reminiscent of the single-qubit example in Sec.~\ref{sec::physSigA}, where the qubit in the M2 model experiences phase rotation due to $S_{1,1}^-(\omega)$, an effect absent in the M1 model. The presence of the quantum self-spectra in the qubit dynamics of the M2 model has implications for the development of our QNS protocol, as we shall show.

We now examine how the spectra can be extracted from the expansion coefficients in Eqs. (\ref{ddc1})-(\ref{ddc6}). \\

\noindent {{\bf Step 1:}} Consider $\mathcal{C}_{\ell,0}(t)$, $\mathcal{C}_{1,2}(t)$, $\mathcal{C}_{2,1}(t)$, $\mathcal{C}_{12,0}(t)$ and $\mathcal{C}_{12,12}(t)$, which contain the spectra $S_{\ell,\ell}^+(\omega)$, $S_{12,12}^+(\omega)$, $S_{\ell,12}^+(\omega)$ and  $S_{1,2}^+(\omega)$.  Because these expansion coefficients only depend on $G_{a,a';b,b'}^+(\omega,t)$  FFs, they can be deconvolved using control repetition, provided that the non-generic cases are avoided. Extracting $S_{\ell,\ell}^+(\omega)$, $S_{12,12}^+(\omega)$, $\text{Re}[S_{\ell,12}^+(\omega)]$ and  $\text{Re}[S_{1,2}^+(\omega)]$ requires a real FF.  We can insure $G_{a,a';b,b'}^+(\omega,T)$ is real by using a control sequence with $y_{a,a'}(t)$ and $y_{b,b'}(t)$ that are both mirror symmetric or antisymmetric on the interval $[0,T/4]$ and satisfy product-displacement $\pm$-symmetry on $[0,T]$ and $[0,T/2]$. Through a set of sequences with these symmetries, we obtain a system of linear equations of the form in \erf{eq::GpInvert}, which can be inverted to obtain the real components of the spectra. To extract $\text{Im}[S_{\ell,12}^+(\omega)]$  and  $\text{Im}[S_{1,2}^+(\omega)]$ from $\mathcal{C}_{1,2}(t)$, $\mathcal{C}_{2,1}(t)$ and $\mathcal{C}_{12,12}(t)$, respectively, we repeat the exact same procedure except we use control sequences where $G_{a,a';b,b'}^+(\omega,T)$ is imaginary.
This can be accomplished when $y_{a,a'}(t)$ and $y_{b,b'}(t)$ that are both mirror symmetric or antisymmetric on the interval $[0,T/4]$ and satisfy product-displacement $-$-symmetry on $[0,T]$ along with with product-displacement antisymmetry on $[0,T/2]$. 

It should be noted that when an expansion coefficient is the sum of multiple convolutions, such as $\mathcal{C}_{\ell,0}(t)$ or $\mathcal{C}_{12,0}(t)$, the individual spectra can still be isolated. Consider $\mathcal{C}_{1,0}(t)$, for example.  After $M$ repetitions of a base sequence, this expansion coefficient becomes
\begin{align*}
\mathcal{C}_{1,0}(MT)\! & =\frac{M}{T} \sum_{k\in\mathcal{K}}  \textrm{Re}[S_{1,1}^{+}(\omega_0) G_{1,1;1,1}^{+} (k \omega_0,T) ] \\ 
& + \frac{M}{T} \sum_{k\in\mathcal{K}} \textrm{Re}[ S_{12,12}^{+}(k \omega_0) G_{12,12;12,12}^{+} (k \omega_0,T) ].
\end{align*}
Note that the base sequences can be chosen so that  $G_{1,1;1,1}^{+} (k \omega_0,T)\neq G_{12,12;12,12}^{+} (k \omega_0,T)$ for all ${k\in\mathcal{K}}$. As long as the number of base sequences is $N_c\geq2|\mathcal{K}|$, this creates a $N_c\times 2|\mathcal{K}|$ linear system that can be inverted to obtain both  $\{S_{12,12}^{+}(k \omega_0)\}$ and $\{ S_{1,1}^{+}(k \omega_0) \}$ for all ${k\in\mathcal{K}}$.  An alternative is taking the difference of two $\mathcal{C}_{1,0}(MT)$ obtained with control sequences that produce identical $G_{1,1;1,1}^{+} (k \omega_0,T)$ and different  $G_{12,12;12,12}^{+} (k \omega_0,T)$ or vice versa, which cancels the terms containing $G_{1,1;1,1}^{+} (k \omega_0,T)$ or those containing  $G_{12,12;12,12}^{+} (k \omega_0,T)$.  When repeated for $N_c\geq|\mathcal{K}|$ pairs of control sequences, this creates a $N_c\times |\mathcal{K}|$ linear system that can be inverted to obtain $\{S_{12,12}^{+}(k \omega_0)\}$ or $S_{1,1}^{+}(k \omega_0) \}$. This procedure is illustrated in detail in Appendix \ref{sec::procedureExciton}.\\

\noindent {{\bf Step 2:}} One can access $S_{1,2}^{-} (\omega)$ by deconvolving the integrals in $\mathcal{C}_{\ell,12}(t)$, which contain \emph{both} $G_{1,1;2,2}^\pm(\omega,t)$. To carry this out, we isolate the terms containing $G_{1,1;2,2}^+(\omega,t)$ by taking
\begin{align*}
\mathcal{C}_{1,12}(t) - \mathcal{C}_{2,12}(t) 
=2 i \int\limits_{-\infty}^\infty \frac{d \omega}{2\pi} \textrm{Im}[S_{1,2}^{-} (\omega) G^{+}_{1,1;2,2} (\omega,t) ].
\end{align*}
We can then use repetition of sequences with real $G^{+}_{1,1;2,2} (\omega,T)$ to extract $\textrm{Im}[S_{1,2}^{-} (\omega)]$ and repetition sequences with imaginary $G^{+}_{1,1;2,2} (\omega,T)$ to extract $\textrm{Re}[S_{1,2}^{-} (\omega)]$. \\

\noindent {{\bf Step 3:}} The remaining contribution is $\mathcal{C}_{\ell,\ell}(t)$, which contains the only power spectrum in Eqs. (\ref{ddc1})-(\ref{ddc6}) that we miss, $S^-_{\ell,0}(\omega)$. This needs to be treated in a case-by-case basis:

\vspace*{1mm}

$\bullet$ In M1, $S_{\ell,0}^{-} (\omega) = 0$ and $\mathcal{C}_{\ell,\ell}(t)$ trivially vanishes. Thus, we can access all the 
power spectra relevant to the dynamics generated by local control for M1. Note that, as was already seen in Eqs.~\eqref{ddc1}-\eqref{ddc6}, the quantum self-spectra $S_{\ell,\ell}^{-}(\omega)$ do not influence local-control dynamics in a M1 model, 
which implies that they are not accessible via local control only.  

$\bullet$ For the M2 case, the situation is more complicated. In  the previous step, we obtained  $\{S_{1,2}^{-}(k\omega_0)\}$. If one assumes that this spectrum is sufficiently smooth so that it is well approximated by an interpolation of $\{S_{1,2}^{-}(k\omega_0)\}$, we can use the interpolation to determine $S_{\ell,0}^{-} (\omega)$.  Let $S_{1,2}^{-,I} (\omega)$ denote the interpolation of $\{S_{1,2}^{-}(k\omega_0)\}$ 
Recall that for M2, $S^{-}_{\ell,0} (\omega) = S_{\ell,1}^{-} (\omega)+ S_{\ell,2}^{-} (\omega)$. Consider 
\begin{align*}
& \mathcal{C}_{\ell,\ell}(t) - 2 i\!\!\! \int\limits_{-\infty}^\infty\!\! \frac{d \omega}{2\pi} \textrm{Im}[S_{1,2}^{-,I} (\omega)(G^{+}_{\ell,\ell;0,0} (\omega,t) + G^{-}_{\ell,\ell;0,0} (\omega,t)) ] \\
&\approx 2 i\!\!\! \int\limits_{-\infty}^\infty \!\!\frac{d \omega}{2\pi} \textrm{Re}[S_{\ell,\ell}^{-} (\omega)] \textrm{Im}[(G^{+}_{\ell,\ell;0,0} (\omega,t) + G^{-}_{\ell,\ell;0,0} (\omega,t))].
\end{align*}
Note that all quantities on the left hand-side are known. We can, thus, obtain $S_{\ell,\ell}^{-} (\omega)$ by deconvolving the righthand side. To do so, let us first note that control repetition and displacement antisymmetry in $[0,T]$ lead to
\begin{align*}
G^{+}_{\ell,\ell;0,0} (\omega,T) &\! =\! -2 i \frac{\sin^2 (\frac{\omega T M}{2})}{\sin (\frac{\omega T}{2})} F^{(1)}_{\ell,\ell} \Big(\!\omega,\frac{T}{2} \Big) F^{(1)}_{0,0}  \Big(\!\!-\omega,\frac{T}{2} \Big) , 
\end{align*}
\begin{align*}
G^{-}_{\ell,\ell;0,0} (\omega,T) &\!=\! -\frac{\sin  (\omega T M )}{\sin (\frac{\omega T}{2})} F^{(1)}_{\ell,\ell}  \Big(\!\omega,\frac{T}{2} \Big) F^{(1)}_{0,0}  \Big(\!\!-\omega,\frac{T}{2} \Big).
\end{align*}
It follows then that, if $y_{\ell,\ell}(s)$ is chosen to be a mirror antisymmetric sequence in $[0,T/2]$, then $F^{(1)}_{\ell,\ell} (\omega,\frac{T}{2})  F^{(1)}_{0,0} (-\omega,\frac{T}{2})$ is purely imaginary. Consequently, $G^{+}_{\ell,\ell;0,0} (\omega,T)$ is real, 
while  $G^{-}_{\ell,\ell;0,0} (\omega,T)$ is imaginary, which implies
\begin{align*}
& 2 i \int \frac{d \omega}{2\pi} \textrm{Re}[S_{\ell,\ell}^{-} (\omega)] \textrm{Im}[(G^{+}_{\ell,\ell;0,0} (\omega,T) + G^{-}_{\ell,\ell;0,0} (\omega,T))]\\
&=2 i \int \frac{d \omega}{2\pi} \textrm{Re}[S_{\ell,\ell}^{-} (\omega)] \textrm{Im}[G^{-}_{\ell,\ell;0,0} (\omega,T)],
\end{align*}
which, as was shown earlier, generates a frequency comb. Unlike M1, local control is sufficient to characterize {\it all} power spectra for M2.

\vspace*{1mm}

{\bf Non-diagonal (non-local) control.} In the case of non-local control, the dynamics are considerably richer. It can be seen from Eq.~\eqref{C2}, that the expansion coefficients depend on additional power spectra that were absent in the case of local control.    
This enables us to obtain the {\em quantum self-spectra} $S_{\ell,\ell}^{-}(\omega)$, which we could {\em not} access for M1. For M2, non-diagonal control provides a means of accessing the quantum self-spectra {\em without} resorting to interpolation. Under non-diagonal control, $S_{\ell,\ell}^{-}(\omega)$ enters the dynamics through the expansion coefficient $C_{\ell,12}(t)$, which takes the form 
\begin{align*}
 \mathcal{C}_{\ell,12}(t)\! = i \!\!\sum_{m,m'=1}^2\int \limits_{-\infty}^\infty \!\frac{d \omega}{2\pi} \textrm{Im} &\{S_{m,m'}^-(\omega)[G_{1,m;2,m'}^{-}(\omega,t)\\
 &- {(-1)^\ell} G_{1,m;2,m'}^{+}(\omega,t)]\}.
\end{align*}
Because the FFs cannot generate combs simultaneously, we isolate $G_{1,a;2,b}^{-}(\omega,t)$ and $G_{1,a;2,b}^{+}(\omega,t)$ by
\begin{align*}
& \mathcal{C}_{1,12}(t)+ \mathcal{C}_{2,12}(t)\!  =2i\sum_{\ell=1}^2\int \limits_{-\infty}^\infty \!\frac{d \omega}{2\pi}S_{\ell,\ell}^-(\omega) \textrm{Im}[G_{1,\ell;2,\ell}^{-}(\omega,t)]\\
 &+\!2i\!\!\!\!\int \limits_{-\infty}^\infty\!\!\! \frac{d \omega}{2\pi}\textrm{Im}[S_{1,2}^-(\omega)G_{1,1;2,2}^{-}(\omega,t)\!+\!S_{2,1}^-(\omega)G_{1,2;2,1}^{-}(\omega,t)] , \\
 & \mathcal{C}_{1,12}(t)- \mathcal{C}_{2,12}(t)\!  =2i\sum_{\ell=1}^2\int \limits_{-\infty}^\infty \!\frac{d \omega}{2\pi}S_{\ell,\ell}^-(\omega) \textrm{Im}[G_{1,\ell;2,\ell}^{+}(\omega,t)]\\
&+\!2i\!\!\!\!\int \limits_{-\infty}^\infty\!\!\! \frac{d \omega}{2\pi}\textrm{Im}[S_{1,2}^-(\omega)G_{1,1;2,2}^{+}(\omega,t)\!+\!S_{2,1}^-(\omega)G_{1,2;2,1}^{+}(\omega,t)].
\end{align*}
Note that the last lines in both of these expressions depend on the quantum cross-spectra, which we have already obtained through diagonal control. Under repetition of suitable sequences (a displacement antisymmetric sequence for the $G_{a,a';b,b'}^{-}(\omega,t)$ FFs), we have 
\begin{align}
\label{eq::delp}
&\Delta^{+}(MT) \equiv \mathcal{C}_{1,12}(MT) + \mathcal{C}_{2,12}(MT)-I^+(MT) \\\notag
&\;\;\;\;\;\;\;\;\approx\frac{2 i}{T} \sum_{\ell=1}^2 \sum_{k\in\mathcal{K}} S^{-}_{\ell,\ell} (k\omega_0) \textrm{Im} [G^{-}_{1,\ell; 2,\ell} (k\omega_0,T)],\\
\label{eq::delm}
&\Delta^{-}(MT) \equiv \mathcal{C}_{1,12}(MT) - \mathcal{C}_{2,12}(MT) -I^-(MT) \\\notag
&\;\;\;\;\;\;\;\;\approx \frac{2 iM}{T} \sum_{\ell=1}^2 \sum_{k\in\mathcal{K}}  S^{-}_{\ell,\ell} (k\omega_0) \textrm{Im} [G^{+}_{1,\ell; 2,\ell} (k\omega_0,T)],
\end{align}
where the terms $I^\pm(MT)$ depend on the reconstruction of the quantum cross-spectrum,
\begin{align*}
I^+(MT)=\frac{2 i}{T} \sum_{k\in\mathcal{K}} \textrm{Im}[&S_{1,2}^-(k\omega_0)G_{1,1;2,2}^{-}(k\omega_0,T)
\\&\;\;+\!S_{2,1}^-(k\omega_0)G_{1,2;2,1}^{-}(k\omega_0,t)],\\
I^-(MT)=\frac{2 iM}{T} \sum_{k\in\mathcal{K}} \textrm{Im}[&S_{1,2}^-(k\omega_0)G_{1,1;2,2}^{+}(k\omega_0,T)
\\&\;\;+\!S_{2,1}^-(k\omega_0)G_{1,2;2,1}^{+}(k\omega_0,t)].
\end{align*}
Determining $\Delta^{\pm}(MT)$ for a set of sequences creates a system of linear equations that can be inverted to obtain $S_{1,1}^-(\omega)$ and  $S_{2,2}^-(\omega)$. We have, thus, obtained all dynamically relevant spectra for both the M1 and M2 models on
$N=2$ qubits.

\subsection{Noise spectroscopy beyond two qubits}
\label{sec::Nqubit}

The above procedure may be extended to $N$ qubits, where the goal  is characterizing the full set of spectra $\mathcal{S}_N\equiv \{S_{a,b}^{\pm}(\omega)|a,b\in\mathcal{I}_N\}$.   Without describing the protocol to the level of detail given in the two-qubit case, 
we show how, through proper application of control symmetries and measurement of qubit observables, it is still possible in principle to access all the spectra governing the $N$-qubit dynamics.

To obtain the spectra in $\mathcal{S}_N$, we consider a tripartite setting: the two-party setting we have treated thus far (consisting, say, of qubits $\ell$ and $\bar{\ell}\,$) plus a third party, $R_{\ell,\bar{\ell}} \equiv \{1,\ldots,N\}-\{\ell,\bar{\ell}\,\}$, which contains the remaining $N-2$ qubits. Once again, we use the convention $\ell\neq\bar{\ell}$. Let $\mathcal{I}_{\ell,\bar{\ell}}\equiv \{\ell,\bar{\ell},\ell\bar{\ell}\,\}$ be the set of indices relevant to qubits $\ell$ and $\bar{\ell}$ alone and  $B_{\ell,\bar{\ell}}\equiv \{\ell \,r,\bar{\ell}\,r|r\in R_{\ell,\bar{\ell}}\}$. By adapting the two-qubit protocol, we can reconstruct $\{S_{a,b}^\pm(\omega),S_{a,c}^+(\omega),S_{c,d}^+(\omega)|a,b\in\mathcal{I}_{\ell,\bar{\ell}},\,c,d\in B_{\ell,\bar{\ell}}\}$ for a fixed pair of qubits $\ell$ and $\bar{\ell}$. By repeating this procedure for every pair $\ell$ and $\bar{\ell}$, all spectra in $\mathcal{S}_N$ can be characterized.

We start by modifying the two-qubit approach introduced in the last section so that it is possible to access the spectra affecting qubits $\ell$ and $\bar{\ell}$ in the presence of the remaining $N-2$ qubits. Let $\vec{s}\;$ be a vector of length $N$ with entries that are either $+$ or $-$.  The first column of Table \ref{tab::Nqubit} describes initial states of the $N$-qubit ensemble in which qubit $\ell$ and/or qubit $\bar{\ell}$ are prepared in $\ket{+}_\ell$ and/or $\ket{+}_{\bar{\ell}}$ with the remaining qubits prepared in $\ket{s_jz}_j$, where $s_j\in\{+,-\}$ is the $j$th component of $\vec{s}$. The second column contains observables on qubits $\ell$ and $\bar{\ell}$. These are the $N$-qubit analogues to the two-qubit state preparations and observables presented in Table \ref{tab::2qubit}. Preparing the qubits in the states specified in Table \ref{tab::Nqubit} and measuring the corresponding observables allows one to obtain the expansion coefficients 
$\{\mathcal{C}_{a,b}(t)|a\in\mathcal{I}_{\ell,\bar{\ell}},b\in\mathcal{I}_N\}$. For $a\in\{\ell,\bar{\ell}\}$ and $R_a=\{1,\ldots,N\}-\{a\}$,
these coefficients read 
\begin{align*}
&\mathcal{C}_{a,0}(t)+\!\!\sum_{j\in R_a}\!\!s_j\,\mathcal{C}_{a,j}(t)\!=\!\frac{-\text{log}\{E_{\psi_a^{\vec{s}}}[X_a(t)]^2\!+\!E_{\psi_a^{\vec{s}}}[Y_a(t)]^2\}}{2}\\
&\mathcal{C}_{a,a}(t)\!+\!\!\sum_{j\in R_a}\!s_j\,\mathcal{C}_{a,aj}(t)=\text{tanh}^{-1}\left\{i\frac{E_{\psi_a^{\vec{s}}}[Y_a(t)]}{E_{\psi_a^{\vec{s}}}[X_a(t)]}\right\}\\
&\mathcal{C}_{\ell\bar{\ell},0}(t)\!\mp\!\mathcal{C}_{\ell\bar{\ell},\ell\bar{\ell}}(t)\!\!=\!\frac{1}{2}\text{log}
\big(\{E_{\psi_{\ell\bar{\ell}}^{\vec{s}}}[X_\ell Y_{\bar{\ell}}(t)]\!\pm\! E_{\psi_{\ell\bar{\ell}}^{\vec{s}}}[Y_\ell X_{\bar{\ell}}(t)]\}^2\\
&\;\;\;\;\;\;\;\;\;\;\;\; \;\;\;\;\;\;\;\;\;\;\;\; \;\;\;\;\;\;\;+\{E_{\psi_{\ell\bar{\ell}}^{\vec{s}}}[X_\ell X_{\bar{\ell}}(t)]\!\mp\! E_{\psi_{\ell\bar{\ell}}^{\vec{s}}}[Y_\ell Y_{\bar{\ell}}(t)]\}^2\big)\\
&\mathcal{C}_{\ell\bar{\ell},\bar{\ell}}(t)\pm\mathcal{C}_{\ell\bar{\ell},\ell}(t)+\sum_{j\in R_{\ell,\bar{\ell}}}\!s_j\,[\mathcal{C}_{\ell\bar{\ell},\bar{\ell} j}(t)\!\pm\!\mathcal{C}_{\ell\bar{\ell},\ell j}(t)]=\\
&\;\;\;\;\;\;\;\;\;\;\;\; \;\;\;\;\;\;\;\;\text{tanh}^{-1}\left\{i\frac{E_{\psi_{\ell\bar{\ell}}^{\vec{s}}}[X_\ell Y_{\bar{\ell}}(t)]\pm E_{\psi_{\ell\bar{\ell}}^{\vec{s}}}[Y_\ell X_{\bar{\ell}}(t)]}{E_{\psi_{\ell\bar{\ell}}^{\vec{s}}}[X_\ell X_{\bar{\ell}}(t)]\mp E_{\psi_{\ell\bar{\ell}}^{\vec{s}}}[Y_\ell Y_{\bar{\ell}}(t)]}\right\}.
\end{align*}
Determining the expectation values of the observables in Table \ref{tab::Nqubit} for all possible state preparations forms systems of linear equations, which can be solved to obtain the $\mathcal{C}_{a,b}(t)$.

\begin{table}
\begin{tabular}{| c | c | }
 \hline
{\bf Initial multiqubit states} & {\bf Observables}  \\[3pt]
\hline 
$\;\ket{\psi_\ell^{\vec{s}}\,}= \ket{+}_\ell \bigotimes_{j \neq \ell} \ket{s_jz}_j\;$ & $\;X_\ell, Y_\ell \;$\\[3pt]
\hline 
$\; \ket{\psi_{\bar{\ell}}^{\vec{s}}\,}= |+\rangle_{\bar{\ell}} \bigotimes_{j \neq \bar{\ell}} \ket{s_jz}_j\;$ & $\;X_{\bar{\ell}}, Y_{\bar{\ell}} \;$\\[3pt]
\hline 
$\;\ket{\psi_{\ell,\bar{\ell}}^{\vec{s}}\,} = \ket{+}_\ell \, |+\rangle_{\bar{\ell}} \bigotimes_{j \neq \ell,\bar{\ell}} \ket{s_jz}_j\; \;$ & $  \begin{array}{c} X_\ell X_{\bar{\ell}},Y_\ell Y_{\bar{\ell}} \\ Y_\ell X_{\bar{\ell}},X_\ell Y_{\bar{\ell}} \end{array}$ \\[5pt]
\hline 
\end{tabular}
\caption{State preparations and observables for noise spectroscopy on $N$ qubits. As in Table \ref{tab::2qubit}, $\ket{\pm}$ denote the eigenstates of $X$, while $\ket{\!+\!z}\equiv\ket{0}$ and
$\ket{\!-\!z}\equiv\ket{1}$  denote eigenstates of $Z$.}
\label{tab::Nqubit}
\end{table}

As in the two-qubit case, we deconvolve the expansion coefficients by generating frequency combs in the FFs. To accomplish this, we apply selective control to two qubits, say $\ell$ and $\bar{\ell}$, and homogeneous control to the remaining $N-2$ qubits. Consider first the {M1 model} in the diagonal control scenario. For a fixed pair  $\ell,\, \bar{\ell}$, the expansion coefficients $\mathcal{C}_{a,b}(t)$ for $a,b \in \mathcal{I}_{\ell,\bar{\ell}}$ depend on convolutions involving the spectra $S^\pm_{a,b}(t)$ for $a,b \in \mathcal{I}_{\ell,\bar{\ell}}$, as if one were dealing with the two qubit case, plus additional classical spectra that arise due to the other $N-2$ qubits: $S^+_{\ell r,\ell r}(\omega),S^+_{\bar{\ell} r, \bar{\ell} r}(\omega), S^+_{{\ell} r, \bar{\ell} r}(\omega)$, $S^+_{\bar{\ell} r, {\ell} r}(\omega)$ for $r\in R_{\ell,\bar{\ell}}$\footnote{The classical spectra $S^+_{\ell r,r}(\omega)$ and $S^+_{\bar{\ell} r,r}(\omega)$, which would contribute to $\mathcal{C}_{\bar{\ell},\ell}$ and $\mathcal{C}_{\ell,\bar{\ell}}$ coefficients, respectively, do not appear as they are necessarily filtered by $(G^{+}_{\cdot r,\cdot r;r,r} (\omega,T) - G^{+}_{\cdot r,\cdot r;r,r}(\omega,T)) =0$, where $\cdot$ stands for 
an index in $\{\ell,\bar{\ell}\}$, as can be seen from Eq.~\eqref{eq::sgn}.}.
Note the absence of quantum spectra involving the indices $\ell r$ or $\bar{\ell} r$, which occurs because these spectra depend on commutators of classical noise operators and therefore vanish. Quantum cross-spectra of the form $S_{\ell,r}^-(\omega)$ do not enter the convolutions present in $\mathcal{C}_{a,b}(t)$ for $a,b \in \mathcal{I}_{\ell,\bar{\ell}}$. The additional classical spectra involving $r$ present in the convolutions pose a complication in that they are always filtered by the same non-vanishing function. For example, consider the expansion coefficient 
\begin{align*}
\mathcal{C}_{\ell\bar{\ell},\ell\bar{\ell}}(t) &=  \frac{1}{\pi}\int \limits_{-\infty}^\infty \!\!\!d\omega\, \textrm{Re}[ S_{\ell,\bar{\ell}}^{+}(\omega) G_{\ell,\ell;\bar{\ell},\bar{\ell}}^{+} (\omega,t)]\\
&+  \frac{1}{\pi} \int \limits_{-\infty}^\infty \!\!\!d\omega\sum_{r \in R_{\ell,\bar{\ell}}}\!\! \textrm{Re}[ S_{\ell r ,\bar{\ell} r }^{+}(\omega) G_{\ell r ,\ell r ;\bar{\ell}r ,\bar{\ell}r }^{+} (\omega,t)].
\end{align*}
Generating a frequency comb  in all convolutions is not a problem, since $\mathcal{C}_{\ell\bar{\ell},\ell\bar{\ell}}(t)$ only involves $G^+_{a,a';b,b'}(\omega,t)$ FFs. Because of the homogeneous control on all $r\in R_{\ell,\bar{\ell}}$, however, the FF $G_{\ell r ,\ell r ;\bar{\ell}r ,\bar{\ell}r }^{+} (\omega,t)$ is the same for all $r$, implying that the power spectra $S_{\ell r ,\bar{\ell}r }^{+}(\omega) $ are not distinguishable. We can break this symmetry using the $\mathcal{C}_{\ell\bar{\ell},\ell\bar{\ell}}(t) $ for all other pairs $\ell$ and $\bar{\ell}$.  For each $\ell$ and $\bar{\ell}$, we measure $\mathcal{C}_{\ell\bar{\ell},\ell\bar{\ell}}(t) $ ensuring that \emph{unique} sequences are applied to each of the three parties $\ell$, $\bar{\ell}$ and $r \in R_{\ell,\bar{\ell}}$, i.e., none of the base sequences applied to the three parties should be the same. Recall that the remaining $N-2$ qubits in $R_{\ell,\bar{\ell}}$ are subjected to homogeneous control at each iteration. As long as these conditions are met, the $\mathcal{C}_{\ell\bar{\ell},\ell\bar{\ell}}(MT)\,\forall\,\ell,\bar{\ell}\in\{1,...,N\}$ form a system of non-degenerate linear equations under repetition, which can be inverted to  obtain all classical spectra $S^+_{\ell,\bar{\ell}}(\omega)$ and $S^+_{\ell r,\bar{\ell}r}(\omega)$ for $r\in R_{\ell,\bar{\ell}}$. As in the two qubit case, the real and imaginary character of the FFs can be controlled using timing symmetries, enabling us to extract both the real and imaginary components. Using a similar procedure for the expansion coefficients $\mathcal{C}_{\ell \bar{\ell},0}(t)$, $\mathcal{C}_{ \bar{\ell}\ell,0}(t)$, $\mathcal{C}_{\ell,0}(t)$ and $\mathcal{C}_{\bar{\ell},0}(t)$ enables us to access the spectra $S^+_{\ell,\ell}(\omega)$, $S^+_{\bar{\ell},\bar{\ell}}(\omega)$, $S^+_{\ell\bar{\ell},\ell\bar{\ell}}(\omega)$, $S^+_{\ell r,\ell r}(\omega)$ and $S^+_{\bar{\ell} r, \bar{\ell} r}(\omega)$ for $r\in R_{\ell,\bar{\ell}}$. The remaining  expansion coefficients $\mathcal{C}_{\bar{\ell},{\ell}}(t)$, $\mathcal{C}_{\ell,\bar{\ell}}(t)$, $\mathcal{C}_{\bar{\ell},\ell \bar{\ell}}(t)$ and $\mathcal{C}_{\ell,\ell \bar{\ell}}(t)$ pose no additional complications in the $N$-qubit case, since spectra involving $r\in R_{\ell,\bar{\ell}}$ do not appear in the convolutions. From these expansion coefficients, therefore, we can obtain $S^-_{\ell,\bar{\ell}}(\omega)$, $S^-_{\bar{\ell},\ell}(\omega)$, $S^+_{\ell,\ell\bar{\ell}}(\omega)$ and $S^+_{\bar{\ell},\ell\bar{\ell}}(\omega)$ for all pairs $\ell$,\,$\bar{\ell}$ using the two-qubit protocol. Diagonal control, thus, gives us access to all spectra relevant to $N$-qubits except for the quantum self-spectra.

Like the two qubit case, accessing the quantum self-spectra for M1 requires non-local control. From the expansion coefficients $\mathcal{C}_{\ell,\ell\bar{\ell}}(t)$ and $\mathcal{C}_{\bar{\ell},\ell\bar{\ell}}(t)$ and the previously reconstructed quantum cross-spectra $S_{\ell,\bar{\ell}}^-(\omega)$ and $S_{\bar{\ell},\ell}^-(\omega)$, we can determine the $N$-qubit analogues to Eqs. (\ref{eq::delp}) and (\ref{eq::delm}),
\begin{align*}
\Delta^{+}(MT) \approx&\frac{2 i}{T} \sum_{\ell'=1}^N \sum_{k\in\mathcal{K}} S^{-}_{\ell',\ell'} (k\omega_0) \textrm{Im} [G^{-}_{\ell,\ell'; \bar{\ell},\ell'} (k\omega_0,T)],\\
\Delta^{-}(MT) \approx& \frac{2 iM}{T}\!\! \sum_{\ell'=1}^N \sum_{k\in\mathcal{K}}  S^{-}_{\ell',\ell'} (k\omega_0) \textrm{Im} [G^{+}_{\ell,\ell'; \bar{\ell},\ell'} (k\omega_0,T)].
\end{align*}
Homogeneous control over all $r\in R_{\ell,\bar{\ell}}$, implies that all $S^{-}_{r,r}(\omega)$ will be filtered by the same function and, hence, indistinguishable. However, if the non-locality of the control is restricted to just the qubits $\ell, \bar{\ell}$, i.e., the only swap gate used in the base sequences is SWAP$_{\ell,\bar{\ell}}$, then $G^{+}_{\ell,\ell'; \bar{\ell},\ell'}(\omega,T) =0$ for $\ell' \neq \ell,\bar{\ell}$ and the problematic contributions disappear. One can then reconstruct all $S_{\ell,\ell}^-(\omega)$ using the two-qubit protocol. With the addition of the quantum self-spectra, we have shown how to reconstruct all spectra for the $N$-qubit M1 model. 

The M2 model is similar to the M1 model, but with the addition of the terms $B_0(t)= \sum_{\ell=1}^N B_\ell(t)$. If we apply the same strategy using diagonal control to M2 as we did for M1, we can obtain all spectra except for the quantum self-spectra. Similar to the two-qubit case, the expansion coefficients $\mathcal{C}_{\ell,\ell}(t)$ enable us to determine the quantum self spectra using purely local control. Using $\mathcal{C}_{\ell,\ell}(t)$ and interpolations of the previously reconstructed quantum cross-spectra, we obtain
\begin{align*}
& \mathcal{C}_{\ell,\ell}(t)\! -\!\! \!\sum_{\ell'\neq \ell} \!\frac{i}{\pi}\!\!\! \int  \limits_{-\infty}^\infty \!\!\!\! d \omega\textrm{Im}[S_{\ell,\ell'}^{-,I} (\omega) (G^{+}_{\ell,\ell;0,0} (\omega,t)\! +\! G^{-}_{\ell,\ell;0,0} (\omega,t)) ] \\
&\;\;\;\;\;\approx\!\frac{i}{\pi}\!\! \int \limits_{-\infty}^\infty \! \!d \omega \textrm{Re}[S_{\ell,\ell}^{-} (\omega)] \textrm{Im}[(G^{+}_{\ell,\ell;0,0} (\omega,t) + G^{-}_{\ell,\ell;0,0} (\omega,t))].
\end{align*}
As in the two-qubit case, we can apply repetitions of  a displacement antisymmetric base sequence with imaginary $G^{-}_{\ell,\ell;0,0} (\omega,T)$ and real $G^{+}_{\ell,\ell;0,0} (\omega,T)$ in order to deconvolve this expression and solve for $S_{\ell,\ell}^{-} (\omega)$. By repeating this procedure for every $\ell$, we obtain all quantum self-spectra.  

\section{Case study: Quantum noise spectroscopy on two exciton qubits}
\label{sec:Exciton}

We demonstrate the use and power of the proposed QNS protocols by focusing on 
the reconstruction of self- and cross-spectra of two exciton qubits in self-assembled quantum dots, 
coupled to a common phonon bath. Physically, 
the interaction between the excitons and the vibrational modes of the host crystal lattice 
is known to be the dominant source of decoherence for typical operating regimes 
\cite{Hodgson,Cotlet:2014,Krzywda:2016}. 

The relevant open-system interaction-picture Hamiltonian is given in Eq.~(\ref{eq::ExBt}), with $N=2$ and 
vanishing inter-qubit coupling, $B_{12}(t)\equiv 0$.
The complex coupling constants may now be taken to be of the form $g_k^\ell=|g_k|e^{i\vec{k}\cdot\vec{r}_\ell}$, 
where $\vec{k}$ is the wave-vector of phonon mode $k$ and $\vec{r}_\ell$ is the position of qubit $\ell$. Assuming linear dispersion, 
the wave-vector satisfies $\vec{k}\cdot (\vec{r}_\ell-\vec{r}_\ell')=\omega\, t_{\ell,\ell'}$, where $t_{\ell,\ell'}$ is referred to as the 
``transit time'' \cite{multiDD}. If $v_s$ is the speed of sound in the bath, $|t_{\ell,\ell'}|=|\vec{r}_\ell-\vec{r}_{\ell'}|/v_s$. 
In order to make contact with experimentally accessible control resources, we shall assume access only to {\em local (diagonal) control}, 
in which case the applied $H_\text{ctrl}^{\ell}(t)$ generates sequences consisting of (nearly-instantaneous) $\pi$-pulses about 
an axis orthogonal to $z$. For exciton qubits, such control sequences can be implemented with femtosecond optical pulses.
In the toggling frame, the Hamiltonian has the form given in Eq.~(\ref{eq::dephasingH}) with $c=1$, i.e., 
\begin{align*} 
\tilde{H}(t)=\hbar \sum_{\ell=1,2} \,(y_{\ell,\ell}(t)Z_\ell+I_\ell) \otimes B_\ell(t), 
\end{align*}
where $B_\ell(t)$ is given in Eq.~(\ref{eq::SBBt})  and the the switching functions toggle between $\pm 1$ with each applied  
$\pi$-pulse. For a phonon bath initially in a thermal state, the operators $B_\ell(t)$ exhibit Gaussian statistics and the relevant spectra 
are obtained from Eqs. (\ref{eq::SBllSpectra})-(\ref{eq::sd}) by letting $J_{\ell,\ell'}(\omega) \equiv  e^{-i \omega t_{\ell, \ell'}} J(\omega)$, 
where  $J(\omega)=\sum_k|g_k|^2 [\delta(\omega-\Omega_k)+\delta(\omega+\Omega_k)] = J(-\omega)$ is the bath spectral density.  Thus, 
the spectra to be reconstructed are
\begin{align}
\label{eq::ExcitonSpectra} 
S_{\ell, \ell'}(\omega)\!=\!\pi e^{-i {\omega |\vec{r}_\ell -\vec{r}_{\ell'} | }{ /v_s}} J(\omega)\! \left\{\! \begin{array}{ll}
         \text{coth}\big(\frac{\beta\omega}{2}\big)\!+\!1, &\omega \geq 0\\
        \text{coth}\big(\!\!-\!\frac{\beta\omega}{2}\big)\!-\!1, &\omega< 0 \end{array} \right..
\end{align}
Explicitly, dephasing dynamics of the exciton qubits described by the following set of real spectra:
\begin{align}
\label{eq::excitonS}
\mathcal{S}=&\{ S_{1,1}^{\pm}(\omega),S_{2,2}^{\pm}(\omega),\text{Re}[S_{1,2}^\pm(\omega)],\text{Im}[S_{1,2}^\pm(\omega)] \}.
\end{align}

\subsection{Spectral reconstruction procedure}
\label{sec::GenProcedure}

Reconstructing the desired spectra requires that we obtain the expansion coefficients $\mathcal{C}_{a,b}(t)$, which depend on the convolutions between the spectra and FFs given in Eqs. (\ref{ddc2})-(\ref{ddc6}) in the case of diagonal control. Recall that the expansion coefficients can also be related to expectation values of observables on the two qubits, as given in Eqs. (\ref{eq::c1})-(\ref{eq::c6}). In an experimental implementation, average measured values of the observables replace the expectation values. With this in mind, 
the general procedure for reconstructing a particular spectrum $s(\omega)\in\mathcal{S}$ entails the following steps: 

\vspace*{2mm}

\noindent {{{\bf Step 1:}} Identify an expansion coefficient, $\mathcal{C}_{a,b}(t)$,
that depends on $s(\omega)$.

\vspace*{2mm}

\noindent {{{\bf Step 2:}} Initialize the two qubits in states that allows one to access $\mathcal{C}_{a,b}(t)$. 

\vspace*{2mm}

\noindent {{{\bf Step 3:}}
Apply repetitions of a control sequence that has the appropriate symmetries to create a frequency comb for a FF 
entering the integral for $\mathcal{C}_{a,b}(t)$. Symmetries can also be utilized to control whether the FF is real or imaginary, 
allowing for the real or imaginary components of $s(\omega)$ to be isolated.

\vspace*{2mm}

\noindent {{{\bf Step 4:}}
Determine $\mathcal{C}_{a,b}(t)$ by measuring observables on one or both qubits.

\vspace*{2mm}

\noindent {{{\bf Step 5:}}
In order to sample $s(\omega)$ at a set of harmonic frequencies $\{ \omega_k \equiv k\omega_0|k\in\mathcal{K}\}$, steps (2)-(5) must be repeated for $N_c \geq |\mathcal{K}|$ control sequences with the desired symmetries. Each set of sequences can have identical cycle times, say $T$, or different cycle times of the form $\{T,T/2,\ldots,T/N_c\}$. 

\vspace*{2mm}

\noindent {{{\bf Step 6:}}
Determining the $\mathcal{C}_{a,b}(t)$ for each control sequence produces a set of linear equations relating $\mathcal{C}_{a,b}(t)$ to the spectrum $s(\omega)$ evaluated at $\{ \omega_k \}$.  In order to isolate the contribution of $s(\omega)$, it may be necessary to take linear combinations of the $N_c$ measured $\mathcal{C}_{a,b}(t)$. The resulting system of linear equations is inverted to obtain 
the reconstructed (R) spectrum $s^{R}(\omega_k)$, an estimate of $\{s(\omega_k) | k\in\mathcal{K}\}$.

\subsection{Spectral reconstruction results}
\label{sec::results}

\begin{figure*}[t]
\includegraphics[width=\textwidth,scale=.8]{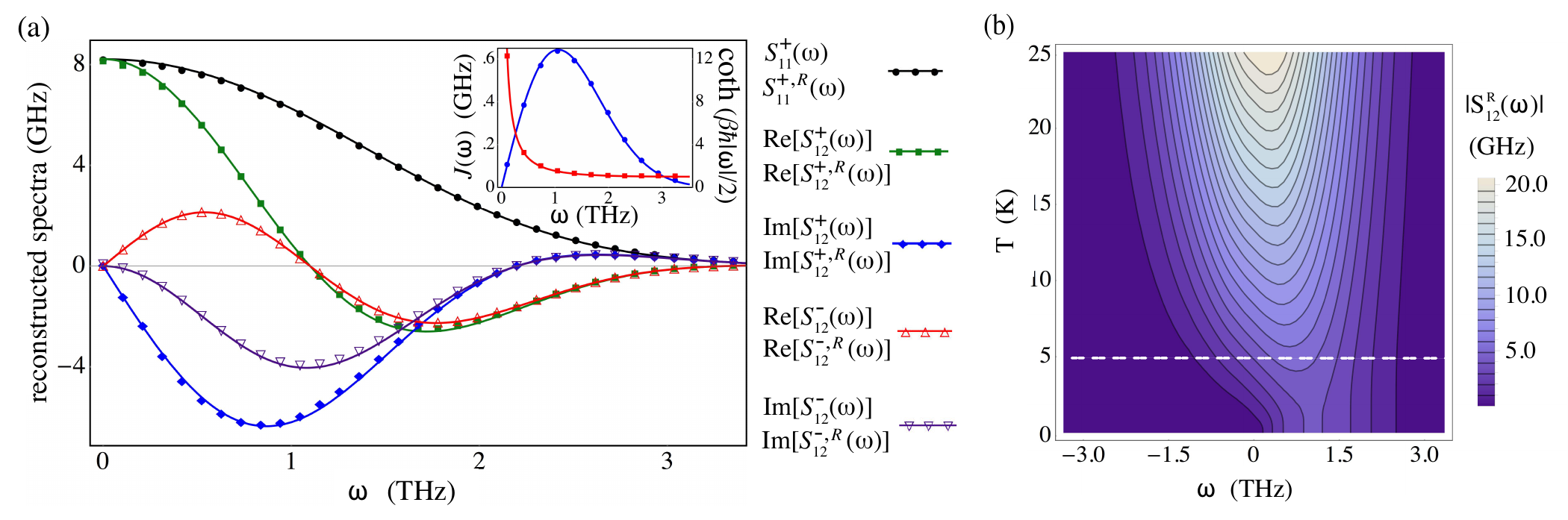}
\vspace*{-5mm}
\caption{(Color online) Spectral reconstructions and estimates of bath parameters for two exciton qubits coupled to a common 1D phonon bath. (a) Spectral reconstructions and actual spectra for an initially thermal bath at $T_B=5$ K. The reconstructed spectra  $S_{11}^{+,R}(\omega)$ (black dots), $\text{Re}[S_{12}^{+,R}(\omega)]$ (green squares), $\text{Im}[S_{12}^{+,R}(\omega)]$ (blue diamonds), $\text{Re}[S_{12}^{-,R}(\omega)]$ (red up arrows) and $\text{Im}[S_{12}^{-,R}(\omega)]$ (purple down arrows) are plotted alongside the actual spectra (solid lines). (Inset) Temperature dependence obtained by spectral reconstruction via $S_{12}^{+,R}(\omega)/S_{12}^{-,R}(\omega)$ (red dots) versus the actual temperature dependence, $\text{coth}(\beta|\omega|\hbar/2)$, for $T_B=5$ K (red line). The bath temperature estimated via Eq.~
(\ref{eq::temp}) is 5.02 K. Spectral density $J(\omega)$ obtained from the spectral reconstructions, Eq.~(\ref{eq::j}) (blue dots) and actual spectral density (blue line). (b) 
Contours of the magnitude of the reconstructed cross-correlation spectrum $|S_{12}^R(\omega)| 
= |S_{12}^{+,R}(\omega) + S_{12}^{-,R}(\omega)|/2$ 
vs. angular frequency and temperature (white dash line corresponding to the estimated bath temperature).
Note how the spectral asymmetry about $\omega=0$, a signature of the quantum bath, becomes more pronounced as the temperature decreases. }
\label{fig::RecContour}
\end{figure*}

In our simulations, we used parameters relevant to exciton quantum dots interacting with phonons in one-dimensional (1D) 
geometries, such as nano-wires or carbon nanotubes \cite{Sasakura:2012,Hofmann:2013,Cotlet:2014}. 
As is characteristic of 1D geometries, the spectral density of the bath was Ohmic, 
\begin{align*}
J(\omega)=\xi|\omega| e^{-{\omega^2}/{\omega_c}^2}, 
\end{align*}
with a dimensionless coupling parameter $\xi=0.001$ and a Gaussian rolloff at high frequencies, with a cutoff $\omega_c=1.5$ THz.  The separation distance between the excitons and the speed of sound in the bath were taken to be $|\vec{r}_1-\vec{r}_2|=10$ nm and 
$v_s=7$ km/s, respectively, corresponding to a transit time $t_{1,2}\approx 1.4$ ps. Unless otherwise stated, the temperature of the bath was $T_B=5$K. For the reconstructions, we used a range of control sequences with maximum cycle time $T=60$ ps and minimum switching time $\tau_0=0.2$ ps. From the FFs of these control sequences and the spectra in Eq.~(\ref{eq::ExcitonSpectra}), we numerically computed expectation values of qubit observables, from which we determined the expansion coefficients. Because, by assumption, we are limited to diagonal control, we could not directly reconstruct the quantum self-spectra.  However, we can still {\em indirectly} infer the quantum self-spectra from reconstructions of $S_{1,1}^{+}(\omega)$, $S_{2,2}^{+}(\omega)$, $\text{Re}[S_{1,2}^\pm(\omega)]$ and $\text{Im}[S_{1,2}^\pm(\omega)]$, as described below. A detailed description of the reconstruction procedure, along with the control sequences used, is included in Appendix \ref{sec::procedureExciton}.

Numerical reconstructions of the spectra $S_{1,1}^{+}(\omega)$, $S_{2,2}^{+}(\omega)$, $\text{Re}[S_{1,2}^\pm(\omega)]$ and $\text{Im}[S_{1,2}^\pm(\omega)]$ at $|{\cal K}|=33$ harmonic frequencies for an initially thermal bath at $T_B=5$K are plotted in Fig. \ref{fig::RecContour}(a), demonstrating excellent agreement with the actual spectra. 
With prior knowledge that the bath is bosonic and assuming equilibrium conditions, we can estimate the temperature by means of 
Eqs. (\ref{eq::cqspectra}) and (\ref{eq::cqspectra2}): that is, 
\begin{equation}
\label{eq::temp}
\frac{S_{12}^{+,R}(\omega)}{S_{12}^{-,R}(\omega)}\approx\text{coth}(\beta\omega/2), \quad  \omega>0.
\end{equation}
The estimated temperature, $T_B \approx 5.02$ K, again indicates excellent agreement with the actual temperature. The spectral density 
of the bath can also be inferred by using 
\begin{equation}
\label{eq::j}
J(\omega)\approx \frac{S_{11}^{+,R}(\omega)}{ 2\pi\text{coth}(\beta\omega/2)}, \quad \omega>0, 
\end{equation}
where we replaced $\text{coth}(\beta\omega/2)$ with its estimate. The inset of Fig. \ref{fig::RecContour}(a) shows the actual and estimated values of $J(\omega)$ and $\text{coth}(\beta\omega/2)$.   Figure \ref{fig::RecContour}(b) depicts reconstructed contours of $|S_{12}(\omega)|$ obtained from $\text{Re}[S_{12}^{\pm,R}(\omega)]$ and $\text{Im}[S_{12}^{\pm,R}(\omega)]$, 
for a range of temperatures and harmonic frequencies. The {\em spectral asymmetry} of positive vs. negative frequencies, 
which becomes more pronounced as the temperature decreases, is clearly evidenced in the contours. 
With prior knowledge of the bosonic nature of the bath, we can also obtain the quantum self-spectra $S_{\ell,\ell}^{-}(\omega)$ from reconstructions of $S_{1,1}^{+}(\omega)$, $S_{2,2}^{+}(\omega)$, $\text{Re}[S_{1,2}^\pm(\omega)]$ and $\text{Im}[S_{1,2}^\pm(\omega)]$. From \erf{eq::cqspectra}, 
\begin{align}
S^{-}_{1,1} (\omega) =S^{-}_{2,2} (\omega) &=  2\pi J(\omega) \textrm{sign} (\omega). 
\end{align}

\begin{figure}[h]
\includegraphics[scale=.9]{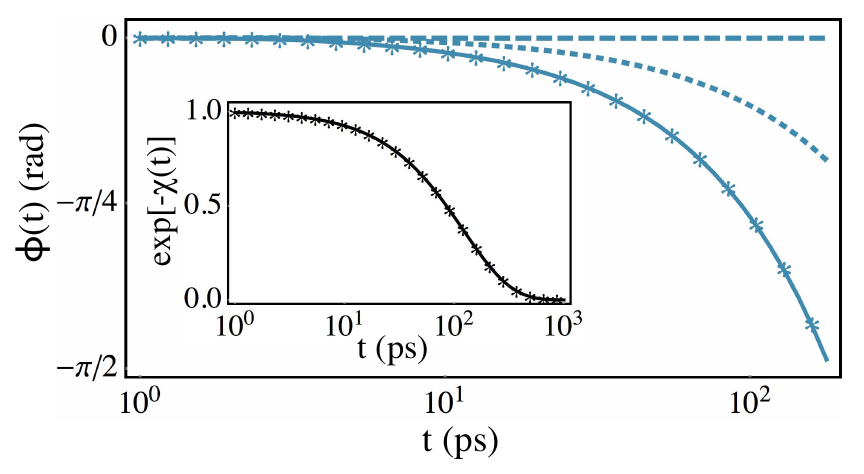}
\vspace*{-2mm}
\caption{(Color online) Qubit dynamics under free evolution for the initial product state $|\psi_1^-\rangle= |+\rangle_1 \otimes |\!-\!z\rangle_2$. 
(a) Phase evolution $\phi(t)=\mathcal{C}_{1,12}(t)-\mathcal{C}_{1,1}(t)$ of qubit 1. Because $\phi(t)$ depends exclusively on the quantum spectra $S_{12}^{-}(\omega)$ and $S_{11}^{-}(\omega)$, it is a signature of the quantum bath. The actual phase evolution under free evolution (solid line) is plotted along with the phase evolution predicted using: (1)  all reconstructed spectra, ${\cal S}$ (asterisks); (2) all reconstructed spectra except the quantum self-spectra (dots), ${\cal S}_r$; and (3) only the classical reconstructed spectra, ${\cal S}_c$ (large dashes). (Inset) Dephasing of qubit 1. Unlike phase evolution, the decay of coherences depends exclusively on the classical spectra. Actual dephasing (solid line) and dephasing predicted by spectral reconstructions (asterisks) show excellent agreement.}
\label{fig::PhaseDephase}
\end{figure}

\begin{figure}[t]
\includegraphics[scale=.85]{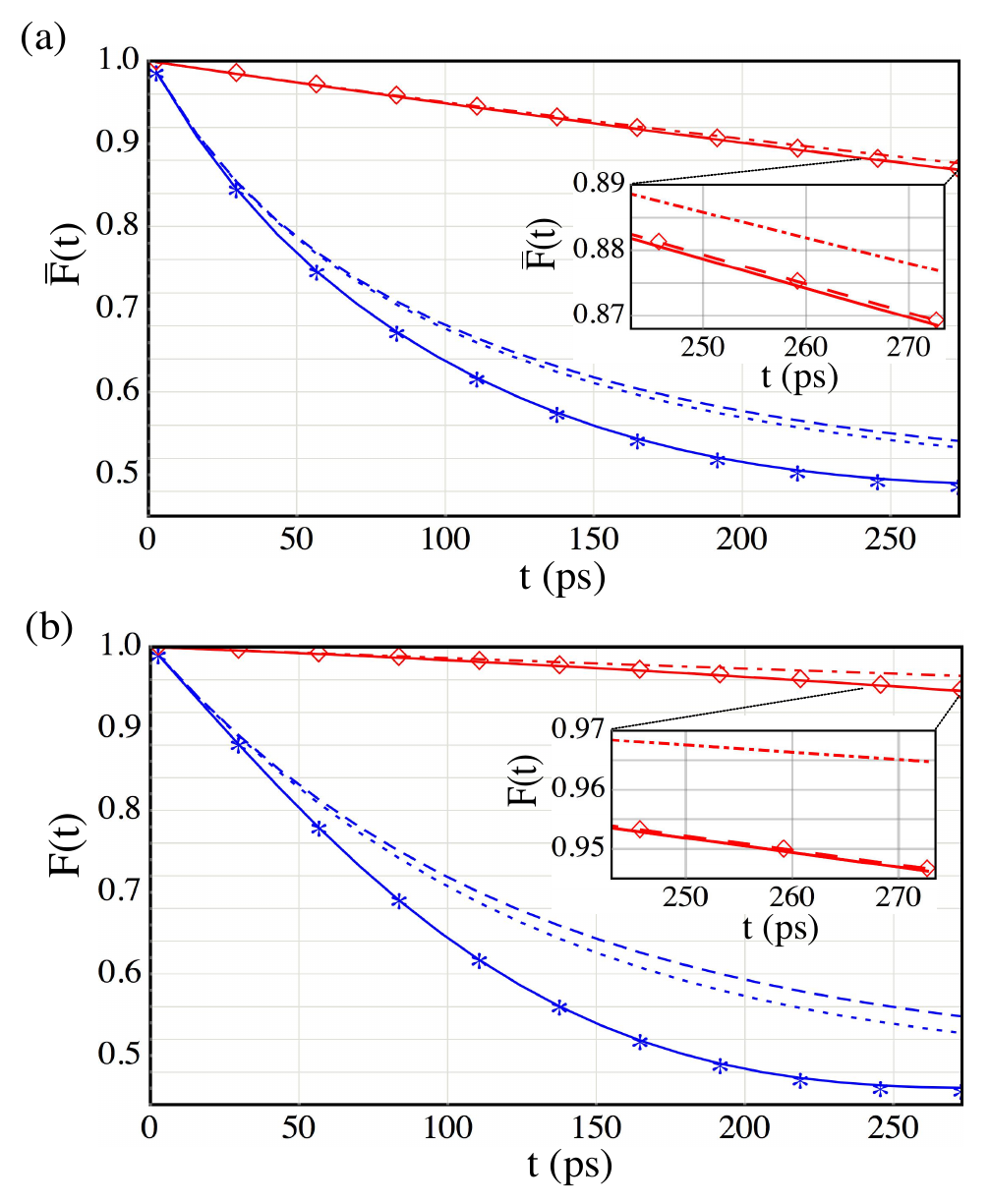}
\vspace*{-5mm}
\caption{(Color online) Two-qubit fidelity decay under free and controlled evolution. 
Controlled evolution consists of 
repetitions of the product-mirror antisymmetric concatenated sequence CDD$_3\times$CDD$_2$, with cycle time $T=2.7$ ps. (a) Estimated average fidelity, $\bar{F}(t)$, obtained by averaging  1000 Haar-random initial pure states of the two qubits. Under free evolution, the actual $\bar{F}(t)$ (blue solid line) and the $\bar{F}(t)$ predicted using all spectral reconstructions in $\mathcal{S}$ (blue asterisks) demonstrate excellent agreement. Predictions produced with $\mathcal{S}_c$ (small blue dashes) and $\mathcal{S}_r$ (blue dots) significantly {\em overestimate} the fidelity. The classical prediction made with $\mathcal{S}_c$ and the actual dynamics differ by 
over $6\%$.  
For the controlled evolution, the actual $\bar{F}(t)$ (red solid line), the predicted $\bar{F}(t)$ using $\mathcal{S}$ (red diamonds), the $\bar{F}(t)$ predicted using $\mathcal{S}_r$ (large red dashes) and the $\bar{F}(t)$ predicted using $\mathcal{S}_c$ (red dot-dashed line) are in closer agreement, as the controlled evolution partially suppresses the contributions of the quantum spectra. However, the classical prediction and the actual $\bar{F}(t)$ still differ by 
$1\%$ (Inset).  (b) Actual and predicted fidelity for the subset of initial states, out of the total sampled 1000 random states, which demonstrate the greatest discrepancies between the classical predictions using $\mathcal{S}_c$ and the actual dynamics. For free evolution, the actual fidelity and the classical prediction differ by as much as 
$11\%$. For the controlled evolution, the maximum difference is 
about $2\%$ (Inset). }
\label{fig::fidelities}
\end{figure}

This full set of spectra in ${\cal S}$ can be used to predict the evolution of the qubits -- specifically, to target particular dynamics stemming from the quantum or classical nature of the bath. An example of this is demonstrated in Fig. \ref{fig::PhaseDephase}, where the two exciton qubits are initially prepared in state $|\psi_1^-\rangle= |+\rangle_1 \otimes |-z\rangle_2$ and undergo free evolution. The coherence element,
 \begin{align*}
\bra{0}\text{Tr}_2 [\rho(t)]\ket{1}&=\frac{1}{2}\left[E_{\psi_1^-}(X_1)+E_{\psi_1^-}(Y_1)\right]\\
&=e^{-\mathcal{C}_{1,0}(t)+i[\mathcal{C}_{1,12}(t)-\mathcal{C}_{1,1}(t)]},
\end{align*}
indicates that qubit 1 undergoes {\em both} decay and phase evolution, similar to the single-qubit dephasing in Sec.~\ref{sec::singleQdephase}. The decay of the coherence is given by the expansion coefficient $\mathcal{C}_{1,0}(t)$, which depends entirely on the classical self-spectrum $S^{+}_{1,1}(\omega)$, as seen in \erf{ddc4a}. The expansion coefficients  $\mathcal{C}_{1,12}(t)$ and $\mathcal{C}_{1,1}(t)$, which determine the phase evolution, depend instead on the quantum spectra $S^{-}_{1,1}(\omega)$ and  $S^{-}_{1,2}(\omega)$, as seen in Eqs. (\ref{ddc2}) and  (\ref{ddc3}). 
The phase evolution is a signature of quantum noise. 
Specifically, Fig. \ref{fig::PhaseDephase} depicts the actual phase evolution and the phase evolution predicted using three different sets of 
spectra: 

(1) {\em all} reconstructed spectra, ${\cal S}$, in Eq.~(\ref{eq::excitonS}); 

(2) all reconstructed spectra {\em except} the quantum self-spectra, 
${\cal S}_r =\{ S_{1,1}^{+}(\omega),S_{2,2}^{+}(\omega), \text{Re}[S_{1,2}^\pm(\omega)],\text{Im}[S_{1,2}^\pm(\omega)] \}$; 

(3) only the {\em classical} reconstructed spectra, 
${\cal S}_c =\{ S_{1,1}^{+}(\omega),S_{1,1}^{+}(\omega), \text{Re}[S_{1,2}^+(\omega)],\text{Im}[S_{1,2}^+(\omega)] \}$. 

\vspace*{1mm}

\noindent 
The phase evolution predicted using all of the reconstructed spectra in $\mathcal{S}$ shows excellent agreement with the actual dynamics. The 
prediction based on ${\cal S}_r$, which ignores the quantum self-spectra, deviates markedly from the actual phase evolution and, unsurprisingly, the prediction made from $\mathcal{S}_c$ fails entirely to capture the phase evolution. This shows how accurately modeling the system's reduced dynamics clearly requires knowledge of the quantum spectra. 

To get an idea of how the quantum bath influences the exciton qubits under more general circumstances, Fig. \ref{fig::fidelities} tracks the average and worst-case fidelity of the qubits versus time for both free evolution and a representative controlled evolution. Specifically, the controlled evolution consists of stroboscopic repetitions of a product-mirror antisymmetric concatenated DD sequence, CDD$_3\times$CDD$_2$ \cite{multiDD},  with cycle time $T=2.7$ ps. As a measure of the fidelity, we use $F(t)=\text{Tr}(\rho(t)\rho_0)$, quantifying the extent to which the qubits have decohered at time $t$ from their initial state. Both the average and worst-case fidelities were determined from 1000 Haar-random initial pure states. Similar to Fig. \ref{fig::PhaseDephase}, the actual fidelities are plotted along with predicted fidelities based on $\mathcal{S}$, $\mathcal{S}_r$ and $\mathcal{S}_c$. Figure \ref{fig::fidelities}(a) shows excellent agreement between the actual average fidelity and the average fidelity predicted using $\mathcal{S}$ for both free and controlled evolution. For free evolution, the average fidelity predicted with $\mathcal{S}_r$ and $\mathcal{S}_c$, in which some or all quantum spectra are ignored, deviates from the actual by as much as 6$\%$. The applied DD sequence suppresses the contributions of both the classical and the quantum spectra. Consequently, the deviation between the actual and predicted average fidelities for  $\mathcal{S}_r$ and $\mathcal{S}_c$ is considerably less, at about $1\%$. 

Figure \ref{fig::fidelities}(b) shows the worst-case fidelity obtained from the sample of 1000 Haar-random initial pure states. Again, the predictions made using $\mathcal{S}$ are in excellent agreement with the actual worst-case fidelities.  Under free evolution, the maximum deviation between the actual worst-case fidelity and those predicted using $\mathcal{S}_r$ and $\mathcal{S}_c$  is considerable, about $11\%$. Under controlled evolution, the maximum deviation is less, at about $2\%$. These plots show that the relative contributions of the quantum and classical spectra can change considerably depending on external control. Ultimately, however, accurate quantitative modeling of both controlled and free dynamics requires properly accounting for the quantum spectra.

\section{Conclusion} 

We have presented a multiqubit quantum noise spectroscopy protocol that utilizes the response in the dynamics of set of qubits to different control symmetries in order to extract information about the bath affecting them. We argue that the ability to characterize the effect of a quantum or classical bath inducing noise on the qubits is not only useful, as it opens the way to exploit the bath as a characterized resource, but is also necessary toward the full deployment of quantum technologies. We further discuss how quantum vs. classical baths lead to distinctive spectral features and dynamical signatures in the qubits being used as probes. The proposed protocols are implemented in a realistic two-exciton system subject to phonon-induced dephasing, and manipulated using only experimentally accessible sequences of 
single-qubit $\pi$ pulses.  Complete reconstruction of all the relevant classical and quantum self- and cross-spectra is demonstrated, allowing in particular for quantum thermometry and quantitative prediction of free and controlled qubit dynamics. Our findings emphasize the central role that the quantum spectra play in influencing the dynamics of the qubits, their accurate characterization being necessary for meeting the requirements of high-fidelity quantum control.

The approach behind the multi-qubit protocol described here can in principle be extended to non-Gaussian (classical or bosonic) dephasing noise in a relatively straightforward fashion, by combining our present results with the ones we presented in Ref.~\cite{Norris2016}. While in most physical qubit implementations, environment-induced dephasing is indeed a dominant source of error, ultimately one would like to design a complete noise spectroscopy protocol that permits the characterization of a general decoherence process for an arbitrary bath -- without invoking Gaussianity or weak coupling limits, and allowing for both environmental and control noise. This would provide a full characterization of a target open quantum system of interest but is a much more complex problem that we leave for future work. We expect that suitably incorporating more general symmetries will remain a key ingredient for control design.

\vspace*{2mm}

{\em Note added.--} 
While this manuscript was being finalized for submission, the experimental observation of quantum noise  in a superconducting flux qubit was reported by C. M. Quintana {\em et al.} in arXiv:1608.08752v1.  In this work, the symmetric and antisymmetric components of the flux noise spectrum were reconstructed by measurement of the relaxation rate and steady-state population of the qubit while tuning the strength of the Josephson term using a method similar to the one we described in Sec.~\ref{sec::singleQdephase} and Appendix \ref{sec::spinlocking}.  Reconstructions of the symmetric and antisymmetric components of the spectrum were used to perform thermometry on the flux noise, 
similar in spirit to our proposed approach in Sec.~\ref{sec::results}. At frequencies below $f\simeq$ 1GHz, where the symmetric component is dominant, the antisymmetric component displayed $1/T_B$ scaling, providing evidence of a paramagnetic environment.

\vspace*{-2mm}

\section*{Acknowledgements}

It is a pleasure to thank Fei Yan and Katarzyna Roszak for valuable input.
Work at Dartmouth was supported from the US ARO under contract No. W911NF-14-1-0682 
and the Constance and Walter Burke Special Projects Fund in Quantum Information Science.
GAPS acknowledges support from the ARC Centre of Excellence grant No. CE110001027 
and the Griffith University Postdoctoral Fellowship program.

\onecolumngrid

\begin{appendix}

\section{Exact solution for time-evolved observables under Gaussian dephasing noise}
\label{App_Gauss_Exact}

We provide here a more general version of the theorem stated in the main text, applicable to an 
arbitrary open quantum system $S$ undergoing controlled dephasing dynamics.  
Specifically, in place of the $N$-qubit toggling-frame Hamiltonian given in Eq.~(\ref{eq::togglingH}), 
our starting point is a time-dependent Hamiltonian of the form
\beq
\label{eq:A0}
\tilde{H}(t) = \sum_{a, a'} y_{a,a'}(t) P_a \otimes B_{a,a'}(t), 
\eneq
in a suitable frame where both the free bath Hamiltonian and the applied control Hamiltonian are 
explicitly removed.  Here, 
$\{ P_a \}$ is a set of Hermitian, mutually commuting operators on $S$, that is, $[P_a, P_{a'}]= 0$, 
$P_0 \equiv {\mathbf 1}$, $B_a(t)$ are bath operators, and $y_{a,a'}(t)$ arbitrary real 
functions, determined by the external control. 
We are interested in evaluating the time-dependent expectation value $E_{\rho_0}(O(t))$ of an operator 
${O}$ starting from an initial state $\rho_0$ on $S$, under the assumption that $O$ is invertible and preserves 
the dephasing character 
of $\{ P_a\}$.  That is, we now require that $O^{-1} P_a O = \sum_{b} V_{a b} P_b$, for all $a$ and suitable 
(generally complex) coefficients $V_{ab}$.  As in the main text, for fixed $t>0$ we define an effective ( generally non-Hermitian) Hamiltonian given by
\beq
\label{eq::effH}
\tilde{H}_O(s) \equiv \begin{cases}  -O^{-1}  \tilde{H}(t- s) O  & \textrm{   for  }  0 < s \leq t  \\ 
\,\,\,\,\,\,\,\,\,\,\tilde{H}(t+s) &\textrm{   for  }  -t \leq s < 0 \end{cases}.
\eneq 

\vspace*{1mm}

{\bf Theorem.} 
{\em The time-dependent expectation value of a dephasing-preserving invertible operator $O$ on an arbitrary 
open quantum system under controlled Gaussian dephasing dynamics is given by}
\beq
\label{Gauss-sol2}
 E_{\rho_0} (O(t)) = \Tr \Big[ e^{-i \mathcal{C}^{(1)}_O(t) - \frac{\mathcal{C}^{(2)}_O(t)}{2!}} \rho_{0} O \Big],
\eneq
{\em where the time-dependent cumulants have formally the same expressions given in 
Eqs. (\ref{eq::CO1})-(\ref{eq::CO2}) in the main text, namely:}
\begin{align}
\mathcal{C}^{(1)}_O(t)  &= \int_{-t}^{t} ds_1 C^{(1)}(\tilde{H}_O(s_1)) = \int_{-t}^t ds \langle 
\tilde{H}_O(s)\rangle_{c,q},  
\notag \\
\mathcal{C}^{(2)}_O(t)  
&=  2 \int_{-t}^{t} ds_1 \int_{-t}^{s_1} ds_2 C^{(2)}(\tilde{H}_O(s_1)\tilde{H}_O(s_2)) \notag \\
& = 2 \int_{-t}^t ds_1 \int_{-t}^{s_1} ds_2 \langle \tilde{H}_O(s_1)\tilde{H}_O(s_2)\rangle_{c,q}
-  \int_{-t}^t ds_1 \langle \tilde{H}_O(s_1)\rangle_{c,q} \int_{-t}^t ds_2 \langle \tilde{H}_O(s_2)
\rangle_{c,q}\notag .
\end{align}

\begin{proof} 
The desired expectation value $E_{\rho_0} (O(t))$ is given by 
\begin{align}
\label{eq::average}
E_{\rho_0} ( O(t) ) &= \langle \Tr[  \rho_{SB}(t) O] \rangle_c
= \Tr_S [ \langle \mathcal{T}_+ e^{- i \int_{-t}^{t} \tilde{H}_O(s) ds}\rangle_{c,q} \, \rho_0 O], 
\end{align}
Under the dephasing-preserving assumption, 
we may write $\tilde{H}_O(s) = \sum_{a,a'} \tilde{y}_{a,a'}(s) P_a \otimes \hat{B}_{a'}(s)$, with effective control functions 
and bath operators given by
$$ \tilde{y}_{a,a'}(s) \equiv 
\begin{cases}  -  \sum_{a''} y_{a'',a'}(t- s)  V_{a a''} & \textrm{   for  }  0 < s \leq t  \\ \,\,\,\,\,\,\,\,\,\,y_{a,a'}(t+s) 
&\textrm{   for  }  -t \leq s < 0 \end{cases}, \quad 
\hat{B}_{a'}(s) \equiv 
\begin{cases}  {B}_{a'}(t- s)  & \textrm{   for  }  0 < s \leq t  \\ B_{a'}(t+s) 
&\textrm{   for  }  -t \leq s < 0 \end{cases}. $$
With reference to Kubo's generalized cumulant expansion approach~\cite{Kubo}, 
note that the time-ordered exponential entering in Eq.~(\ref{eq::average}) 
is a valid generalized exponential function and, similarly, the operation 
$$ \langle \cdot \rangle_{c,q} \equiv \langle \Tr_B [\cdot \rho_B]\rangle_{c}  $$
defines a valid normalized ``average''. Introducing the compact notation $\int\limits_{-t}^{t} d \vec{t}_{[k]} \equiv \int_{-t}^t dt_1 \int_{-t}^{t_1}dt_2 \cdots \int_{-t}^{t_{k-1}}dt_k$, it then follows [from Theorem V and Eq.~(6.4) therein, or direct calculation] that 
\begin{equation} 
\label{cumexpa}
\langle \mathcal{T}_+ e^{- i \int_{-t}^{t} \tilde{H}_O(s) ds}\rangle_{c,q} = e^{ \sum_{k=1}^\infty 
(-i)^k \int\limits_{-t}^{t} d \vec{t}_{[k]} C^{(k)} (\tilde{H}_O(t_{1}), \cdots, \tilde{H}_O(t_{k}))},
\end{equation}
where $C^{(k)} (\tilde{H}_O(t_{1}), \cdots, \tilde{H}_O(t_{k})) \equiv C^{(k)}(\{\tilde{H}_O(t_u)\})$ is a generalized cumulant. Crucially, since the $\{ P_a\}$ commute, then
\begin{align*}
{C}^{(k)}(\{\tilde{H}_O(t_u)\} ) = \sum_{\vec{a},\vec{a}'} \Big(\prod_{r=1}^k \tilde{y}_{a_r,a'_r}(t_r) P_{a_r}\Big) 
  C^{(k)} (\hat{B}_{a'_1}(t_1),\hat{B}_{a'_2}(t_2),\cdots,\hat{B}_{a'_k}(t_k)), 
\end{align*}
and one has that $[{C}^{(r_j)}(\{\tilde{H}_O(t_j)\}), {C}^{(r_{j'})}(\{\tilde{H}_O(t_{j'})\} )] =0$. Notice that ${C}^{(k)}(\{\tilde{H}_O(t_u)\} )$ depends explicitly on the $k$th-order cumulants of bath operators $\hat{B}_{a'}(s)$ and that the Gaussianity assumption, ${C}^{(k)}(\{{B}_{a'_j}(t_j)\} )=0$ for $k>2$, also implies that ${C}^{(k)}(\{\hat{B}_{a'_j}(t_j)\} )=0$ for $k>2$. Then, it follows that ${C}^{(k>2)}(\{\tilde{H}_O(t_u)\} ) =0$ and truncating Eq.~\eqref{cumexpa} to $k=2$ yields  
$$E_{\rho_0} ( O(t) ) = \Tr_S [ e^{ -i \int_{-t}^{t} ds_1 C^{(1)}(\tilde{H}_O(s_1)) - \int_{-t}^t ds_1 \int_{-t}^{s_1} ds_2 C^{(2)} (\tilde{H}_O(s_1),\tilde{H}_O(s_2))} \, \rho_0 O],$$
which, using the above definitions of ${\cal C}^{(k)}_O(t)$ for $k=1,2$, is equivalent to Eq.~(\ref{Gauss-sol2}). 
\end{proof}


\newpage

We stress that while the theorem applies only to dephasing-preserving invertible operators, any operator on $S$ can always be decomposed in terms of an orthogonal basis consisting of dephasing-preserving unitary operators, by making use of 
{\em Weyl operators} $X$ and $Z$ that generalize those of the qubit case [H. Weyl, Zeitschr. Phys. {\bf 46}, 1 (1927)].
That is, for a $d$-dimensional (qudit) system 
with basis $\{ |m \rangle\} \equiv \{\ket{0}, \cdots, \ket{d-1}\}$, let $X$ and $Z$ be defined by
$$   X \ket{m} \equiv  | (m + 1) \; \text{mod}\,d \rangle  
\quad \textrm{    and      } \quad Z \ket{m} = \zeta^m \ket{m}, 
\quad \zeta \equiv e^{ i  \frac{2 \pi}{d} }. $$ 
The desired operator basis may then be constructed by considering the set of generalized Pauli operators 
$\{ \sigma_{(a,b)} \} \equiv \{ Z^a X^b \}$, with 
$a,b \in  \{0, \ldots, d-1\}$, leading to $Z^a X^b = \zeta^{a b} X^b Z^a$. 
In infinite dimension, Weyl operators may be similarly defined by letting
$Z\equiv e^{\frac{i}{\hbar} \hat{q}}$, $X\equiv e^{- \frac{i}{\hbar} \hat{p}}$, where $\hat{q}$ and $\hat{p}$ are the 
canonical position and momentum operators ($[\hat{q},\hat{p}] = i \hbar$) and 
$\zeta = e^\frac{i}{\hbar}$, respectively. 
Regardless of dimensionality, then, if $\{ P_a\} \equiv \{Z^a\}$ or $\{X^a\}$ for example, 
each element of the basis is a dephasing-preserving (invertible) operator, as claimed.

\vspace*{1mm}

$\bullet$ {\bf Frequency domain.} 
Fourier-transforming to the frequency domain, the generalized cumulants of the relevant Hamiltonian, i.e., 
$C^{(1)}(\tilde{H}_O(s_1))$ and $C^{(1)}(\tilde{H}_O(s_1)\tilde{H}_O(s_2))$, can be written in terms of the FF formalism, 
specifically:

\begin{align*}
\int_{-t}^t d s_1 C^{(1)}(\tilde{H}_O(s_1)) &= \int_{-t}^t d s_1 \langle \tilde{H}_O(s_1) \rangle_{c,q}\\
& = \sum_{a,a'} P_a \int_{0}^t ds_1 \, \Big [{y}_{a,a'} (s_1) - \sum_{b} V_{a,a''}  {y}_{a'',a'} (s_1)\Big] \langle B_{a'(s_1)}\rangle_{c,q}\\
&= \sum_a P_a  \int_{-\infty}^\infty  \frac{d\omega}{ 2 \pi} \Big[ {F}^{(1)}_{a,a'} (\omega,t)  - \sum_b V_{a,a''} {F}^{(1)}_{a'',a'} (\omega,t) \Big] C^{(1)} (B_{a'} (\omega)),\\
\int_{-t}^t d s_1 \int_{-t}^{s_1} d s_2 C^{(2)}(\tilde{H}_O(s_1)\tilde{H}_O(s_2)) &= \int_{-t}^t d s_1 \int_{-t}^{s_1} d s_2 \Big( \langle \tilde{H}_O(s_1)\tilde{H}_O(s_2)\rangle_{c,q} - \langle \tilde{H}_O(s_1)\rangle_{c,q} \langle \tilde{H}_O(s_2)\rangle_{c,q} \Big)\\ 
& = \sum_{a,b, a',b'} P_a P_b \Big \{ \int_{-\infty}^\infty  \frac{d\omega}{ 2 \pi}  \Big [ F^{(2)}_{a,a'; b,b'} (\omega,T) \\
& + \sum_{a'',b''}  V_{a,a''} V_{b,b''}  F^{(2)}_{b'',b'; a'',a'} (-\omega,T) \\
& - \sum_{a''} V_{a, a''} F^{(1)}_{a'',a'} (\omega,T) F^{(1)}_{b,b'} (-\omega,T) \Big] S_{a',b'} (\omega)\Big \} ,
 \end{align*}
where the relevant FFs have expressions similar to Eqs. (\ref{eq::F1})-(\ref{eq::F2}) in the main text. Specializing to the $N$-qubit setting considered in the main text, for the case where $\{P_a\} \equiv \{Z_a\}$, operators $O$ such that $V_{a,a''} = \delta_{a,a''} \textrm{sign}(O,a,0)$, and Gaussian stationary noise with $C^{(1)}(t) \equiv 0$, leads directly to Eqs. (\ref{eq::CO22}) and \eqref{C2} quoted therein.

\vspace*{1mm}

$\bullet$ {\bf Expansion coefficients. } Equations (\ref{ddc1})-(\ref{ddc6}) of the main text gave explicit forms of the expansion coefficients in the two-qubit case for diagonal control. Here we give the most general form of these expansion coefficients, valid for non-diagonal control in the two-qubit case. Writing $\bar{\ell} =\{1,2\} - \{\ell\}$ and using shorthand notation 
$\text{sign}(O,a,0) \equiv f_a^O \in \{-1,1\}$, we find   
\begin{align*}
\mathcal{C}^{}_{0,O}(t) &= -\frac{1}{2} \sum_{\ell,a',b'\ignore{ \in \{1,2,12\}}}   \int_{-\infty}^\infty \frac{d\omega}{2 \pi} S^{(f^{O}_{\ell}f^{O}_{\ell})}_{a',b'} (\omega) \Big( {f^{O}_{\ell}} G_{\ell,a';\ell,b'}^{(1)} (\omega,t) - G_{\ell,a';\ell,b'}^{(2, +)} (\omega,t) \Big), \\
\mathcal{C}^{}_{\ell,O}(t) &= -\frac{1}{2} \sum_{a',b'\ignore{ \in \{1,2,12\}}}   \int_{-\infty}^\infty \frac{d\omega}{2 \pi} S^{(f^{O}_{\ell})}_{a',b'} (\omega) \Big( {f^{O}_{\ell}} G_{\ell,a';0,b'}^{(1)} (\omega,t) - G_{\ell,a';0,b'}^{(2, f^{O}_{\ell})} (\omega,t) + {} G_{0,a';\ell,b'}^{(1)} (\omega,t) - G_{0,a';\ell,b'}^{(2, f^{O}_{\ell})} (\omega,t)\Big)\\
& \,\,\,\,-\frac{1}{2} \sum_{a',b'\ignore{ \in \{1,2,12\}}}   \int_{-\infty}^\infty \frac{d\omega}{2 \pi} S^{(f^{O}_{\bar{\ell}} f^{O}_{12})}_{a',b'} (\omega) \Big( {f^{O}_{\bar{\ell}}} G_{\bar{\ell},a';12,b'}^{(1)} (\omega,t) - G_{\bar{\ell},a';12,b'}^{(2, f^{O}_{\bar{\ell}}f^{O}_{12})} (\omega,t) + {f^{O}_{12}} G_{12,a';\bar{\ell},b'}^{(1)} (\omega,t) - G_{12,a';\bar{\ell},b'}^{(2, f^{O}_{\bar{\ell}}f^{O}_{12})} (\omega,t)\Big),\\
\mathcal{C}^{}_{12,O}(t) &= -\frac{1}{2} \sum_{a',b'\ignore{ \in \{1,2,12\}}}   \int_{-\infty}^\infty \frac{d\omega}{2 \pi} S^{(f^{O}_{1}f^{O}_{2})}_{a',b'} (\omega) \Big( {f^{O}_{1}} G_{1,a';2,b'}^{(1)} (\omega,t) - G_{1,a';2,b'}^{(2, f^{O}_{1}f^{O}_{2})} (\omega,t) +{f^{O}_{2}} G_{2,a';1,b'}^{(1)} (\omega,t) - G_{2,a';1,b'}^{(2, f^{O}_{1}f^{O}_{2})} (\omega,t)\Big) \\
&\,\,\,\, -\frac{1}{2} \sum_{a',b'\ignore{ \in \{1,2,12\}}}   \int_{-\infty}^\infty \frac{d\omega}{2 \pi} S^{(f^{O}_{12})}_{a',b'} (\omega) \Big( {f^{O}_{12}} G_{12,a';0,b'}^{(1)} (\omega,t) - G_{12,a';0,b'}^{(2, f^{O}_{12})} (\omega,t) +{} G_{0,a';12,b'}^{(1)} (\omega,t) - G_{0,a';12,b'}^{(2, f^{O}_{12})} (\omega,t)\Big),
\end{align*}
where we used $f^{O}_0 =1$. These expressions can be simplified further, as we did to obtain Eqs.~\eqref{ddc1}-\eqref{ddc6}, by using the symmetry properties obeyed by spectra [Eq.~(\ref{eq::spsymmetry})]
and by noting that  
\begin{align*}
(G^{\pm}_{a,a';b,b'} (\omega,T))^* & = \pm G^{\pm}_{b,b';a,a'} (\omega,T) = G^{\pm}_{a,a';b,b'} (-\omega,T)\\
G^{+}_{a,a';b,b'} (\omega,T) \pm  G^{+}_{a,a';b,b'} (\omega,T) &= ( 1 \pm 1) G^{+}_{a,a';b,b'} (\omega,T),\\
G^{+}_{a,a';b,b'} (\omega,T) \pm  G^{-}_{a,a';b,b'} (\omega,T) &= ( 1 \pm 1) F^{(2)}_{a,a';b,b'} (\omega,T) + (1 \mp 1) F^{(2)}_{a,a';b,b'} (-\omega,T).
\end{align*}

\section{Noise spectroscopy via continuous driving}
\label{sec::spinlocking}

In the steady-state solution of Eqs. (\ref{eq::rhodotplus}) and (\ref{eq::rhodotminus}), the dependence of the populations on the bath spectrum suggests the possibility of performing noise spectroscopy with an off-axis driving term. In fact, the Hamiltonian in Eq.~(\ref{eq::relaxationH}) with $c=0$ is related to the spin-locking Hamiltonian utilized for spectroscopy and sensing applications in NMR and other platforms \cite{SpinLocking}.
Spin-locking techniques have also been employed for noise spectroscopy, though mainly on classical noise sources \cite{Yan2013,Bylander2011}. Reference \cite{Yan2013}, for example, uses spin locking to characterize the flux and tunnel-coupling noise affecting a superconducting qubit in a temperature regime where the noise is effectively classical. Spin-locking approaches can be extended to quantum noise sources, which we outline below. First note, however, that the qubit dynamics under equation Eq.~(\ref{eq::relaxationH}) are not exactly solvable. This necessitates the assumption of weak coupling or other approximations, which may not be applicable to the system at hand. With one additional qubit, our protocols can characterize the quantum spectra in a pure dephasing setting, without an off-axis driving term. The advantage of this setting is that we can solve for the reduced dynamics of the qubits \emph{exactly}, without relying on approximations that limit the portability of the protocol.

Consider a qubit with the internal Hamiltonian $H_0=\hbar\omega_0 Z/2$. In a spin-locking experiment, the qubit is subject to continuous driving along an axis that rotates about $Z$ with frequency $\omega_0$, resonant with the qubit's internal energy splitting.  When the qubit is transformed into the interaction picture associated with $H_0$, the traditional spin-locking setting is equivalent to Eq.~(\ref{eq::relaxationH}) with $c=0$. To make use of our previous results, we choose to work with Eq.~(\ref{eq::relaxationH}), rather than the spin-locking Hamiltonian in the lab frame. At long times, the equations of motion for the populations in Eqs. (\ref{eq::rhodotplus}) and (\ref{eq::rhodotminus}) become 
\begin{align}
\label{eq::rhoppSL}
\dot{\rho}_{++} &=-S_{1,1}(-g)\rho_{++}+S_{1,1}(g)\rho_{--},\\
\label{eq::rhommSL}
\dot{\rho}_{--}&=S_{1,1}(-g)\rho_{++}-S_{1,1}(g)\rho_{--}.
\end{align}
The spectral asymmetry dictates the difference between the rate of emission, $\Gamma_{+-}=S_{1,1}(-g)$, and absorption, $\Gamma_{-+}=S_{1,1}(g)$. If the qubit is initially prepared in $\ket{+}$, we can solve Eqs. (\ref{eq::rhoppSL}) and (\ref{eq::rhommSL}) to obtain
\begin{align}\label{eq::popplus}
\rho_{--}(t)&=\frac{-\Gamma_{+-}\text{exp}[-(\Gamma_{+-}+\Gamma_{-+})t]+\Gamma_{+-}}{\Gamma_{+-}+\Gamma_{-+}},\\
\rho_{++}(t)&=\frac{\Gamma_{+-}\text{exp}[-(\Gamma_{+-}+\Gamma_{-+})t]+\Gamma_{-+}}{\Gamma_{+-}+\Gamma_{-+}}.\label{eq::popminus}
\end{align}
Experimentally measuring the populations at different $t$ and fitting the results to the population curves in Eqs. (\ref{eq::popplus}) and (\ref{eq::popminus}), determines the rates of emission and absorption, producing estimates of $S_{1,1}(g)$ and $S_{1,1}(-g)$. Repeating this process for different values of the drive amplitude $g$ gives access to the spectrum at a range of frequencies.

\section{Protocol for noise spectroscopy of exciton qubits} 
\label{sec::procedureExciton}

The general spectroscopy procedure described in Sec.~\ref{sec::GenProcedure} is readily adapted to exciton qubits coupled to a phonon bath. Before delving into details, however, it should be emphasized that there is substantial freedom in how the procedure is implemented. The state preparation of the qubits, the control sequences to be applied, the number of repetitions and the measured observables should be selected according to the system in consideration.  In particular, the control sequences presented in this example are not intended to be a ``one size fits all" solution. In some platforms, for instance, the strength of the coupling between the qubits and the bath may so large that higher-order control sequences are required. Ultimately, the specifics of the spectroscopy protocol will vary from platform to platform.

Here, we show how our QNS protocol can be used to reconstruct the spectra 
$S^+_{11}(\omega)$, $S^+_{22}(\omega)$, $\text{Re}[S_{12}^+(\omega)]$, $\text{Im}[S_{12}^+(\omega)]$, $\text{Re}[S_{12}^-(\omega)]$ 
and $\text{Im}[S_{12}^-(\omega)]$, using only local control.
Recall from Sec.~\ref{sec::results} that the quantum self-spectra, $S_{1,1}^-(\omega)$ and 
$S_{2,2}^-(\omega)$, can be estimated from reconstructions of the other spectra with prior knowledge that the bath is bosonic. In the first stage of the procedure, we reconstruct the spectra at 32 non-zero harmonics, $\{\omega_0,\ldots,32\omega_0\}$, where $\omega_0=2\pi/T$. To accomplish this, we apply repetitions of base control sequences with cycle times $\{T, T/2,\ldots,T/32\}$, where the maximum cycle time is chosen to be $T=60$ ps.  All sequences are constrained by the minimal switching time $\tau_0=0.2$ ps. The shorter the cycle time, the more repetitions of the base sequence can be applied before the qubit significantly decoheres. We apply 7 repetitions for sequences with the largest cycle time, $T$. For sequences with the cycle times $T/2$ and $T/3$, we apply 15 repetitions. For the remainder of the cycle times, we apply 20 repetitions.  

In the second stage of the procedure, we estimate the spectra at $\omega=0$. The spectra $\text{Im}[S_{12}^+(\omega)]$, $\text{Re}[S_{12}^-(\omega)]$, $S_{1,1}^-(\omega)$ and $S_{2,2}^-(\omega)$ are odd functions, necessarily zero at $\omega=0$. The remaining even spectra can be reconstructed at $\omega=0$ by using a base sequence with zero filter order \cite{Paz2014}, which produces a FF that is non-zero at 
$\omega=0$. Measuring specific qubit observables after repetitions of this base sequence combined with knowledge of the non-zero harmonics enables us infer the spectra at $\omega=0$.

$\bullet$ Consider first the classical self-spectra $S_{11}^+(\omega)$ and $S_{22}^+(\omega)$, which enter the qubit dynamics through
\begin{align*}
\mathcal{C}_{{}{12},0}(t) =\frac{1}{2\pi}\int_{-\infty}^\infty d\omega&\big[G_{1,1;1,1}^{+}(\omega,t)S^+_{11}(\omega)+G_{2,2;2,2}^{+}(\omega,t)S^+_{22}(\omega)\big].
\end{align*}
From \erf{eq::c5}, this expansion coefficient can be obtained experimentally by preparing the qubits in the initial state 
$|\psi_{12}\rangle_= |+\rangle_1 \otimes |+\rangle_2$ 
and measuring the observables $X_1X_2$, $Y_1Y_2$, $X_1Y_2$ and $Y_1X_2$. After state preparation, the following 
$64$ control sequences are applied:
\\\\(1) CPMG on qubit 1 and CPMG on qubit 2 with cycle times $\{T,T/2,\ldots,T/32\}$
\\\\(2) CDD$_3$ on qubit 1 and CPMG on qubit 2 with cycle times $\{T,T/2,\ldots,T/32\}$.
\\\\
Let $\mathcal{C}_{{}{12},0}^{(i,n)}(t)$ denote the expansion coefficient $\mathcal{C}_{{}{12},0}(t)$ measured after $M_n$ repetitions of the sequence $i$ with cycle time $T/n$. From this point forward, the superscript $(i,n)$  will always denote an expansion coefficient measured after $M_n$ repetitions of sequence $i$ with cycle time $T/n$, i.e., at time $t=M_n T/n$. By invoking the frequency comb approximation, we can write the $\mathcal{C}_{{}{12},0}^{(i,n)}(M_nT/n)$ as linear equations 
\begin{align*}
\mathcal{C}_{{}{12},0}^{(1,n)}(M_nT/n)\!\simeq
&\frac{nM_n}{T}\sum_{k=1}^{32}\Big[|F^{(1)}_{\text{\scriptsize{CPMG}}}(k\omega_0,T/n)|^2S_{1,1}^+(k\omega_0)
+|F^{(1)}_{\text{\scriptsize{CPMG}}}(k\omega_0,T_p/n)|^2S_{2,2}^+(k\omega_0)\Big] ,\notag\\\notag\\
\mathcal{C}_{{}{12},0}^{(2,n)}(M_nT/n)\!\simeq
&\frac{nM_n}{T}\sum_{k=1}^{32}\Big[|F^{(1)}_{\text{\scriptsize{CDD}}_3}(k\omega_0,T/n)|^2S_{1,1}^+(k\omega_0)
+|F^{(1)}_{\text{\scriptsize{CPMG}}}(k\omega_0,T/n)|^2S_{2,2}^+(k\omega_0)\Big]\notag.
\end{align*}
The values of the $\mathcal{C}_{{}{12},0}^{(i,n)}(M_nT/n)$ for each of the 32 cycle times $\{T, T/2,\ldots,T/32\}$ form a system of linear equations. By taking $\mathcal{C}_{{}{12},0}^{(1,n)}(M_nT/n)-\mathcal{C}_{{}{12},0}^{(2,n)}(M_nT/n)$ and inverting the resulting linear system, we can solve for $S_{1,1}^{+}(\omega)$ at the 32 harmonics $\{\omega_0,\ldots,32\omega_0\}$. The classical self-spectrum, $S_{2,2}^{+}(\omega)$, is obtained either by repeating the procedure with sequence (2)  replaced by CDD$_3$ on qubit 2 and CPMG on qubit 1, or by substituting the reconstruction of $S_{1,1}^{+}(\omega)$ into one of the equations above and solving for $S_{2,2}^{+}(\omega)$.

$\bullet$ The classical cross-spectra $\text{Re}[S_{1,2}^+(\omega)]$ and $\text{Im}[S_{1,2}^+(\omega)]$ enter the dynamics through
\begin{align*}
\mathcal{C}_{{}{12},12}(t) =\frac{1}{2\pi}\int_{-\infty}^\infty d\omega&\big[G_{1,1;2,2}^{+}(\omega,t)S_{1,2}^+(\omega)
+G_{2,2;1,1}^{+}(\omega,t)S_{2,1}^+(\omega)\big].
\end{align*}
Like $\mathcal{C}_{{}{12},0}(t)$, this expansion coefficient can also be obtained by preparing the qubits in the initial state $|\psi\rangle_{12}$ and measuring the observables $X_1X_2$, $Y_1Y_2$, $X_1Y_2$ and $Y_1X_2$. We use the control sequences:
\\\\(1) CDD$_3$ on qubit 1 and CDD$_1$ on qubit 2 with cycle times $\{T,T/2,\ldots,T/32\}$,
\\\\(2) CDD$_3$ on qubit 1 and CPMG on qubit 2 with cycle times $\{T,T/2,\ldots,T/32\}$.
\\\\
Note that sequence (1) above is product-displacement $-$-symmetric in $[0,T/n]$ but product-displacement $-$-antisymmetric in $[0,T/2n]$ for each $n\in\{1,\ldots,32\}$. Sequence 2, on the other hand, is product-displacement $-$-symmetric in both $[0,T/n]$ and $[0,T/2n]$ for each $n$. As a consequence of Eqs. (\ref{eq::prodsym1}) and (\ref{eq::prodsym2}), the FF produced by (1) is purely real, while the FF produced by (2) is purely imaginary, enabling us to easily obtain both the real and imaginary components of $S_{1,2}^+(\omega)$. Determining $\mathcal{C}_{{}{12},12}^{(1,n)}(M_nT/n)$ and $\mathcal{C}_{{}{12},12}^{(2,n)}(M_nT/n)$ for all $n$, and making the frequency comb approximation produces now the linear equations 
\begin{align*}
\mathcal{C}_{{}{12},12}^{(1,n)}(M_nT/n)\simeq
&\frac{2nM_n}{T}\sum_{k=1}^{32}F^{(1)}_{\text{\scriptsize{CDD}}_3}(k\omega_0,T/n)
 F^{(1)}_{\text{\scriptsize{CDD}}_1}(-k\omega_0,T/n)\text{Re}[S_{1,2}^+(k\omega_0)], \\\\
\mathcal{C}_{{}{12},12}^{(2,n)}(M_nT/n)\simeq
&\frac{2nM_n}{T}\sum_{k=1}^{32}F^{(1)}_{\text{\scriptsize{CDD}}_3}(k\omega_0,T/n)
 F^{(1)}_{\text{\scriptsize{CPMG}}}(-k\omega_0,T/n) \text{Im}[S_{1,2}^+(k\omega_0)],
\end{align*}
which are inverted to obtain $\text{Im}[S_{1,2}^{+}(\omega)]$ and $\text{Re}[S_{1,2}^{+}(\omega)]$ at the harmonics 
$\{\omega_0,\ldots,32\omega_0\}$.

$\bullet$ Next, we turn to the quantum cross-spectra, $\text{Re}[S_{1,2}^{-}(\omega)]$ and $\text{Im}[S_{1,2}^{-}(\omega)]$, which 
enter the dynamics through 
\begin{align*}
\mathcal{C}_{{}1,12}(t)=\frac{1}{4\pi}\int_{-\infty}^\infty d\omega\big[G_{1,1;2,2}^{+}(\omega,t)
+G_{1,1;2,2}^{-}(\omega,t)\big]S_{12}^-(\omega)+\frac{1}{4\pi}\int_{-\infty}^\infty d\omega\big[-G_{2,2;1,1}^{+}(\omega,t)
+G_{2,2;1,1}^{-}(\omega,t)\big]S_{21}^-(\omega),
\end{align*}
\begin{align*}
\mathcal{C}_{{}2,12}(t)=\frac{1}{4\pi}\int_{-\infty}^\infty d\omega\big[-G_{1,1;2,2}^{+}(\omega,t)
+G_{1,1;2,2}^{-}(\omega,t)\big]S_{12}^-(\omega)+\frac{1}{4\pi}\int_{-\infty}^\infty d\omega\big[G_{2,2;1,1}^{+}(\omega,t)
+G_{2,2;1,1}^{-}(\omega,t)\big]S_{21}^-(\omega)
\end{align*}
From \erf{eq::c3}, the expansion coefficient $\mathcal{C}_{{}1,12}(t)$ can be obtained by preparing the qubits in 
$|\psi_1^\pm \rangle  = |+\rangle_1 \otimes |\pm z\rangle_2$
and measuring $X_1$. Similarly, $\mathcal{C}_{{}2,12}(t)$ can be extracted by preparing the qubits in 
$|\psi_2^\pm\rangle = |\pm z\rangle_1 \otimes |+\rangle_2$
and measuring $X_2$. Alternatively, these expansion coefficients can be accessed by preparing the qubits in $|\psi_{12}\rangle
=|++\rangle$ 
and measuring $Y_1Z_2$, $Z_1Y_2$, $X_1$ and $X_2$, since
\begin{align}
&\mathcal{C}_{{}1,12}(t)=\text{tan}^{-1}\left[\frac{E(Y_1Z_2)}{E(X_1)}\right]\;\;\;\text{and}\;\;\;
\mathcal{C}_{{}2,12}(t)=\text{tan}^{-1}\left[\frac{E(Z_1Y_2)}{E(X_2)}\right].
\end{align}
To reconstruct $\text{Im}[S^{-}_{1,2}(\omega)]$, we apply the control sequence
\\\\(1) CPMG on qubit 1 and CPMG on qubit 2 with cycle times $\{T,T/2,\ldots,T/32\}$.\\\\
This sequence creates a frequency comb in the FF $G^{+}$, producing the system of linear equations  
\begin{align*}
\mathcal{C}_{{}1,12}^{(1,n)}(M_nT/n)\simeq&\frac{2nM_n}{T}\sum_{k=1}^{32}|F^{(1)}_{\text{\scriptsize{CPMG}}}(k\omega_0,T/n)|^2\text{Im}[S^{-}_{1,2}(k\omega_0)]+C[G^{-}],\\
\mathcal{C}_{{}2,12}^{(1,n)}(M_nT/n)\simeq&-\frac{2nM_n}{T}\sum_{k=1}^{32}|F^{(1)}_{\text{\scriptsize{CPMG}}}(k\omega_0,T/n)|^2\text{Im}[S^{-}_{1,2}(k\omega_0)]+C[G^{-}],
\end{align*}
where $C[G^{-}]$ denotes the contribution from the second-order FFs. By inverting the system of linear equations formed by 
taking $\mathcal{C}_{{}1,12}^{(1,n)}(M_nT/n)-\mathcal{C}_{{}2,12}^{(1,n)}(M_nT/n)$, we can solve for $\text{Im}[S^{-}_{1,2}(\omega)]$ 
at $\{\omega_0,\ldots,32\omega_0\}$. Next, we apply the control sequence
\\\\(2) Two repetitions of CDD$_1$ (CDD$_1\!\!\times\!2$) on qubit 1 and a single repetition of CDD$_1$ on qubit 2 with cycle times 
$\{T,T/2,\ldots,T/32\}$.\\\\
This sequence, which is displacement-product $+$-antisymmetric in $[0,T/n]$, creates a frequency comb in the $G^{-}$ FFs. 
Through the comb, we obtain the system of linear equations 
\begin{align*}
\mathcal{C}_{{}1,12}^{(2,n)}(M_nT/n)\simeq&\frac{2n}{T}\sum_{k=1}^{32}(-1)^kF^{(1)}_{\text{\scriptsize{CDD}}_1\!\times\!2}(k\omega_0,T/n)F^{(1)}_{\text{\scriptsize{CDD}}_1}(-k\omega_0,T_p/n)\text{Re}[S^{-}_{1,2}(k\omega_0)]+C[G^{+}] ,\\
\mathcal{C}_{{}2,12}^{(2,n)}(M_nT/n)\simeq&\frac{2n}{T}\sum_{k=1}^{32}(-1)^kF^{(1)}_{\text{\scriptsize{CDD}}_1\!\times\!2}(k\omega_0,T/n)F^{(1)}_{\text{\scriptsize{CDD}}_1}(-k\omega_0,T_p/n)\text{Re}[S^{-}_{1,2}(k\omega_0)]
-C[G^{+}],
\end{align*}
where $C[G^{+}]$ denotes the contribution from the first-order 
FFs. Taking $\mathcal{C}_{{}1,12}^{(1,n)}+\mathcal{C}_{{}2,12}^{(1,n)}$ and inverting the resulting system of linear equations 
determines $\text{Re}[S^{-}_{1,2}(\omega)]$ at $\{\omega_0,\ldots,32\omega_0\}$.

$\bullet$ Our final task is reconstructing  $S_{11}^+(\omega)$, $S_{22}^+(\omega)$, $\text{Re}[S_{12}^+(\omega])$ and $\text{Im}[S_{12}^-(\omega)]$ at $\omega=0$. This requires a control sequence with FO$=0$, which produces a FF that is nonzero at $\omega=0$.
This may be achieved by appending segments of free evolution to DD sequences with non-zero FO, as in 
\cite{Norris2016}. We use such a sequence, which we term  ``uneven-CDD$_1$" or ``$\neq$CDD$_1$".  This sequence is described by the control propagator $U_{\text{ctrl}}(T)=U_f(31T/32)X_\ell U_f(T/32)$, where $U_f$ denotes free evolution. Thus, $\neq$CDD$_1$
is CDD$_1$ with a time duration $T/16$ followed by free evolution for a time $15T/16$. 

To reconstruct $S_{11}^+(\omega=0)$, we again consider the expansion coefficient $\mathcal{C}_{{}{12},0}(t)$.  We apply $M=35$ repetitions of the sequences
\\\\(1) CPMG on qubit 1 and CPMG on qubit 2 with cycle time $T/16$.
\\\\(2) $\neq$CDD$_1$ on qubit 1 and CPMG on qubit 2 with cycle time $T/16$.\\\\
The frequency comb approximation produces the linear equations
\begin{align}
\mathcal{C}_{{}{12},0}^{(1,16)}(M_nT/n)\!\simeq
&\frac{16M}{T}\sum_{k=1}^{32}\Big[|F^{(1)}_{\text{\scriptsize{CPMG}}}(k\omega_0,T/16)|^2S_{1,1}^+(k\omega_0)+|F^{(1)}_{\text{\scriptsize{CPMG}}}(k\omega_0,T/16)|^2S_{2,2}^+(k\omega_0)\Big]\notag,\\
\mathcal{C}_{{}{12},0}^{(2,16)}(M_nT/n)\!\simeq
&\frac{16M}{T}\sum_{k=0}^{32}\Big[|F^{(1)}_{\neq\text{\scriptsize{CDD}}_1}(k\omega_0,T/16)|^2S_{1,1}^+(k\omega_0)
+|F^{(1)}_{\text{\scriptsize{CPMG}}}(k\omega_0,T/16)|^2S_{2,2}^+(k\omega_0)\Big]\notag.
\end{align}
By taking $\mathcal{C}_{{}{12},0}^{(1,16)}(M_nT/n)-\mathcal{C}_{{}{12},0}^{(2,16)}(M_nT/n)$ and substituting the previously estimated values of $S_{1,1}^+(k\omega_0)$ for $1\leq k\leq 32$, we solve for $S_{1,1}^+(0)$.  Similarly, by applying $\neq$CDD$_1$ to qubit 2 and CPMG to qubit 1 in sequence (2), we can obtain $S_{2,2}^+(0)$.

Reconstructing $\text{Re}[S_{12}^+(\omega=0)]$ requires the expansion coefficient $\mathcal{C}_{{}{12},12}(t)$. We use $M=35$ repetitions of the control sequence
\\\\(1) $\neq$CDD$_1$ on qubit 1 and $\neq$CDD$_1$  on qubit 2 with cycle time $T/16$.\\\\
Through the frequency comb approximation, we obtain the linear equation
\begin{align*}
\mathcal{C}_{{}{12},12}^{(1,16)}(M_nT/n)\simeq
&\frac{32M}{T}\sum_{k=0}^{32}|F^{(1)}_{\neq\text{\scriptsize{CDD}}_1}(k\omega_0,T/16)|^2\text{Re}[S_{1,2}^+(k\omega_0)].
\end{align*}
By substituting into this equation the previously estimated values of $\text{Re}[S_{1,2}^+(k\omega_0)]$ for $1\leq k\leq 32$, we can solve for $\text{Re}[S_{1,2}^+(0)]$.

Lastly, we reconstruct $\text{Im}[S_{12}^-(\omega=0)]$ through $\mathcal{C}_{{}1,12}(t)$ and $\mathcal{C}_{{}2,12}(t)$. We again apply $M=35$ repetitions of the control sequence 
\\\\(1) $\neq$CDD$_1$ on qubit 1 and $\neq$CDD$_1$  on qubit 2 with cycle time $T/16$.\\\\
This produces the linear equations
\begin{align*}
\mathcal{C}_{{}1,12}^{(1,16)}(M_nT/n)\simeq&\frac{32M}{T}\sum_{k=0}^{32}|F^{(1)}_{\neq\text{\scriptsize{CDD}}_1}(k\omega_0,T/16)|^2\text{Im}[S^{-}_{1,2}(k\omega_0)]+C[G^{-}],\\
\mathcal{C}_{{}2,12}^{(1,16)}(M_nT/n)\simeq&-\frac{32M}{T}\sum_{k=0}^{32}|F^{(1)}_{\neq\text{\scriptsize{CDD}}_1}(k\omega_0,T/16)|^2\text{Im}[S^{-}_{1,2}(k\omega_0)]+C[G^{-}],
\end{align*}
By taking $\mathcal{C}_{{}1,12}^{(1,16)}(M_nT/n)-\mathcal{C}_{{}2,12}^{(1,16)}(M_nT/n)$ and substituting the estimates for $\text{Im}[S^{-}_{1,2}(k\omega_0)]$ for $1\leq k\leq 32$ into this expression, we obtain $\text{Im}[S^{-}_{1,2}(0)]$.

\end{appendix}

\twocolumngrid


\begin{thebibliography}{58}%
\makeatletter
\providecommand \@ifxundefined [1]{%
 \@ifx{#1\undefined}
}%
\providecommand \@ifnum [1]{%
 \ifnum #1\expandafter \@firstoftwo
 \else \expandafter \@secondoftwo
 \fi
}%
\providecommand \@ifx [1]{%
 \ifx #1\expandafter \@firstoftwo
 \else \expandafter \@secondoftwo
 \fi
}%
\providecommand \natexlab [1]{#1}%
\providecommand \enquote  [1]{``#1''}%
\providecommand \bibnamefont  [1]{#1}%
\providecommand \bibfnamefont [1]{#1}%
\providecommand \citenamefont [1]{#1}%
\providecommand \href@noop [0]{\@secondoftwo}%
\providecommand \href [0]{\begingroup \@sanitize@url \@href}%
\providecommand \@href[1]{\@@startlink{#1}\@@href}%
\providecommand \@@href[1]{\endgroup#1\@@endlink}%
\providecommand \@sanitize@url [0]{\catcode `\\12\catcode `\$12\catcode
  `\&12\catcode `\#12\catcode `\^12\catcode `\_12\catcode `\%12\relax}%
\providecommand \@@startlink[1]{}%
\providecommand \@@endlink[0]{}%
\providecommand \url  [0]{\begingroup\@sanitize@url \@url }%
\providecommand \@url [1]{\endgroup\@href {#1}{\urlprefix }}%
\providecommand \urlprefix  [0]{URL }%
\providecommand \Eprint [0]{\href }%
\providecommand \doibase [0]{http://dx.doi.org/}%
\providecommand \selectlanguage [0]{\@gobble}%
\providecommand \bibinfo  [0]{\@secondoftwo}%
\providecommand \bibfield  [0]{\@secondoftwo}%
\providecommand \translation [1]{[#1]}%
\providecommand \BibitemOpen [0]{}%
\providecommand \bibitemStop [0]{}%
\providecommand \bibitemNoStop [0]{.\EOS\space}%
\providecommand \EOS [0]{\spacefactor3000\relax}%
\providecommand \BibitemShut  [1]{\csname bibitem#1\endcsname}%
\let\auto@bib@innerbib\@empty
\bibitem [{\citenamefont {Lidar}\ and\ \citenamefont
  {T.~A.~Brun}(2013)}]{Lidar:book}%
  \BibitemOpen
  \bibfield  {author} {\bibinfo {author} {\bibfnamefont {D.~A.}\ \bibnamefont
  {Lidar}}\ and\ \bibinfo {author} {\bibfnamefont {E.}~\bibnamefont
  {T.~A.~Brun}},\ }\href@noop {} {\emph {\bibinfo {title} {Quantum Error
  Correction}}}\ (\bibinfo  {publisher} {Oxford University Press},\ \bibinfo
  {address} {Oxford},\ \bibinfo {year} {2013})\BibitemShut {NoStop}%
\bibitem [{\citenamefont {Khodjasteh}\ and\ \citenamefont {Lidar}(2005)}]{CDD}%
  \BibitemOpen
  \bibfield  {author} {\bibinfo {author} {\bibfnamefont {K.}~\bibnamefont
  {Khodjasteh}}\ and\ \bibinfo {author} {\bibfnamefont {D.~A.}\ \bibnamefont
  {Lidar}},\ }\href {\doibase 10.1103/PhysRevLett.95.180501} {\bibfield
  {journal} {\bibinfo  {journal} {Phys. Rev. Lett.}\ }\textbf {\bibinfo
  {volume} {95}},\ \bibinfo {pages} {180501} (\bibinfo {year}
  {2005})}\BibitemShut {NoStop}%
\bibitem [{\citenamefont {Khodjasteh}\ \emph {et~al.}(2010)\citenamefont
  {Khodjasteh}, \citenamefont {Lidar},\ and\ \citenamefont {Viola}}]{DCG}%
  \BibitemOpen
  \bibfield  {author} {\bibinfo {author} {\bibfnamefont {K.}~\bibnamefont
  {Khodjasteh}}, \bibinfo {author} {\bibfnamefont {D.~A.}\ \bibnamefont
  {Lidar}}, \ and\ \bibinfo {author} {\bibfnamefont {L.}~\bibnamefont
  {Viola}},\ }\href {\doibase 10.1103/PhysRevLett.104.090501} {\bibfield
  {journal} {\bibinfo  {journal} {Phys. Rev. Lett.}\ }\textbf {\bibinfo
  {volume} {104}},\ \bibinfo {pages} {090501} (\bibinfo {year}
  {2010})}\BibitemShut {NoStop}%
\bibitem [{\citenamefont {Brif}\ \emph {et~al.}(2010)\citenamefont {Brif},
  \citenamefont {Chakrabarti},\ and\ \citenamefont {Rabitz}}]{OptCtrl}%
  \BibitemOpen
  \bibfield  {author} {\bibinfo {author} {\bibfnamefont {C.}~\bibnamefont
  {Brif}}, \bibinfo {author} {\bibfnamefont {R.}~\bibnamefont {Chakrabarti}}, \
  and\ \bibinfo {author} {\bibfnamefont {H.}~\bibnamefont {Rabitz}},\ }\href
  {http://stacks.iop.org/1367-2630/12/i=7/a=075008} {\bibfield  {journal}
  {\bibinfo  {journal} {New J. Phys.}\ }\textbf {\bibinfo {volume} {12}},\
  \bibinfo {pages} {075008} (\bibinfo {year} {2010})}\BibitemShut {NoStop}%
\bibitem [{\citenamefont {Glaser}\ \emph {et~al.}(2015)\citenamefont {Glaser},
  \citenamefont {Boscain}, \citenamefont {Calarco}, \citenamefont {Koch},
  \citenamefont {K\"{o}ckenberger}, \citenamefont {Kosloff}, \citenamefont
  {Kuprov}, \citenamefont {Luy}, \citenamefont {Schirmer}, \citenamefont
  {Schulte-Herbr\"{u}ggen}, \citenamefont {Sugny},\ and\ \citenamefont
  {Wilhelm}}]{Calarco}%
  \BibitemOpen
  \bibfield  {author} {\bibinfo {author} {\bibfnamefont {S.~J.}\ \bibnamefont
  {Glaser}}, \bibinfo {author} {\bibfnamefont {U.}~\bibnamefont {Boscain}},
  \bibinfo {author} {\bibfnamefont {T.}~\bibnamefont {Calarco}}, \bibinfo
  {author} {\bibfnamefont {C.~P.}\ \bibnamefont {Koch}}, \bibinfo {author}
  {\bibfnamefont {W.}~\bibnamefont {K\"{o}ckenberger}}, \bibinfo {author}
  {\bibfnamefont {R.}~\bibnamefont {Kosloff}}, \bibinfo {author} {\bibfnamefont
  {I.}~\bibnamefont {Kuprov}}, \bibinfo {author} {\bibfnamefont
  {B.}~\bibnamefont {Luy}}, \bibinfo {author} {\bibfnamefont {S.}~\bibnamefont
  {Schirmer}}, \bibinfo {author} {\bibfnamefont {T.}~\bibnamefont
  {Schulte-Herbr\"{u}ggen}}, \bibinfo {author} {\bibfnamefont {D.}~\bibnamefont
  {Sugny}}, \ and\ \bibinfo {author} {\bibfnamefont {F.~K.}\ \bibnamefont
  {Wilhelm}},\ }\href@noop {} {\bibfield  {journal} {\bibinfo  {journal} {Eur.
  Phys. J. D}\ }\textbf {\bibinfo {volume} {69}},\ \bibinfo {pages} {279}
  (\bibinfo {year} {2015})}\BibitemShut {NoStop}%
\bibitem [{\citenamefont {Giovannetti}\ \emph {et~al.}(2006)\citenamefont
  {Giovannetti}, \citenamefont {Lloyd},\ and\ \citenamefont
  {Maccone}}]{QMetro}%
  \BibitemOpen
  \bibfield  {author} {\bibinfo {author} {\bibfnamefont {V.}~\bibnamefont
  {Giovannetti}}, \bibinfo {author} {\bibfnamefont {S.}~\bibnamefont {Lloyd}},
  \ and\ \bibinfo {author} {\bibfnamefont {L.}~\bibnamefont {Maccone}},\ }\href
  {\doibase 10.1103/PhysRevLett.96.010401} {\bibfield  {journal} {\bibinfo
  {journal} {Phys. Rev. Lett.}\ }\textbf {\bibinfo {volume} {96}},\ \bibinfo
  {pages} {010401} (\bibinfo {year} {2006})}\BibitemShut {NoStop}%
\bibitem [{\citenamefont {Wasilewski}\ \emph {et~al.}(2010)\citenamefont
  {Wasilewski}, \citenamefont {Jensen}, \citenamefont {Krauter}, \citenamefont
  {Renema}, \citenamefont {Balabas},\ and\ \citenamefont {Polzik}}]{QMagneto}%
  \BibitemOpen
  \bibfield  {author} {\bibinfo {author} {\bibfnamefont {W.}~\bibnamefont
  {Wasilewski}}, \bibinfo {author} {\bibfnamefont {K.}~\bibnamefont {Jensen}},
  \bibinfo {author} {\bibfnamefont {H.}~\bibnamefont {Krauter}}, \bibinfo
  {author} {\bibfnamefont {J.~J.}\ \bibnamefont {Renema}}, \bibinfo {author}
  {\bibfnamefont {M.~V.}\ \bibnamefont {Balabas}}, \ and\ \bibinfo {author}
  {\bibfnamefont {E.~S.}\ \bibnamefont {Polzik}},\ }\href {\doibase
  10.1103/PhysRevLett.104.133601} {\bibfield  {journal} {\bibinfo  {journal}
  {Phys. Rev. Lett.}\ }\textbf {\bibinfo {volume} {104}},\ \bibinfo {pages}
  {133601} (\bibinfo {year} {2010})}\BibitemShut {NoStop}%
\bibitem [{\citenamefont {DeVience}\ \emph {et~al.}(2015)\citenamefont
  {DeVience}, \citenamefont {Pham}, \citenamefont {Lovchinsky}, \citenamefont
  {Sushkov}, \citenamefont {Bar-Gill}, \citenamefont {Belthangady},
  \citenamefont {Casola}, \citenamefont {Corbett}, \citenamefont {Zhang},
  \citenamefont {Lukin}, \citenamefont {Park}, \citenamefont {Yacoby},\ and\
  \citenamefont {Walsworth}}]{Misha}%
  \BibitemOpen
  \bibfield  {author} {\bibinfo {author} {\bibfnamefont {S.~J.}\ \bibnamefont
  {DeVience}}, \bibinfo {author} {\bibfnamefont {L.~M.}\ \bibnamefont {Pham}},
  \bibinfo {author} {\bibfnamefont {I.}~\bibnamefont {Lovchinsky}}, \bibinfo
  {author} {\bibfnamefont {A.~O.}\ \bibnamefont {Sushkov}}, \bibinfo {author}
  {\bibfnamefont {N.}~\bibnamefont {Bar-Gill}}, \bibinfo {author}
  {\bibfnamefont {C.}~\bibnamefont {Belthangady}}, \bibinfo {author}
  {\bibfnamefont {F.}~\bibnamefont {Casola}}, \bibinfo {author} {\bibfnamefont
  {M.}~\bibnamefont {Corbett}}, \bibinfo {author} {\bibfnamefont
  {H.}~\bibnamefont {Zhang}}, \bibinfo {author} {\bibfnamefont
  {M.}~\bibnamefont {Lukin}}, \bibinfo {author} {\bibfnamefont
  {H.}~\bibnamefont {Park}}, \bibinfo {author} {\bibfnamefont {A.}~\bibnamefont
  {Yacoby}}, \ and\ \bibinfo {author} {\bibfnamefont {R.~L.}\ \bibnamefont
  {Walsworth}},\ }\href@noop {} {\bibfield  {journal} {\bibinfo  {journal}
  {Nature Nanotech.}\ }\textbf {\bibinfo {volume} {10}},\ \bibinfo {pages}
  {129} (\bibinfo {year} {2015})}\BibitemShut {NoStop}%
\bibitem [{Dat()}]{DattaReview}%
  \BibitemOpen
  \href@noop {} {}\bibinfo {note} {M. Szczykulska, T. Baumgratz, and A. Datta,
  e-print arXiv:1604.02615.}\BibitemShut {Stop}%
\bibitem [{\citenamefont {Schoelkopf}\ \emph {et~al.}(2003)\citenamefont
  {Schoelkopf}, \citenamefont {Clerk}, \citenamefont {Girvin}, \citenamefont
  {Lehnert},\ and\ \citenamefont {Devoret}}]{QSpectroOld}%
  \BibitemOpen
  \bibfield  {author} {\bibinfo {author} {\bibfnamefont {R.~J.}\ \bibnamefont
  {Schoelkopf}}, \bibinfo {author} {\bibfnamefont {A.~A.}\ \bibnamefont
  {Clerk}}, \bibinfo {author} {\bibfnamefont {S.~M.}\ \bibnamefont {Girvin}},
  \bibinfo {author} {\bibfnamefont {K.~W.}\ \bibnamefont {Lehnert}}, \ and\
  \bibinfo {author} {\bibfnamefont {M.~H.}\ \bibnamefont {Devoret}},\ }\enquote
  {\bibinfo {title} {Qubits as spectrometers of quantum noise},}\ in\ \href
  {\doibase 10.1007/978-94-010-0089-5_9} {\emph {\bibinfo {booktitle} {Quantum
  Noise in Mesoscopic Physics}}},\ \bibinfo {editor} {edited by\ \bibinfo
  {editor} {\bibfnamefont {Y.~V.}\ \bibnamefont {Nazarov}}}\ (\bibinfo
  {publisher} {Springer Netherlands},\ \bibinfo {address} {Dordrecht},\
  \bibinfo {year} {2003})\ pp.\ \bibinfo {pages} {175--203}\BibitemShut
  {NoStop}%
\bibitem [{\citenamefont {Faoro}\ and\ \citenamefont {Viola}(2004)}]{Lara2004}%
  \BibitemOpen
  \bibfield  {author} {\bibinfo {author} {\bibfnamefont {L.}~\bibnamefont
  {Faoro}}\ and\ \bibinfo {author} {\bibfnamefont {L.}~\bibnamefont {Viola}},\
  }\href {\doibase 10.1103/PhysRevLett.92.117905} {\bibfield  {journal}
  {\bibinfo  {journal} {Phys. Rev. Lett.}\ }\textbf {\bibinfo {volume} {92}},\
  \bibinfo {pages} {117905} (\bibinfo {year} {2004})}\BibitemShut {NoStop}%
\bibitem [{Old()}]{OlderRefs}%
  \BibitemOpen
  \href@noop {} {}\bibinfo {note} {T. Yuge, S. Sasaki, and Y. Hirayama, {\em
  ibid.} {\bf 107}, 170504 (2011); K. C. Young and K. B. Whaley, Phys. Rev. A
  {\bf 86}, 012314 (2012).}\BibitemShut {Stop}%
\bibitem [{\citenamefont {\'Alvarez}\ and\ \citenamefont
  {Suter}(2011)}]{Alvarez2011}%
  \BibitemOpen
  \bibfield  {author} {\bibinfo {author} {\bibfnamefont {G.~A.}\ \bibnamefont
  {\'Alvarez}}\ and\ \bibinfo {author} {\bibfnamefont {D.}~\bibnamefont
  {Suter}},\ }\href@noop {} {\bibfield  {journal} {\bibinfo  {journal} {Phys.
  Rev. Lett.}\ }\textbf {\bibinfo {volume} {107}},\ \bibinfo {pages} {230501}
  (\bibinfo {year} {2011})}\BibitemShut {NoStop}%
\bibitem [{\citenamefont {Norris}\ \emph {et~al.}(2016)\citenamefont {Norris},
  \citenamefont {Paz-Silva},\ and\ \citenamefont {Viola}}]{Norris2016}%
  \BibitemOpen
  \bibfield  {author} {\bibinfo {author} {\bibfnamefont {L.~M.}\ \bibnamefont
  {Norris}}, \bibinfo {author} {\bibfnamefont {G.~A.}\ \bibnamefont
  {Paz-Silva}}, \ and\ \bibinfo {author} {\bibfnamefont {L.}~\bibnamefont
  {Viola}},\ }\href@noop {} {\bibfield  {journal} {\bibinfo  {journal} {Phys.
  Rev. Lett.}\ }\textbf {\bibinfo {volume} {116}},\ \bibinfo {pages} {150503}
  (\bibinfo {year} {2016})}\BibitemShut {NoStop}%
\bibitem [{\citenamefont {Kofman}\ and\ \citenamefont
  {Kurizki}(2001)}]{Kurizki}%
  \BibitemOpen
  \bibfield  {author} {\bibinfo {author} {\bibfnamefont {A.~G.}\ \bibnamefont
  {Kofman}}\ and\ \bibinfo {author} {\bibfnamefont {G.}~\bibnamefont
  {Kurizki}},\ }\href {\doibase 10.1103/PhysRevLett.87.270405} {\bibfield
  {journal} {\bibinfo  {journal} {Phys. Rev. Lett.}\ }\textbf {\bibinfo
  {volume} {87}},\ \bibinfo {pages} {270405} (\bibinfo {year}
  {2001})}\BibitemShut {NoStop}%
\bibitem [{Mik()}]{MikeFF}%
  \BibitemOpen
  \href@noop {} {}\bibinfo {note} {T. J. Green, H. Uys, and M. J. Biercuk,
  Phys. Rev. Lett. {\bf 109}, 020501 (2012); A. Soare, H. Ball, D. Hayes, J.
  Sastrawan, M. C. Jarratt, J. J. McLoughlin, X. Zhen, T. J. Green, and M. J.
  Biercuk, Nature Phys. {\bf 10}, 825 (2014).}\BibitemShut {Stop}%
\bibitem [{\citenamefont {Paz-Silva}\ and\ \citenamefont
  {Viola}(2014)}]{Paz2014}%
  \BibitemOpen
  \bibfield  {author} {\bibinfo {author} {\bibfnamefont {G.~A.}\ \bibnamefont
  {Paz-Silva}}\ and\ \bibinfo {author} {\bibfnamefont {L.}~\bibnamefont
  {Viola}},\ }\href@noop {} {\bibfield  {journal} {\bibinfo  {journal} {Phys.
  Rev. Lett.}\ }\textbf {\bibinfo {volume} {113}},\ \bibinfo {pages} {250501}
  (\bibinfo {year} {2014})}\BibitemShut {NoStop}%
\bibitem [{\citenamefont {Breuer}\ and\ \citenamefont
  {Petruccione}(2002)}]{Breuer:book}%
  \BibitemOpen
  \bibfield  {author} {\bibinfo {author} {\bibfnamefont {H.-P.}\ \bibnamefont
  {Breuer}}\ and\ \bibinfo {author} {\bibfnamefont {F.}~\bibnamefont
  {Petruccione}},\ }\href@noop {} {\emph {\bibinfo {title} {The Theory of Open
  Quantum Systems}}}\ (\bibinfo  {publisher} {Oxford University Press},\
  \bibinfo {address} {Oxford},\ \bibinfo {year} {2002})\BibitemShut {NoStop}%
\bibitem [{\citenamefont {Bylander}\ \emph {et~al.}(2011)\citenamefont
  {Bylander}, \citenamefont {Gustavsson}, \citenamefont {Yan}, \citenamefont
  {Yoshihara}, \citenamefont {Harrabi}, \citenamefont {Fitch}, \citenamefont
  {Cory}, \citenamefont {Nakamura}, \citenamefont {Tsai},\ and\ \citenamefont
  {Oliver}}]{Bylander2011}%
  \BibitemOpen
  \bibfield  {author} {\bibinfo {author} {\bibfnamefont {J.}~\bibnamefont
  {Bylander}}, \bibinfo {author} {\bibfnamefont {S.}~\bibnamefont
  {Gustavsson}}, \bibinfo {author} {\bibfnamefont {F.}~\bibnamefont {Yan}},
  \bibinfo {author} {\bibfnamefont {F.}~\bibnamefont {Yoshihara}}, \bibinfo
  {author} {\bibfnamefont {K.}~\bibnamefont {Harrabi}}, \bibinfo {author}
  {\bibfnamefont {G.}~\bibnamefont {Fitch}}, \bibinfo {author} {\bibfnamefont
  {D.~G.}\ \bibnamefont {Cory}}, \bibinfo {author} {\bibfnamefont
  {Y.}~\bibnamefont {Nakamura}}, \bibinfo {author} {\bibfnamefont {J.-S.}\
  \bibnamefont {Tsai}}, \ and\ \bibinfo {author} {\bibfnamefont {W.~D.}\
  \bibnamefont {Oliver}},\ }\href@noop {} {\bibfield  {journal} {\bibinfo
  {journal} {Nature Phys.}\ }\textbf {\bibinfo {volume} {7}},\ \bibinfo {pages}
  {565} (\bibinfo {year} {2011})}\BibitemShut {NoStop}%
\bibitem [{\citenamefont {Muhonen}\ \emph {et~al.}(2014)\citenamefont
  {Muhonen}, \citenamefont {Dehollain}, \citenamefont {Laucht}, \citenamefont
  {Hudson}, \citenamefont {Kalra}, \citenamefont {Sekiguchi}, \citenamefont
  {Itoh}, \citenamefont {Jamieson}, \citenamefont {McCallum}, \citenamefont
  {Dzurak},\ and\ \citenamefont {Morello}}]{Muhonen2014}%
  \BibitemOpen
  \bibfield  {author} {\bibinfo {author} {\bibfnamefont {J.~T.}\ \bibnamefont
  {Muhonen}}, \bibinfo {author} {\bibfnamefont {J.~P.}\ \bibnamefont
  {Dehollain}}, \bibinfo {author} {\bibfnamefont {A.}~\bibnamefont {Laucht}},
  \bibinfo {author} {\bibfnamefont {F.~E.}\ \bibnamefont {Hudson}}, \bibinfo
  {author} {\bibfnamefont {R.}~\bibnamefont {Kalra}}, \bibinfo {author}
  {\bibfnamefont {T.}~\bibnamefont {Sekiguchi}}, \bibinfo {author}
  {\bibfnamefont {K.~M.}\ \bibnamefont {Itoh}}, \bibinfo {author}
  {\bibfnamefont {D.~N.}\ \bibnamefont {Jamieson}}, \bibinfo {author}
  {\bibfnamefont {J.~C.}\ \bibnamefont {McCallum}}, \bibinfo {author}
  {\bibfnamefont {A.~S.}\ \bibnamefont {Dzurak}}, \ and\ \bibinfo {author}
  {\bibfnamefont {A.}~\bibnamefont {Morello}},\ }\href@noop {} {\bibfield
  {journal} {\bibinfo  {journal} {Nature Nanotech.}\ }\textbf {\bibinfo
  {volume} {9}},\ \bibinfo {pages} {986} (\bibinfo {year} {2014})}\BibitemShut
  {NoStop}%
\bibitem [{Oli()}]{Oliver}%
  \BibitemOpen
  \href@noop {} {}\bibinfo {note} {F. Yoshihara, Y. Nakamura, F. Yan, S.
  Gustavsson, J. Bylander, W. D. Oliver, and J.-S. Tsai, Phys. Rev. B {\bf 89},
  020503 (2014).}\BibitemShut {Stop}%
\bibitem [{\citenamefont {Yan}\ \emph {et~al.}(2013)\citenamefont {Yan},
  \citenamefont {Gustavsson}, \citenamefont {Bylander}, \citenamefont {Jin},
  \citenamefont {Yoshihara}, \citenamefont {Cory}, \citenamefont {Nakamura},
  \citenamefont {Orlando},\ and\ \citenamefont {Oliver}}]{Yan2013}%
  \BibitemOpen
  \bibfield  {author} {\bibinfo {author} {\bibfnamefont {F.}~\bibnamefont
  {Yan}}, \bibinfo {author} {\bibfnamefont {S.}~\bibnamefont {Gustavsson}},
  \bibinfo {author} {\bibfnamefont {J.}~\bibnamefont {Bylander}}, \bibinfo
  {author} {\bibfnamefont {X.}~\bibnamefont {Jin}}, \bibinfo {author}
  {\bibfnamefont {F.}~\bibnamefont {Yoshihara}}, \bibinfo {author}
  {\bibfnamefont {D.~G.}\ \bibnamefont {Cory}}, \bibinfo {author}
  {\bibfnamefont {Y.}~\bibnamefont {Nakamura}}, \bibinfo {author}
  {\bibfnamefont {T.~P.}\ \bibnamefont {Orlando}}, \ and\ \bibinfo {author}
  {\bibfnamefont {W.~D.}\ \bibnamefont {Oliver}},\ }\href@noop {} {\bibfield
  {journal} {\bibinfo  {journal} {Nature Commun.}\ }\textbf {\bibinfo {volume}
  {4}} (\bibinfo {year} {2013})}\BibitemShut {NoStop}%
\bibitem [{\citenamefont {Dial}\ \emph {et~al.}(2013)\citenamefont {Dial},
  \citenamefont {Shulman}, \citenamefont {Harvey}, \citenamefont {Bluhm},
  \citenamefont {Umansky},\ and\ \citenamefont {Yacoby}}]{SpinQubits}%
  \BibitemOpen
  \bibfield  {author} {\bibinfo {author} {\bibfnamefont {O.~E.}\ \bibnamefont
  {Dial}}, \bibinfo {author} {\bibfnamefont {M.~D.}\ \bibnamefont {Shulman}},
  \bibinfo {author} {\bibfnamefont {S.~P.}\ \bibnamefont {Harvey}}, \bibinfo
  {author} {\bibfnamefont {H.}~\bibnamefont {Bluhm}}, \bibinfo {author}
  {\bibfnamefont {V.}~\bibnamefont {Umansky}}, \ and\ \bibinfo {author}
  {\bibfnamefont {A.}~\bibnamefont {Yacoby}},\ }\href {\doibase
  10.1103/PhysRevLett.110.146804} {\bibfield  {journal} {\bibinfo  {journal}
  {Phys. Rev. Lett.}\ }\textbf {\bibinfo {volume} {110}},\ \bibinfo {pages}
  {146804} (\bibinfo {year} {2013})}\BibitemShut {NoStop}%
\bibitem [{NVs()}]{NVs}%
  \BibitemOpen
  \href@noop {} {}\bibinfo {note} {C. A. Meriles, L. Jiang, G. Goldstein, J. S.
  Hodges, J. Maze, M. D. Lukin, and P. Cappellaro, J. Chem. Phys. {\bf 133},
  124105 (2010); N. Zhao, J. Wrachtrup, and R.-B. Liu, Phys. Rev. A {\bf 90},
  032319 (2014); Y. Romach, C. Muller, T. Unden, L. J. Rogers, T. Isoda, K. M.
  Itoh, M. Markham, A. Stacey, J. Meijer, S. Pezzagna, B. Naydenov, L. P.
  McGuinness, N. Bar-Gill, and F. Jelezko, Phys. Rev. Lett. {\bf 114}, 017601
  (2015).}\BibitemShut {Stop}%
\bibitem [{\citenamefont {Kotler}\ \emph {et~al.}(2013)\citenamefont {Kotler},
  \citenamefont {Akerman}, \citenamefont {Glickman},\ and\ \citenamefont
  {Ozeri}}]{Kotler}%
  \BibitemOpen
  \bibfield  {author} {\bibinfo {author} {\bibfnamefont {S.}~\bibnamefont
  {Kotler}}, \bibinfo {author} {\bibfnamefont {N.}~\bibnamefont {Akerman}},
  \bibinfo {author} {\bibfnamefont {Y.}~\bibnamefont {Glickman}}, \ and\
  \bibinfo {author} {\bibfnamefont {R.}~\bibnamefont {Ozeri}},\ }\href@noop {}
  {\bibfield  {journal} {\bibinfo  {journal} {Phys. Rev. Lett.}\ }\textbf
  {\bibinfo {volume} {110}},\ \bibinfo {pages} {110503} (\bibinfo {year}
  {2013})}\BibitemShut {NoStop}%
\bibitem [{\citenamefont {Percival}\ and\ \citenamefont
  {Walden}(1993)}]{Percival}%
  \BibitemOpen
  \bibfield  {author} {\bibinfo {author} {\bibfnamefont {D.~B.}\ \bibnamefont
  {Percival}}\ and\ \bibinfo {author} {\bibfnamefont {A.~T.}\ \bibnamefont
  {Walden}},\ }\href@noop {} {\emph {\bibinfo {title} {Spectral Analysis for
  Physical Applications}}}\ (\bibinfo  {publisher} {Cambridge University
  Press},\ \bibinfo {address} {Cambridge},\ \bibinfo {year} {1993})\BibitemShut
  {NoStop}%
\bibitem [{\citenamefont {Sza$\acute{\textrm{n}}$kowski}\ \emph
  {et~al.}(2016)\citenamefont {Sza$\acute{\textrm{n}}$kowski}, \citenamefont
  {Trippenbach},\ and\ \citenamefont
  {Cywi$\acute{\textrm{n}}$ski}}]{Cywinski2}%
  \BibitemOpen
  \bibfield  {author} {\bibinfo {author} {\bibfnamefont {P.}~\bibnamefont
  {Sza$\acute{\textrm{n}}$kowski}}, \bibinfo {author} {\bibfnamefont
  {M.}~\bibnamefont {Trippenbach}}, \ and\ \bibinfo {author} {\bibfnamefont
  {L.}~\bibnamefont {Cywi$\acute{\textrm{n}}$ski}},\ }\href@noop {} {\bibfield
  {journal} {\bibinfo  {journal} {Phys. Rev. A}\ }\textbf {\bibinfo {volume}
  {94}},\ \bibinfo {pages} {012109} (\bibinfo {year} {2016})}\BibitemShut
  {NoStop}%
\bibitem [{\citenamefont {Mostame}\ \emph {et~al.}(2012)\citenamefont
  {Mostame}, \citenamefont {Rebentrost}, \citenamefont {Eisfeld}, \citenamefont
  {Kerman}, \citenamefont {Tsomokos},\ and\ \citenamefont
  {Aspuru-Guzik}}]{Mostame}%
  \BibitemOpen
  \bibfield  {author} {\bibinfo {author} {\bibfnamefont {S.}~\bibnamefont
  {Mostame}}, \bibinfo {author} {\bibfnamefont {P.}~\bibnamefont {Rebentrost}},
  \bibinfo {author} {\bibfnamefont {A.}~\bibnamefont {Eisfeld}}, \bibinfo
  {author} {\bibfnamefont {A.}~\bibnamefont {Kerman}}, \bibinfo {author}
  {\bibfnamefont {D.~I.}\ \bibnamefont {Tsomokos}}, \ and\ \bibinfo {author}
  {\bibfnamefont {A.}~\bibnamefont {Aspuru-Guzik}},\ }\href@noop {} {\bibfield
  {journal} {\bibinfo  {journal} {New J. Phys.}\ }\textbf {\bibinfo {volume}
  {14}},\ \bibinfo {pages} {105013} (\bibinfo {year} {2012})}\BibitemShut
  {NoStop}%
\bibitem [{\citenamefont {Haeberlein}\ \emph {et~al.}(2015)\citenamefont
  {Haeberlein}, \citenamefont {Deppe}, \citenamefont {Kurcz}, \citenamefont
  {Goetz}, \citenamefont {Baust}, \citenamefont {Eder}, \citenamefont
  {Fedorov}, \citenamefont {Fischer}, \citenamefont {Menzel}, \citenamefont
  {Schwarz}, \citenamefont {Wulschner}, \citenamefont {Xie}, \citenamefont
  {Zhong}, \citenamefont {Solano}, \citenamefont {Marx}, \citenamefont
  {Garc�a-Ripoll},\ and\ \citenamefont {Gross}}]{Gross}%
  \BibitemOpen
  \bibfield  {author} {\bibinfo {author} {\bibfnamefont {M.}~\bibnamefont
  {Haeberlein}}, \bibinfo {author} {\bibfnamefont {F.}~\bibnamefont {Deppe}},
  \bibinfo {author} {\bibfnamefont {A.}~\bibnamefont {Kurcz}}, \bibinfo
  {author} {\bibfnamefont {J.}~\bibnamefont {Goetz}}, \bibinfo {author}
  {\bibfnamefont {A.}~\bibnamefont {Baust}}, \bibinfo {author} {\bibfnamefont
  {P.}~\bibnamefont {Eder}}, \bibinfo {author} {\bibfnamefont {K.}~\bibnamefont
  {Fedorov}}, \bibinfo {author} {\bibfnamefont {M.}~\bibnamefont {Fischer}},
  \bibinfo {author} {\bibfnamefont {E.~P.}\ \bibnamefont {Menzel}}, \bibinfo
  {author} {\bibfnamefont {M.~J.}\ \bibnamefont {Schwarz}}, \bibinfo {author}
  {\bibfnamefont {F.}~\bibnamefont {Wulschner}}, \bibinfo {author}
  {\bibfnamefont {E.}~\bibnamefont {Xie}}, \bibinfo {author} {\bibfnamefont
  {L.}~\bibnamefont {Zhong}}, \bibinfo {author} {\bibfnamefont
  {E.}~\bibnamefont {Solano}}, \bibinfo {author} {\bibfnamefont
  {A.}~\bibnamefont {Marx}}, \bibinfo {author} {\bibfnamefont {J.-J.}\
  \bibnamefont {Garc�a-Ripoll}}, \ and\ \bibinfo {author} {\bibfnamefont
  {R.}~\bibnamefont {Gross}},\ }\href@noop {} {\bibfield  {journal} {\bibinfo
  {journal} {arXiv:1506.09114}\ } (\bibinfo {year} {2015})}\BibitemShut
  {NoStop}%
\bibitem [{\citenamefont {Braun}(2002)}]{Braun2002}%
  \BibitemOpen
  \bibfield  {author} {\bibinfo {author} {\bibfnamefont {D.}~\bibnamefont
  {Braun}},\ }\href {\doibase 10.1103/PhysRevLett.89.277901} {\bibfield
  {journal} {\bibinfo  {journal} {Phys. Rev. Lett.}\ }\textbf {\bibinfo
  {volume} {89}},\ \bibinfo {pages} {277901} (\bibinfo {year}
  {2002})}\BibitemShut {NoStop}%
\bibitem [{\citenamefont {Benatti}\ \emph {et~al.}(2003)\citenamefont
  {Benatti}, \citenamefont {Floreanini},\ and\ \citenamefont
  {Piani}}]{Benatti2003}%
  \BibitemOpen
  \bibfield  {author} {\bibinfo {author} {\bibfnamefont {F.}~\bibnamefont
  {Benatti}}, \bibinfo {author} {\bibfnamefont {R.}~\bibnamefont {Floreanini}},
  \ and\ \bibinfo {author} {\bibfnamefont {M.}~\bibnamefont {Piani}},\
  }\href@noop {} {\bibfield  {journal} {\bibinfo  {journal} {Phys. Rev. Lett.}\
  }\textbf {\bibinfo {volume} {91}},\ \bibinfo {pages} {070402} (\bibinfo
  {year} {2003})}\BibitemShut {NoStop}%
\bibitem [{\citenamefont {Oh}\ and\ \citenamefont {Kim}(2006)}]{Oh2006}%
  \BibitemOpen
  \bibfield  {author} {\bibinfo {author} {\bibfnamefont {S.}~\bibnamefont
  {Oh}}\ and\ \bibinfo {author} {\bibfnamefont {J.}~\bibnamefont {Kim}},\
  }\href {http://link.aps.org/doi/10.1103/PhysRevA.73.062306} {\bibfield
  {journal} {\bibinfo  {journal} {Phys. Rev. A}\ }\textbf {\bibinfo {volume}
  {73}},\ \bibinfo {pages} {062306} (\bibinfo {year} {2006})}\BibitemShut
  {NoStop}%
\bibitem [{\citenamefont {An}\ \emph {et~al.}(2007)\citenamefont {An},
  \citenamefont {Wang},\ and\ \citenamefont {Luo}}]{An2007}%
  \BibitemOpen
  \bibfield  {author} {\bibinfo {author} {\bibfnamefont {J.-H.}\ \bibnamefont
  {An}}, \bibinfo {author} {\bibfnamefont {S.-J.}\ \bibnamefont {Wang}}, \ and\
  \bibinfo {author} {\bibfnamefont {H.-G.}\ \bibnamefont {Luo}},\ }\href@noop
  {} {\bibfield  {journal} {\bibinfo  {journal} {Physica A}\ }\textbf {\bibinfo
  {volume} {382}},\ \bibinfo {pages} {753} (\bibinfo {year}
  {2007})}\BibitemShut {NoStop}%
\bibitem [{\citenamefont {McCutcheon}\ \emph {et~al.}(2009)\citenamefont
  {McCutcheon}, \citenamefont {Nazir}, \citenamefont {Bose},\ and\
  \citenamefont {Fisher}}]{McCutcheon2009}%
  \BibitemOpen
  \bibfield  {author} {\bibinfo {author} {\bibfnamefont {D.~P.~S.}\
  \bibnamefont {McCutcheon}}, \bibinfo {author} {\bibfnamefont
  {A.}~\bibnamefont {Nazir}}, \bibinfo {author} {\bibfnamefont
  {S.}~\bibnamefont {Bose}}, \ and\ \bibinfo {author} {\bibfnamefont {A.~J.}\
  \bibnamefont {Fisher}},\ }\href
  {http://link.aps.org/doi/10.1103/PhysRevA.80.022337} {\bibfield  {journal}
  {\bibinfo  {journal} {Phys. Rev. A}\ }\textbf {\bibinfo {volume} {80}},\
  \bibinfo {pages} {022337} (\bibinfo {year} {2009})}\BibitemShut {NoStop}%
\bibitem [{\citenamefont {Bhaktavatsala~Rao}\ \emph {et~al.}(2011)\citenamefont
  {Bhaktavatsala~Rao}, \citenamefont {Bar-Gill},\ and\ \citenamefont
  {Kurizki}}]{Bhaktavatsala2011}%
  \BibitemOpen
  \bibfield  {author} {\bibinfo {author} {\bibfnamefont {D.~D.}\ \bibnamefont
  {Bhaktavatsala~Rao}}, \bibinfo {author} {\bibfnamefont {N.}~\bibnamefont
  {Bar-Gill}}, \ and\ \bibinfo {author} {\bibfnamefont {G.}~\bibnamefont
  {Kurizki}},\ }\href {http://link.aps.org/doi/10.1103/PhysRevLett.106.010404}
  {\bibfield  {journal} {\bibinfo  {journal} {Phys. Rev. Lett.}\ }\textbf
  {\bibinfo {volume} {106}},\ \bibinfo {pages} {010404} (\bibinfo {year}
  {2011})}\BibitemShut {NoStop}%
\bibitem [{\citenamefont {Paz-Silva}\ \emph {et~al.}(2016)\citenamefont
  {Paz-Silva}, \citenamefont {Lee}, \citenamefont {Green},\ and\ \citenamefont
  {Viola}}]{multiDD}%
  \BibitemOpen
  \bibfield  {author} {\bibinfo {author} {\bibfnamefont {G.~A.}\ \bibnamefont
  {Paz-Silva}}, \bibinfo {author} {\bibfnamefont {S.-W.}\ \bibnamefont {Lee}},
  \bibinfo {author} {\bibfnamefont {T.~J.}\ \bibnamefont {Green}}, \ and\
  \bibinfo {author} {\bibfnamefont {L.}~\bibnamefont {Viola}},\ }\href
  {http://stacks.iop.org/1367-2630/18/i=7/a=073020} {\bibfield  {journal}
  {\bibinfo  {journal} {New J. Phys.}\ }\textbf {\bibinfo {volume} {18}},\
  \bibinfo {pages} {073020} (\bibinfo {year} {2016})}\BibitemShut {NoStop}%
\bibitem [{\citenamefont {Krzywda}\ and\ \citenamefont
  {Roszak}(2016)}]{Krzywda:2016}%
  \BibitemOpen
  \bibfield  {author} {\bibinfo {author} {\bibfnamefont {J.}~\bibnamefont
  {Krzywda}}\ and\ \bibinfo {author} {\bibfnamefont {K.}~\bibnamefont
  {Roszak}},\ }\href {http://dx.doi.org/10.1038/srep23753} {\bibfield
  {journal} {\bibinfo  {journal} {Sci. Rep.}\ }\textbf {\bibinfo {volume}
  {6}},\ \bibinfo {pages} {23753} (\bibinfo {year} {2016})}\BibitemShut
  {NoStop}%
\bibitem [{\citenamefont {Arenz}\ \emph {et~al.}(2016)\citenamefont {Arenz},
  \citenamefont {Burgarth}, \citenamefont {Facchi}, \citenamefont
  {Giovannetti}, \citenamefont {Nakazato}, \citenamefont {Pascazio},\ and\
  \citenamefont {Yuasa}}]{Arenz2016}%
  \BibitemOpen
  \bibfield  {author} {\bibinfo {author} {\bibfnamefont {C.}~\bibnamefont
  {Arenz}}, \bibinfo {author} {\bibfnamefont {D.}~\bibnamefont {Burgarth}},
  \bibinfo {author} {\bibfnamefont {P.}~\bibnamefont {Facchi}}, \bibinfo
  {author} {\bibfnamefont {V.}~\bibnamefont {Giovannetti}}, \bibinfo {author}
  {\bibfnamefont {H.}~\bibnamefont {Nakazato}}, \bibinfo {author}
  {\bibfnamefont {S.}~\bibnamefont {Pascazio}}, \ and\ \bibinfo {author}
  {\bibfnamefont {K.}~\bibnamefont {Yuasa}},\ }\href
  {http://link.aps.org/doi/10.1103/PhysRevA.93.062308} {\bibfield  {journal}
  {\bibinfo  {journal} {Phys. Rev. A}\ }\textbf {\bibinfo {volume} {93}},\
  \bibinfo {pages} {062308} (\bibinfo {year} {2016})}\BibitemShut {NoStop}%
\bibitem [{\citenamefont {Aharonov}\ and\ \citenamefont
  {Ben-Or}(2008)}]{FaultTolerance}%
  \BibitemOpen
  \bibfield  {author} {\bibinfo {author} {\bibfnamefont {D.}~\bibnamefont
  {Aharonov}}\ and\ \bibinfo {author} {\bibfnamefont {M.}~\bibnamefont
  {Ben-Or}},\ }\href@noop {} {\bibfield  {journal} {\bibinfo  {journal} {SIAM
  J. Comput.}\ }\textbf {\bibinfo {volume} {38}},\ \bibinfo {pages} {1207}
  (\bibinfo {year} {2008})}\BibitemShut {NoStop}%
\bibitem [{\citenamefont {Preskill}(2013)}]{PreskillSufficient}%
  \BibitemOpen
  \bibfield  {author} {\bibinfo {author} {\bibfnamefont {J.}~\bibnamefont
  {Preskill}},\ }\href@noop {} {\bibfield  {journal} {\bibinfo  {journal}
  {Quantum Inf. Comput.}\ }\textbf {\bibinfo {volume} {13}},\ \bibinfo {pages}
  {181} (\bibinfo {year} {2013})}\BibitemShut {NoStop}%
\bibitem [{\citenamefont {Ng}\ and\ \citenamefont
  {Preskill}(2009)}]{NgPreskill}%
  \BibitemOpen
  \bibfield  {author} {\bibinfo {author} {\bibfnamefont {H.~K.}\ \bibnamefont
  {Ng}}\ and\ \bibinfo {author} {\bibfnamefont {J.}~\bibnamefont {Preskill}},\
  }\href {\doibase 10.1103/PhysRevA.79.032318} {\bibfield  {journal} {\bibinfo
  {journal} {Phys. Rev. A}\ }\textbf {\bibinfo {volume} {79}},\ \bibinfo
  {pages} {032318} (\bibinfo {year} {2009})}\BibitemShut {NoStop}%
\bibitem [{Nov()}]{Novais}%
  \BibitemOpen
  \href@noop {} {}\bibinfo {note} {E. Novais and E. R. Mucciolo, Phys. Rev.
  Lett. {\bf 110}, 010502 (2013); P. Jouzdani, E. Novais, I. S. Tupitsyn, and
  E. R. Mucciolo, Phys. Rev. A {\bf 90}, 042315 (2014).}\BibitemShut {Stop}%
\bibitem [{\citenamefont {Hutter}\ and\ \citenamefont
  {Loss}(2014)}]{Hutter2014}%
  \BibitemOpen
  \bibfield  {author} {\bibinfo {author} {\bibfnamefont {A.}~\bibnamefont
  {Hutter}}\ and\ \bibinfo {author} {\bibfnamefont {D.}~\bibnamefont {Loss}},\
  }\href {\doibase 10.1103/PhysRevA.89.042334} {\bibfield  {journal} {\bibinfo
  {journal} {Phys. Rev. A}\ }\textbf {\bibinfo {volume} {89}},\ \bibinfo
  {pages} {042334} (\bibinfo {year} {2014})}\BibitemShut {NoStop}%
\bibitem [{\citenamefont {Jeske}\ \emph {et~al.}(2014)\citenamefont {Jeske},
  \citenamefont {Cole},\ and\ \citenamefont {Huelga}}]{Jeske2014}%
  \BibitemOpen
  \bibfield  {author} {\bibinfo {author} {\bibfnamefont {J.}~\bibnamefont
  {Jeske}}, \bibinfo {author} {\bibfnamefont {J.~H.}\ \bibnamefont {Cole}}, \
  and\ \bibinfo {author} {\bibfnamefont {S.~F.}\ \bibnamefont {Huelga}},\
  }\href {http://stacks.iop.org/1367-2630/16/i=7/a=073039} {\bibfield
  {journal} {\bibinfo  {journal} {New Journal of Physics}\ }\textbf {\bibinfo
  {volume} {16}},\ \bibinfo {pages} {073039} (\bibinfo {year}
  {2014})}\BibitemShut {NoStop}%
\bibitem [{\citenamefont {Sabin}\ \emph {et~al.}(2014)\citenamefont {Sabin},
  \citenamefont {White}, \citenamefont {Hackermuller},\ and\ \citenamefont
  {Fuentes}}]{Ivette}%
  \BibitemOpen
  \bibfield  {author} {\bibinfo {author} {\bibfnamefont {C.}~\bibnamefont
  {Sabin}}, \bibinfo {author} {\bibfnamefont {A.}~\bibnamefont {White}},
  \bibinfo {author} {\bibfnamefont {L.}~\bibnamefont {Hackermuller}}, \ and\
  \bibinfo {author} {\bibfnamefont {I.}~\bibnamefont {Fuentes}},\ }\href@noop
  {} {\bibfield  {journal} {\bibinfo  {journal} {Sci. Rep.}\ }\textbf {\bibinfo
  {volume} {4}},\ \bibinfo {pages} {6436} (\bibinfo {year} {2014})}\BibitemShut
  {NoStop}%
\bibitem [{\citenamefont {Correa}\ \emph {et~al.}(2015)\citenamefont {Correa},
  \citenamefont {Mehboudi}, \citenamefont {Adesso},\ and\ \citenamefont
  {Sanpera}}]{BosonicThermo}%
  \BibitemOpen
  \bibfield  {author} {\bibinfo {author} {\bibfnamefont {L.~A.}\ \bibnamefont
  {Correa}}, \bibinfo {author} {\bibfnamefont {M.}~\bibnamefont {Mehboudi}},
  \bibinfo {author} {\bibfnamefont {G.}~\bibnamefont {Adesso}}, \ and\ \bibinfo
  {author} {\bibfnamefont {A.}~\bibnamefont {Sanpera}},\ }\href {\doibase
  10.1103/PhysRevLett.114.220405} {\bibfield  {journal} {\bibinfo  {journal}
  {Phys. Rev. Lett.}\ }\textbf {\bibinfo {volume} {114}},\ \bibinfo {pages}
  {220405} (\bibinfo {year} {2015})}\BibitemShut {NoStop}%
\bibitem [{\citenamefont {Doll}\ \emph {et~al.}(2008)\citenamefont {Doll},
  \citenamefont {Zueco}, \citenamefont {Wubs}, \citenamefont {Kohler},\ and\
  \citenamefont {H{\"a}nggi}}]{Doll2008}%
  \BibitemOpen
  \bibfield  {author} {\bibinfo {author} {\bibfnamefont {R.}~\bibnamefont
  {Doll}}, \bibinfo {author} {\bibfnamefont {D.}~\bibnamefont {Zueco}},
  \bibinfo {author} {\bibfnamefont {M.}~\bibnamefont {Wubs}}, \bibinfo {author}
  {\bibfnamefont {S.}~\bibnamefont {Kohler}}, \ and\ \bibinfo {author}
  {\bibfnamefont {P.}~\bibnamefont {H{\"a}nggi}},\ }\href@noop {} {\bibfield
  {journal} {\bibinfo  {journal} {Chem. Phys.}\ }\textbf {\bibinfo {volume}
  {347}},\ \bibinfo {pages} {243} (\bibinfo {year} {2008})}\BibitemShut
  {NoStop}%
\bibitem [{\citenamefont {Leggett}\ \emph {et~al.}(1987)\citenamefont
  {Leggett}, \citenamefont {Chakravarty}, \citenamefont {Dorsey}, \citenamefont
  {Fisher}, \citenamefont {Garg},\ and\ \citenamefont {Zwerger}}]{Leggett}%
  \BibitemOpen
  \bibfield  {author} {\bibinfo {author} {\bibfnamefont {A.~J.}\ \bibnamefont
  {Leggett}}, \bibinfo {author} {\bibfnamefont {S.}~\bibnamefont
  {Chakravarty}}, \bibinfo {author} {\bibfnamefont {A.~T.}\ \bibnamefont
  {Dorsey}}, \bibinfo {author} {\bibfnamefont {M.~P.~A.}\ \bibnamefont
  {Fisher}}, \bibinfo {author} {\bibfnamefont {A.}~\bibnamefont {Garg}}, \ and\
  \bibinfo {author} {\bibfnamefont {W.}~\bibnamefont {Zwerger}},\ }\href
  {\doibase 10.1103/RevModPhys.59.1} {\bibfield  {journal} {\bibinfo  {journal}
  {Rev. Mod. Phys.}\ }\textbf {\bibinfo {volume} {59}},\ \bibinfo {pages} {1}
  (\bibinfo {year} {1987})}\BibitemShut {NoStop}%
\bibitem [{Hod()}]{Hodgson}%
  \BibitemOpen
  \href@noop {} {}\bibinfo {note} {T. E. Hodgson, L. Viola, and I. D'Amico,
  Phys. Rev. B {\bf 78}, 165311(2008); Phys. Rev. A {\bf 81}, 062321
  (2010).}\BibitemShut {Stop}%
\bibitem [{\citenamefont {Cotlet}\ and\ \citenamefont
  {Lovett}(2014)}]{Cotlet:2014}%
  \BibitemOpen
  \bibfield  {author} {\bibinfo {author} {\bibfnamefont {O.}~\bibnamefont
  {Cotlet}}\ and\ \bibinfo {author} {\bibfnamefont {B.~W.}\ \bibnamefont
  {Lovett}},\ }\href {http://stacks.iop.org/1367-2630/16/i=10/a=103016}
  {\bibfield  {journal} {\bibinfo  {journal} {New J. Phys.}\ }\textbf {\bibinfo
  {volume} {16}},\ \bibinfo {pages} {103016} (\bibinfo {year}
  {2014})}\BibitemShut {NoStop}%
\bibitem [{\citenamefont {Kubo}(1962)}]{Kubo}%
  \BibitemOpen
  \bibfield  {author} {\bibinfo {author} {\bibfnamefont {R.}~\bibnamefont
  {Kubo}},\ }\href@noop {} {\bibfield  {journal} {\bibinfo  {journal} {J. Phys.
  Soc. Japan}\ }\textbf {\bibinfo {volume} {17}},\ \bibinfo {pages} {1100}
  (\bibinfo {year} {1962})}\BibitemShut {NoStop}%
\bibitem [{\citenamefont {Kardar}(2007)}]{Kardar:book}%
  \BibitemOpen
  \bibfield  {author} {\bibinfo {author} {\bibfnamefont {M.}~\bibnamefont
  {Kardar}},\ }\href@noop {} {\emph {\bibinfo {title} {Statistical Physics of
  Fields}}}\ (\bibinfo  {publisher} {Oxford University Press},\ \bibinfo {year}
  {2007})\BibitemShut {NoStop}%
\bibitem [{Spi()}]{SpinLocking}%
  \BibitemOpen
  \href@noop {} {}\bibinfo {note} {C. P. Slichter and D. C. Ailion, Phys. Rev.
  {\bf 135}, A1099 (1964); D. C. Ailion and C. P. Slichter, {\em ibid.} {\bf
  137}, A235 (1965); D.C. Look and I. J. Lowe, J. Chem. Phys. {\bf 44}, 2995
  (1966); G. Ithier, E. Collin, P. Joyez, P. J. Meeson, D. Vion, D. Esteve, F.
  Chiarello, A. Shnirman, Y. Makhlin, J. Schriefl, and G. Sch\"on, Phys. Rev. B
  {\bf 72}, 134519 (2005); M. Loretz, T. Rosskopf, and C. L. Degen, Phys. Rev.
  Lett. {\bf 110}, 017602 (2013).}\BibitemShut {Stop}%
\bibitem [{\citenamefont {Viola}\ \emph {et~al.}(2000)\citenamefont {Viola},
  \citenamefont {Knill},\ and\ \citenamefont {Lloyd}}]{ViolaSymm}%
  \BibitemOpen
  \bibfield  {author} {\bibinfo {author} {\bibfnamefont {L.}~\bibnamefont
  {Viola}}, \bibinfo {author} {\bibfnamefont {E.}~\bibnamefont {Knill}}, \ and\
  \bibinfo {author} {\bibfnamefont {S.}~\bibnamefont {Lloyd}},\ }\href
  {\doibase 10.1103/PhysRevLett.85.3520} {\bibfield  {journal} {\bibinfo
  {journal} {Phys. Rev. Lett.}\ }\textbf {\bibinfo {volume} {85}},\ \bibinfo
  {pages} {3520} (\bibinfo {year} {2000})}\BibitemShut {NoStop}%
\bibitem [{\citenamefont {Wang}\ and\ \citenamefont {Liu}(2011)}]{NUDD}%
  \BibitemOpen
  \bibfield  {author} {\bibinfo {author} {\bibfnamefont {Z.-Y.}\ \bibnamefont
  {Wang}}\ and\ \bibinfo {author} {\bibfnamefont {R.-B.}\ \bibnamefont {Liu}},\
  }\href {\doibase 10.1103/PhysRevA.83.022306} {\bibfield  {journal} {\bibinfo
  {journal} {Phys. Rev. A}\ }\textbf {\bibinfo {volume} {83}},\ \bibinfo
  {pages} {022306} (\bibinfo {year} {2011})}\BibitemShut {NoStop}%
\bibitem [{\citenamefont {Khodjasteh}\ \emph {et~al.}(2011)\citenamefont
  {Khodjasteh}, \citenamefont {Dobrovitski},\ and\ \citenamefont
  {Viola}}]{Pointer}%
  \BibitemOpen
  \bibfield  {author} {\bibinfo {author} {\bibfnamefont {K.}~\bibnamefont
  {Khodjasteh}}, \bibinfo {author} {\bibfnamefont {V.~V.}\ \bibnamefont
  {Dobrovitski}}, \ and\ \bibinfo {author} {\bibfnamefont {L.}~\bibnamefont
  {Viola}},\ }\href {\doibase 10.1103/PhysRevA.84.022336} {\bibfield  {journal}
  {\bibinfo  {journal} {Phys. Rev. A}\ }\textbf {\bibinfo {volume} {84}},\
  \bibinfo {pages} {022336} (\bibinfo {year} {2011})}\BibitemShut {NoStop}%
\bibitem [{\citenamefont {Sasakura}\ \emph {et~al.}(2012)\citenamefont
  {Sasakura}, \citenamefont {Hermannst{\"a}dter}, \citenamefont {Dorenbos},
  \citenamefont {Akopian}, \citenamefont {van Kouwen}, \citenamefont
  {Motohisa}, \citenamefont {Kobayashi}, \citenamefont {Kumano}, \citenamefont
  {Kondo}, \citenamefont {Tomioka}, \citenamefont {Fukui}, \citenamefont
  {Suemune},\ and\ \citenamefont {Zwiller}}]{Sasakura:2012}%
  \BibitemOpen
  \bibfield  {author} {\bibinfo {author} {\bibfnamefont {H.}~\bibnamefont
  {Sasakura}}, \bibinfo {author} {\bibfnamefont {C.}~\bibnamefont
  {Hermannst{\"a}dter}}, \bibinfo {author} {\bibfnamefont {S.~N.}\ \bibnamefont
  {Dorenbos}}, \bibinfo {author} {\bibfnamefont {N.}~\bibnamefont {Akopian}},
  \bibinfo {author} {\bibfnamefont {M.~P.}\ \bibnamefont {van Kouwen}},
  \bibinfo {author} {\bibfnamefont {J.}~\bibnamefont {Motohisa}}, \bibinfo
  {author} {\bibfnamefont {Y.}~\bibnamefont {Kobayashi}}, \bibinfo {author}
  {\bibfnamefont {H.}~\bibnamefont {Kumano}}, \bibinfo {author} {\bibfnamefont
  {K.}~\bibnamefont {Kondo}}, \bibinfo {author} {\bibfnamefont
  {K.}~\bibnamefont {Tomioka}}, \bibinfo {author} {\bibfnamefont
  {T.}~\bibnamefont {Fukui}}, \bibinfo {author} {\bibfnamefont
  {I.}~\bibnamefont {Suemune}}, \ and\ \bibinfo {author} {\bibfnamefont
  {V.}~\bibnamefont {Zwiller}},\ }\href
  {http://link.aps.org/doi/10.1103/PhysRevB.85.075324} {\bibfield  {journal}
  {\bibinfo  {journal} {Phys. Rev. B}\ }\textbf {\bibinfo {volume} {85}},\
  \bibinfo {pages} {075324} (\bibinfo {year} {2012})}\BibitemShut {NoStop}%
\bibitem [{\citenamefont {Hofmann}\ \emph {et~al.}(2013)\citenamefont
  {Hofmann}, \citenamefont {Gluckert}, \citenamefont {Noe}, \citenamefont
  {Bourjau}, \citenamefont {Dehmel},\ and\ \citenamefont
  {Hogele}}]{Hofmann:2013}%
  \BibitemOpen
  \bibfield  {author} {\bibinfo {author} {\bibfnamefont {M.~S.}\ \bibnamefont
  {Hofmann}}, \bibinfo {author} {\bibfnamefont {J.~T.}\ \bibnamefont
  {Gluckert}}, \bibinfo {author} {\bibfnamefont {J.}~\bibnamefont {Noe}},
  \bibinfo {author} {\bibfnamefont {C.}~\bibnamefont {Bourjau}}, \bibinfo
  {author} {\bibfnamefont {R.}~\bibnamefont {Dehmel}}, \ and\ \bibinfo {author}
  {\bibfnamefont {A.}~\bibnamefont {Hogele}},\ }\href
  {http://dx.doi.org/10.1038/nnano.2013.119} {\bibfield  {journal} {\bibinfo
  {journal} {Nature Nanotech.}\ }\textbf {\bibinfo {volume} {8}},\ \bibinfo
  {pages} {502} (\bibinfo {year} {2013})}\BibitemShut {NoStop}%
\end{thebibliography}

%

\end{document}